\documentclass[%
 reprint,
 amsmath,amssymb,
 onecolumn,aps
]{revtex4-2}

\usepackage{xcolor}        % Per i colori
\usepackage{graphicx}      % Per includere immagini
\usepackage{tcolorbox}     % Per creare riquadri colorati
\usepackage{dcolumn}% Align table columns on decimal point
\usepackage{bm}% bold math
\usepackage{hyperref}% add hypertext capabilities
\usepackage{xcolor}
\usepackage[export]{adjustbox}
\usepackage{overpic}
\usepackage{tikz}
\usepackage{caption}
\usepackage{setspace}

\usepackage{newfloat} % Required for declaring new floating environments
\usepackage{soul} % Protegge comandi come \em e \emph
\newcommand{\R}{\mathbb{R}}

\newcommand{\rt}{\rho^{\text{T}}}
\newcommand{\rh}{\rho^{\text{H}}}
\newcommand{\hrt}{\hat{\rho}^{\text{T}}}
\newcommand{\hrh}{\hat{\rho}^{\text{H}}}
\newcommand{\drt}{\delta\rho^{\text{T}}}
\newcommand{\drh}{\delta\rho^\text{H}}
\newcommand{\ktt}{\tilde{k}^{\text{T}}}

\begin{document}

\preprint{APS/123-QED}
\definecolor{myspecialcolor}{RGB}{230, 230, 230} % Definisci un colore personalizzato
\title{Nonreciprocal field theory for decision-making in multi-agent control systems}

%%%%%% ALTERNATIVE TITLES

%Nonreciprocal field theory for decision-making in multi-agent control systems

%Nonreciprocal field theory for decision-making in multi-agent systems with application to shepherding control 

%Nonreciprocal continuum framework for control based decision-making in multi-agent systems

%(A) nonreciprocal continuum framework for decision-making (in multi-agent control systems)

%An interpretable continuum framework for decision-making with application to shepherding control problems

\author{Andrea Lama}
\email{andrea.lama-ssm@unina.it}
\affiliation{Modeling and Engineering Risk and Complexity, Scuola Superiore Meridionale, Naples, 80125, Italy}
\author{Mario di Bernardo}
\email{Correspondence should be addressed to S.K. (sabine.klapp@tu-berlin.de) and M.d.B. (mario.dibernardo@unina.it)}
\affiliation{Department of Electrical Engineering and ICT, 
University of Naples Federico II, Naples, 80125, Italy}
\author{Sabine.~H.~L.~Klapp$^\dag$}

\affiliation{Institute for Theoretical Physics, 
Technical University Berlin, 10623 Berlin, Germany}

\date{\today}%
%It is always \today, today,
%\onecolumngrid

\begin{abstract}
Field theories for complex systems traditionally focus on collective behaviors emerging from simple, reciprocal pairwise interaction rules. However, many natural and artificial systems exhibit behaviors driven by microscopic decision-making processes that introduce both nonreciprocity and many-body interactions, challenging these conventional approaches. We develop a theoretical framework to incorporate 
decision-making into field theories using the shepherding control problem as a paradigmatic example of a multi-agent control system, where agents (herders) must coordinate to confine another group of agents (targets) within a prescribed region. By introducing continuous approximations of two key decision-making elements - target selection and trajectory planning - we derive field equations that capture the essential features of this distributed control problem. Our theory reveals that different decision-making strategies emerge at the continuum level, from average attraction to highly selective choices, and from undirected to goal-oriented motion, driving transitions between homogeneous and confined configurations. The resulting nonreciprocal field theory not only describes the shepherding problem but provides a general framework for incorporating decision-making into continuum theories of collective behavior, with implications for applications ranging from robotic swarms to crowd management systems.
% Shepherding is a paradigmatic example of swarm intelligence, where a group of herders adaptively controls the dynamics of a group of targets to push them into a certain region in space.
% However, a systematic understanding of the emergent collective dynamics and pattern formation in case of many herders is, so far, elusive.
% Here we start from a stochastic agent-based model to derive a hydrodynamic field theory that allows us to explore the conditions for the appearance of a stationary shepherding state in the absence of any cohesive couplings.
% A central piece of our approach are build-in decision-making strategies where the herders {\em select} certain targets and push them along a desired direction. The resulting agent-based and hydrodynamic dynamical equations involve markedly asymmetric couplings, yielding an important conceptual link to recent research on nonreciprocal soft and active systems. Our results elucidate the role and crucial interplay of interactions, on one side, 
% and control, on the other side, to achieve the desired shepherding goal. The results are of relevance for a broad range of macroscopic biological and social systems, as well as in robotics.%
\end{abstract}

\maketitle

\section{Introduction}
\label{sec:intro}
Decision-making is a fundamental process underlying the dynamics of many complex systems in science and technology. At the macroscopic level, the emergent collective behavior of these systems is often shaped by decisions made by individual agents at the microscopic scale. Examples span from animal groups coordinating their movements \cite{sumpter2010collective,ballerini2008interaction}, to autonomous vehicles navigating traffic \cite{kuru2021planning}, to robotic swarms performing distributed tasks \cite{brambilla2013swarm,bullo2009distributed}
and decision processes in economics \cite{PhysRevLett.134.047402}.
Often these processes involve a predefined control goal. Understanding how such microscopic decisions translate into macroscopic behavior is crucial for the analysis, design and control of such systems \cite{d2023controlling}.
A versatile, and in many contexts highly efficient, tool to study emergent behaviors on large length and time scales are hydrodynamic continuum equations comprising a set of coupled partial different equations (PDEs). These not only allow to elucidate the type of instabilities and emerging dynamical states, they can also help to identify relevant mechanisms and parameters ranges in dynamical systems with complex microscopic interaction rules. 
Yet, the precise translation of microscopic decision rules into continuum descriptions remains a significant challenge in physics and control theory. 

Traditional field theories for complex systems often focus on collective behaviors arising from simple, typically
reciprocal pairwise interaction rules. Major examples from the physics community are critical phenomena and phase separation in fluid and magnetic systems \cite{hohenberg1977theory,cross_hohenberg_1993_pattern_formation_outside_equilibrium} and, more recently, active \cite{marchetti_simha_2013_hydrodynamics_soft_active_matter} and bio-matter \cite{juelicher2007active}. In the last years, much attention has been devoted to {\em nonreciprocal } generalizations \cite{you2020nonreciprocity,fruchart_2021_non-reciprocal_phase_transitions} characterized by asymmetrical couplings occuring, e.g., in 
heterogeneous bacterial systems \cite{xiong_tsimring_2020_flower-like_patterns_multi-species_bacteria,theveneau_mayor_2013_chase-and-run_cells,meredith_zarzar_2020_predator-prey_droplets},
synthetic colloidal mixtures in nonequilibrium environments \cite{Ivlev_2015_statistical_mechanics_where_newtons_third_law_is_broken,saha_golestanian_2019_pairing_waltzing_scattering_chemotactic_active_colloids},
macroscopic predator-prey systems \cite{PhysRevLett.91.218102}, systems with vision cones \cite{lavergne2019group,loos_martynec_2023_long-range_order_non-reciprocal_XY_model}, but also neural \cite{sompolinsky_kanter_1986_temporal_association_asymmetric_neural_networks} and social \cite{Hong_Strogatz_Kuramoto_model_positive_negative_coupling_parameters,helbing_molnar_1995_social_force_model_pedestrian_dynamics} networks, and in quantum optics \cite{metelmann_2015_nonreciprocal_photon_transmission,zhang_2018_non-reciprocal_quantum_optical_system,mcdonald_2022_nonequilibrium_quantum_non-Hermitian_lattice,reisenbauer_delic_2024_non-hermitian_dynamics_optically_coupled_particles,livska_brzobohat_2024_pt-like_phase_transition_optomechanical_oscillators}. These theories have demonstrated intriguing collective dynamics \cite{you2020nonreciprocity,fruchart_2021_non-reciprocal_phase_transitions,saha2020scalar,dinelli_tailleur_2023_non-reciprocity_across_scales,frohoff_thiele_2023_nonreciprocal_cahn-hilliard,brauns2024nonreciprocal,suchanek_loos_2023_irreversible_fluctuations_emergence_dynamical_phases,PhysRevLett.133.258303},
and unusual material properties 
\cite{scheibner2020odd}; however, they do not take into account crucial ingredients of decision-making.

Incorporating decision-making with predefined control objectives introduces distinct challenges: agents must make decisions based on local observations to achieve some desired control goal, a fundamental characteristic of distributed feedback control strategies \cite{d2023controlling}.
Here, we demonstrate that the decision-making process induces inherent nonreciprocity and non-pairwise couplings into the system dynamics by explicitly incorporating control objectives,  e.g. considering the  positions of agents relative to their distance from the goal region.  This presents novel challenges for conventional field-theoretic approaches. Here, we propose a framework to address these challenges, using the paradigmatic example of the {\em shepherding} control problem, where a group of agents, the {\em herders}, must coordinate themselves to control the collective dynamics of a second group of agents, the {\em targets}, in a desired way \cite{lama2024shepherding}.

\begin{figure}[tb]
\begin{center}
    \begin{overpic}[width=.42\textwidth]{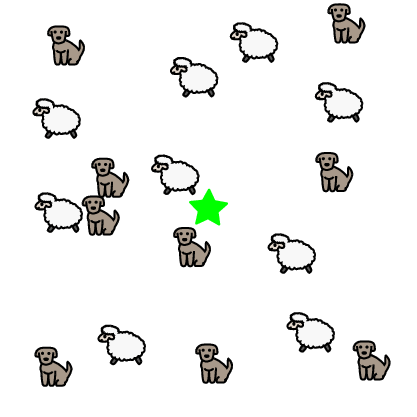}
    \put(3,95){\makebox(0,0){(a)}}
    % \put(50,-3){\makebox(0,0){\textbf{Simplest coupling to describe $\nabla\cdot[v_2(x)\rh\nabla\rt]$}}}
\end{overpic}
\begin{overpic}[width=.42\textwidth]{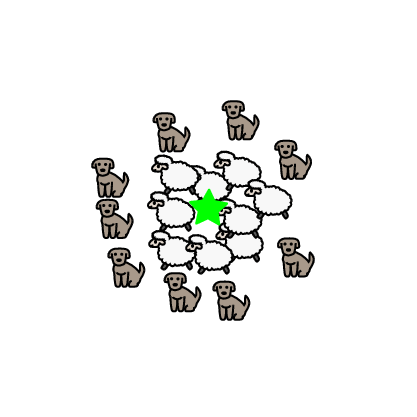}
    \put(3,95){\makebox(0,0){(b)}}
    % \put(50,-3){\makebox(0,0){\textbf{Simplest coupling to describe $\nabla\cdot[v_1(x)\rh\rt]$}}}
\end{overpic}
\end{center}
\captionof{figure}{Schematic illustration of the shepherding task. (a) Initial random configuration of herders (dogs) and targets (sheep), with herders and targets randomly distributed in space and the goal region marked by a green star. (b) Final configuration showing successful containment of targets around the goal region (green star) by the coordinated action of the herders.}
\label{fig:task}
\end{figure}

The shepherding task considered here exemplifies key features of distributed control problems: (i) reliance on local observations and (ii) real-time decision-making to achieve global objectives \cite{nalepka2017herd,zhang2024outmost}. In our setup, we analyze a representative shepherding task where $N_H$ herders must collect and contain $N_T$ targets within a pre-assigned goal region, which we assume to be a circle centered at the origin (see Fig. \ref{fig:task}). To focus on the fundamental aspects of decision-making, we assume that targets  exhibit no intrinsic collective properties, such as cohesion,
velocity alignment, or coherent response \cite{attanasi2014information}, which would otherwise simplify the task \cite{ko2020asymptotic}.  We model the targets as Brownian particles that experience a physical (and, thus, reciprocal) volume exclusion between themselves (and with the herders) through soft repulsion within a distance $\sigma$, and are repelled by herders within a distance $\lambda$ (see Methods).
%%%% OLD
% Instead, we model the targets as non-interacting Brownian particles that only experience physical volume exclusion between themselves and are repelled by herders within a distance $\lambda$ (see Methods).
%%%%%%

The key challenge in solving the shepherding control problem lies in the distributed decision-making that determines the dynamics of the herders that must achieve a collective task through local observations. Each herder faces two fundamental decisions that mirror challenges in many real-world control scenarios
\cite{d2023controlling,liu2011controllability}: what target(s) to influence, and how to influence them. We incorporate these decision-making elements into the dynamics of each herder $i$ through a control input, denoted $\textbf{u}_i$ (see Methods). This control input takes the form of a feedback term that determines the herder's action based on the relative positions of the observed targets with respect to the goal region.

Following established approaches \cite{lama2024shepherding,zhang2024outmost,nalepka2017herd}, each herder $i$ selects one target with position $\textbf{T}_i^*$ at each time; specifically, it chooses the target furthest from the origin (i.e., the goal region) within its circular sensing region of radius $\xi$.
To facilitate the later derivation of a continuum description that explicitly incorporates decision-making, we employ an approximation of this selection mechanism that was first proposed in \cite{lama2024shepherding}. 
Specifically, we express the position of the selected target $\textbf{T}_i^*$ as a weighted average of the position of the targets within the sensing region of the herder $i$. The approximation reads
\begin{equation}
\textbf{T}_i^* = \frac{\sum_{a\in N_{i,\xi}}e^{\gamma (|\textbf{T}_a|-|\textbf{H}_i|)}\textbf{T}_{a}}{\sum_{a\in N_{i,\xi}}e^{\gamma (|\textbf{T}_a|-|\textbf{H}_i|)}},\label{eqn:selection_rule}
\end{equation}
where $N_{i,\xi}$ represents the set of targets within the sensing region of the $i$-th herder, $\textbf{T}_a$ and $\textbf{H}_i$ are the two-dimensional (2D) Cartesian coordinates of targets and herders respectively, and $\gamma\geq0$ is a parameter that controls the selection specificity.

%%%%%%%%%%% OLD %%%%%%%%%%%%%%%%%%%
%  To facilitate the derivation of a continuum description that explicitly incorporates decision-making, we employ a continuous approximation of this selection mechanism first proposed in \cite{lama2024shepherding}. 

% Specifically, we express $\textbf{T}_i^*$ as
% \begin{equation}
% \textbf{T}_i^* = \frac{\sum_{a\in N_{i,\xi}}e^{\gamma (|\textbf{T}_a|-|\textbf{H}_i|)}\textbf{T}_{a}}{\sum_{a\in N_{i,\xi}}e^{\gamma (|\textbf{T}_a|-|\textbf{H}_i|)}},\label{eqn:selection_rule}
% \end{equation}
% where $N_{i,\xi}$ represents the set of targets within herder $i$'s sensing region, $\textbf{T}_a$ and $\textbf{H}_i$ are the two-dimensional (2D) Cartesian coordinates of targets and herders respectively, and $\gamma\geq0$ is a parameter that controls the selection specificity.
%%%%%%%%%%% OLD %%%%%%%%%%%%%%%%%%%

We wish to highlight that the selection rule approximation given by Eq. (1) for continuum descriptions was first introduced in \cite{lama2024shepherding} but its application was limited to stationary target distributions, with no explicit form derived for the resulting continuum dynamics. Our work significantly extends the results presented in \cite{lama2024shepherding} by developing a complete and explicit theoretical framework for dynamic multi-agent systems with decision-making capabilities. 
%We derive explicit expressions for the nonreciprocal coupling functions that emerge from decision-making processes and demonstrate how these lead to rich collective behaviors including confinement, pattern formation, and traveling waves. This framework provides a systematic approach to predict and control collective behavior through microscopic decision rules, with broad applications to robotics, biology, and active matter physics.

Also, notice that both the selection rule in Eq.~(\ref{eqn:selection_rule}) and the resulting control input, given in Eq.~(\ref{eq:feedback_control}) below,
incorporate not only pairwise distances between agents, but also each agent's position relative to the goal. The resulting control has, therefore, a three-body character.
The specific expression in Eq.~(\ref{eqn:selection_rule}), known as a {\em soft-max} function and widely used in neural networks for output selection \cite{krizhevsky2012imagenet}, enables continuous tuning of the selection rule: as $\gamma\to\infty$, it recovers the selection of the furthest target, while for $\gamma\to 0$, $\textbf{T}_i^*$ approaches the center of mass
%barycenter 
of all observed targets.

Once a target is selected,  the herder positions itself behind it at a distance $\delta$ to push it towards the goal. While more sophisticated approaches exist \cite{van2023steering,auletta2022herding}, in our minimal shepherding model the dynamics of herder $i$ is subject to an instantaneous  feedback control input that can be expressed as:
\begin{equation}
\textbf{u}_i=-\left(\textbf{H}_i-\textbf{T}^*_i-\delta\widehat{\textbf{T}^*}_i\right)
\label{eq:feedback_control}
\end{equation}
where $\widehat{\textbf{T}^*}_i=\textbf{T}_i^*/|\textbf{T}_i^*|$ is the unit vector pointing from the origin to $\textbf{T}_i^*$. If no targets are within the sensing radius of herder $i$, then $\textbf{u}_i=0$. This feedback law linearly attracts the herder to the position $\textbf{T}^*_i+\delta\widehat{\textbf{ T}^*}_i$, enabling it to push the target towards the origin when $\delta$ is smaller than the target's repulsion range $\lambda$. Similar to how $\gamma$ controls target selection, $\delta$ provides continuous control over trajectory planning: when $\delta=0$, the herder approaches $\textbf{T}_i^*$ from an arbitrary direction, generally failing to push it towards the origin, while for $\delta>0$, the herder systematically pushes the target towards the origin. Notice that also the trajectory planning, like the selection rule, has a three-body character, depending on the positions of the herder, the selected target, and the goal region.

These two parameters, $\gamma$ and $\delta$, thus enable us to continuously tune both aspects of decision-making in our system. As demonstrated later in Fig.~\ref{fig:radial_snapshots}, for sufficiently large $\gamma$ and appropriate values of $\delta$, the herders successfully confine the targets around the origin, achieving the shepherding task. Conversely, when $\gamma=0$ and $\delta=0$, the system evolves to a homogeneous configuration (on average).
As detailed in the Methods, our model includes white uncorrelated Gaussian noise for both herders and targets, as well as a reciprocal soft repulsion term between all agents within a distance $\sigma$, representing the agents' diameter, which we assume equal for herders and targets.

A key feature of our system, which will be fundamental in the derivation of the field equations, is its \textit{nonreciprocal} nature. The dynamics violates the action-reaction principle of Newtonian mechanics, which is permissible since this principle only needs to hold at the microscopic level, not at the coarse-grained level of the agents \cite{ivlev2015statistical}. While nonreciprocal effects have been extensively studied in physics \cite{fruchart2021non,you2020nonreciprocity}, our model reveals a profound connection between nonreciprocity and decision-making: the very act of making decisions introduces fundamental asymmetries into the dynamics. Not only are targets {\em repelled} by herders while herders are {\em attracted} to targets, but herders also exhibit decision-making capabilities - selecting targets ($\gamma$) and choosing their positions with respect  to a goal region ($\delta$) - that have no counterpart in the targets' dynamics. This intrinsic connection between decision-making and nonreciprocity provides a new perspective on both phenomena, suggesting that nonreciprocal field theories are a natural framework for describing systems with distributed decision-making.

In this work, we show how the continuous tuning of decision-making rules through parameters $\gamma$ and $\delta$ enables the derivation of field equations that capture the macroscopic properties of the shepherding problem.
Our analysis reveals that different decision-making strategies, from random to highly selective target choice and from undirected to goal-oriented pushing, emerge in the continuum description. The resulting nonreciprocal field theory 
is distinctly different from other nonreciprocal scalar theories such as
%goes significantly beyond
%previous approaches such 
the nonreciprocal Cahn-Hillard (NRCH) theory for phase-separation in asymmetric systems \cite{you2020nonreciprocity,saha2020scalar,brauns2024nonreciprocal}
and predator-prey models \cite{PhysRevLett.91.218102}.Our approach does not only successfully describes the shepherding dynamics, including the transition from homogeneous to configurations where targets are confined, but also provides a general framework for incorporating decision-making in the presence of a pre-defined control goal into continuum theories of collective behavior, allowing us to reproduce a variety of collective states. In our system, the pre-defined goal corresponds to a spatially fixed region at the center of the system. This results in an ``external field'' that breaks translational invariance: an aspect that, as shown in the paper, is of key importance for the structure of the resulting field equations. The presence of a predefined goal region is indeed a common feature in a variety of distributed control problems \cite{franklin2002feedback}, from bacterial systems to robotic swarm coordination \cite{VAUGHAN2000109}, crowd management and autonomous transportation systems. As such, our framework opens new perspectives for analyzing and designing distributed control strategies across diverse fields of applications.
%%%%%%%%%%%%OLD
% it not only successfully describes the shepherding dynamics, including the transition from homogeneous to configurations where targets are confined, but also provides a general framework for incorporating decision-making in presence of a pre-defined control goal into continuum theories of collective behavior. This framework opens new perspectives for analyzing and designing distributed control strategies across diverse applications, from robotic swarm coordination and crowd management to understanding collective animal behavior and developing autonomous transportation systems - scenarios where local decisions fundamentally drive global behavior.
%%%%%%% OLD
\section{Results}
\label{sec:results}

\subsection{A field theory for decision-driven shepherding}
\label{sec:key_result}

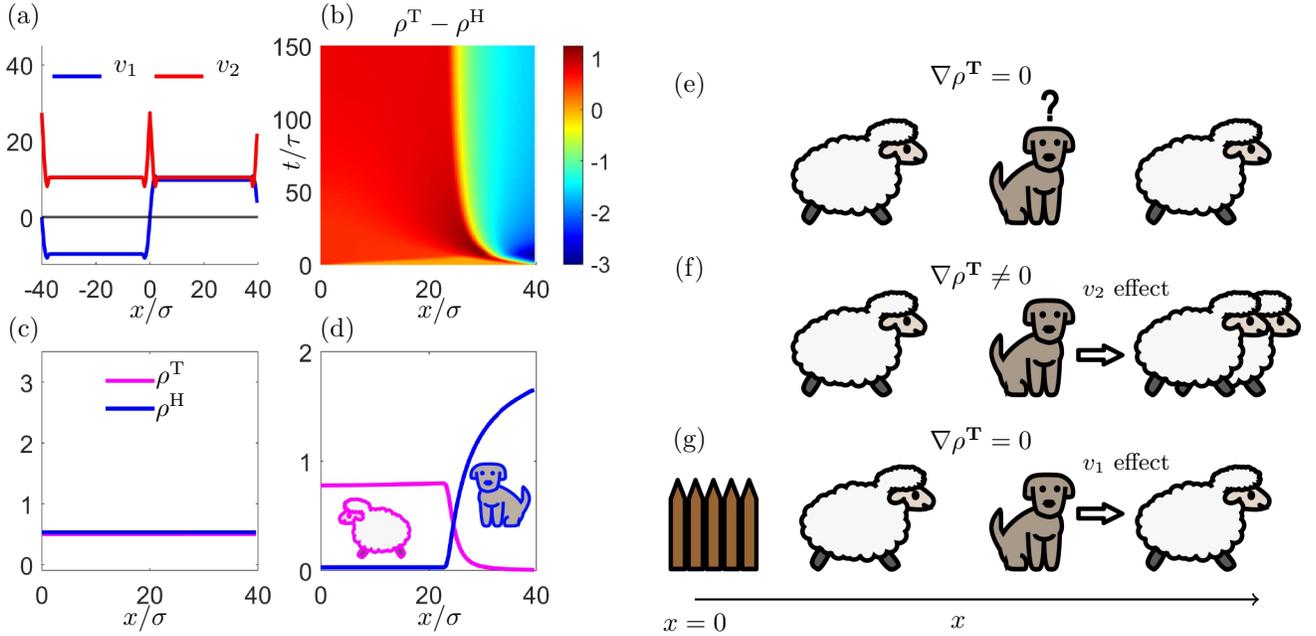
\begin{figure*}[!tb]
\begin{overpic}[width=.475\textwidth]{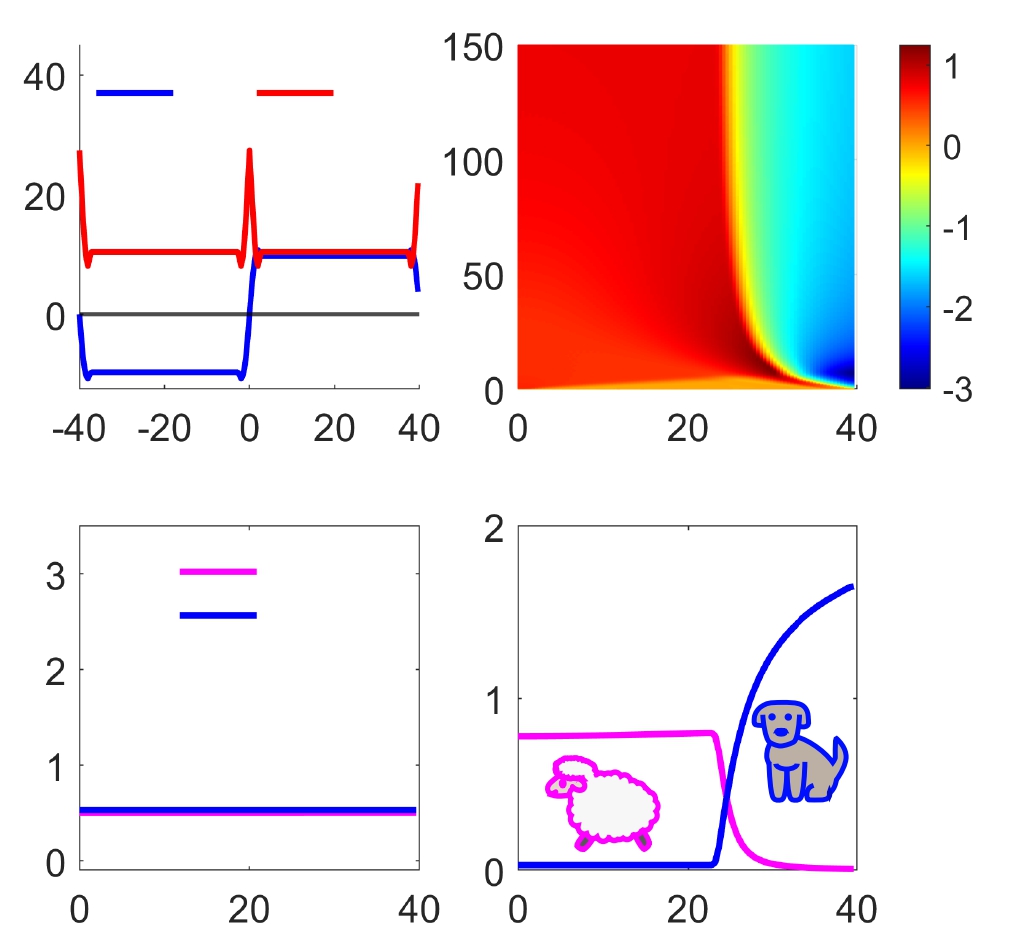}
    \put(5,94){\makebox(0,0){(a)}}
    \put(54,94){\makebox(0,0){(b)}}
    \put(5,45){\makebox(0,0){(c)}}
    \put(54,45){\makebox(0,0){(d)}} 
    % \put(55,50){\makebox(0,0){(t)}} 
    \put(25,48){\makebox(0,0){$x/\sigma$}}
    \put(21,86){\makebox(0,0){$v_1$}}
    \put(37,86){\makebox(0,0){$v_2$}}  
    \put(28,38){\makebox(0,0){$\rt$}}
    \put(28,33){\makebox(0,0){$\rh$}}
    \put(45,70){{\rotatebox{90}{$t/\tau$}}}
    \put(70,93){\makebox(0,0){$\rt-\rh$}}
    \put(70,48){\makebox(0,0){$x/\sigma$}}    
    \put(70,0){\makebox(0,0){$x/\sigma$}}
    \put(24,0){\makebox(0,0){$x/\sigma$}}
\end{overpic}
\hspace{.1cm}
\begin{overpic}[width=.48\textwidth]{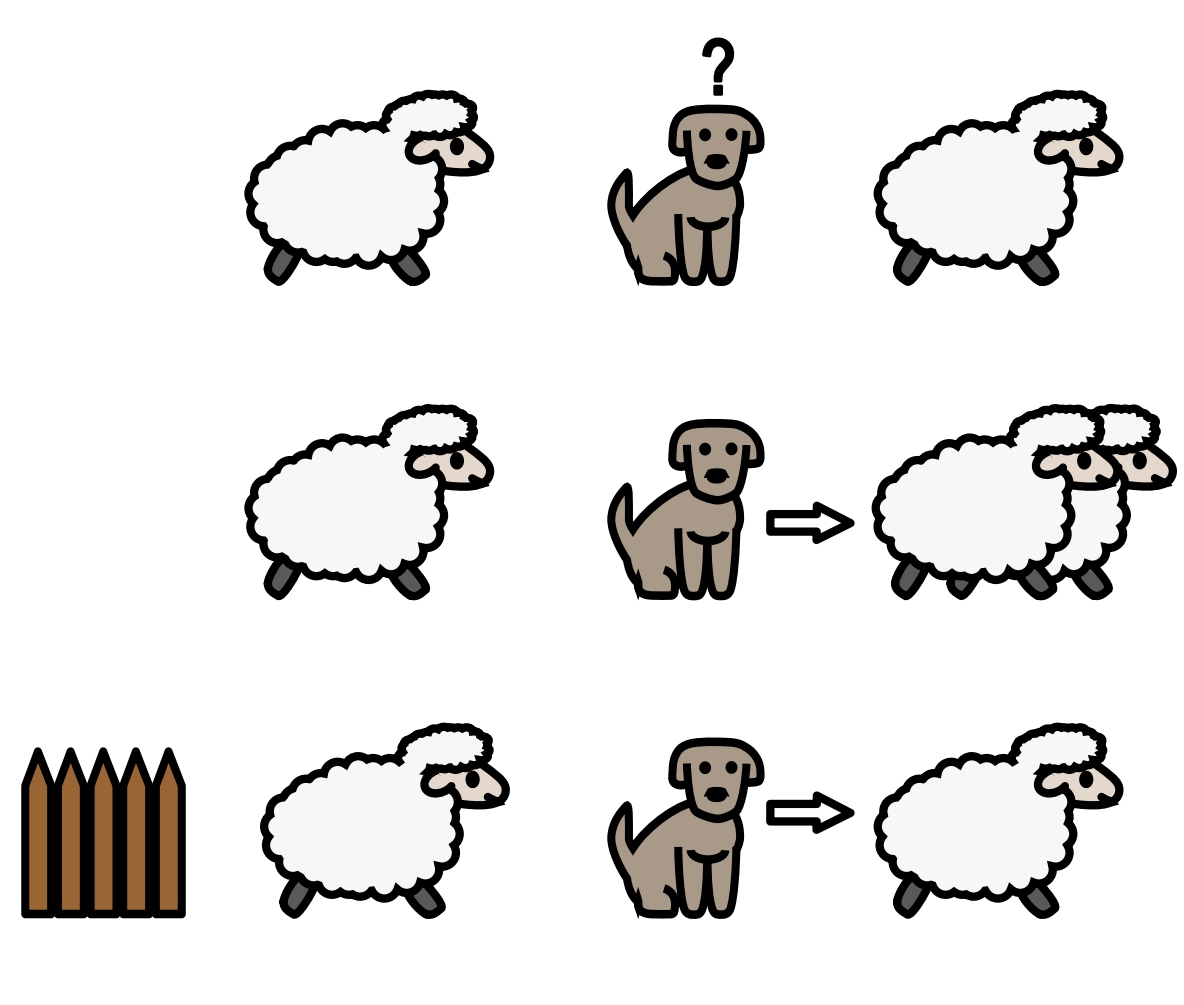}
    \put(5,82.5){\makebox(0,0){(e)}}
    \put(5,54){\makebox(0,0){(f)}}
    \put(5,27.5){\makebox(0,0){(g)}}
    \put(50,84){\makebox(0,0){\textbf{$\nabla\rt= 0$}}}
    \put(50,53){\makebox(0,0){\textbf{$\nabla\rt\neq 0$}}}
    \put(50,27.5){\makebox(0,0){\textbf{$\nabla\rt= 0$}}}
    \put(72.5,51){\makebox(0,0){\small{$v_2$ effect}}}
    \put(72.5,24){\makebox(0,0){\small{$v_1$ effect}}}
    \begin{tikzpicture}[overlay]
                    \draw[->,thick] (.5,.25) -- (8,.25);  % Simple arrow from (1,1) to (5,4)
                    \node at (.5,-.05) [] {$x=0$};  % Label above the arrow
                    \node at (4,-.05) [] {$x$};
    \end{tikzpicture}
    % \put(50,-3){\makebox(0,0){\textbf{Simplest coupling to describe $\nabla\cdot[v_1(x)\rh\rt]$}}}
\end{overpic}
\vspace{.5cm}
\caption{Field theory description of shepherding dynamics. (a) The coupling functions $v_1$ and $v_2$ generating shepherding dynamics; $v_2$ maintains a constant sign reflecting consistent attraction between species, whereas $v_1$ changes sign depending on the position $x$ of the herders with respect to the goal region $x=0$, encoding goal-directed behaviour.
%dashed lines depict the (constant) values of $v_1=v_1^0$ and $v_2=v_2^0$ in the case $\gamma=\delta=0$. 
(b) Spatiotemporal evolution of the density difference $\rt-\rh$: starting from a homogeneous distribution, the density profiles evolve and saturate to a ``shepherding configuration'' where $\rh$ effectively confines $\rt$ in a bounded region around the origin. (c-d) Steady state values of $\rt$ and $\rh$ for $\gamma=0$, $\delta=0$ (c) showing homogeneous distribution and for $\delta>0$ and $\gamma>0$ (d) showing confined configuration.(e-g) Representative one-dimensional configurations of targets (sheep) and herders (dog) illustrating the corresponding nonreciprocal couplings in Eq.~(\ref{eqn:H_field_1}). (e) The herder observes a symmetric distribution of targets ($\nabla\rt=0$), which generates no motion for the herder. (f) The herder observes an asymmetric distribution of targets ($\nabla\rt\neq0$), and moves towards higher targets concentration; this is captured by the coupling $-v_2\rh\nabla\rt$. In (g), the target distribution is symmetric again ($\nabla\rt=0$), however, the herder also possesses the additional information of the position of the goal region (fence) where to collect the targets. The herder will then move so as to complete the task; this is captured by the coupling term $-v_1(x)\rt\rh$ in Eq.~(\ref{eqn:H_field_1}).}
\label{fig:pde}
\end{figure*}

Our main achievement is a first-principles continuum theory that captures the essential physics of shepherding, derived from microscopic rules of agents' behavior to predict the emergence of macroscopic patterns. 
%%Our framework reveals how local decision rules and goal-oriented behavior combine to create robust collective states, providing a universal description that applies to different physical scenarios from sheep-dog interactions to robotic swarm control \textcolor{red}{(a bit redundant)}.
%
Specifically, we propose a mean field framework for shepherding dynamics based on two coupled partial differential equations (PDEs) that describe the spatio-temporal dynamics of the two species of herders and targets. Each species is represented by scalar, conserved density fields $\rho^{\text{A}}$, where ${\text{A}} = {\text{H}}$ denotes herders and  ${\text{A}} = {\text{T}}$ denotes targets. The dynamics incorporate diffusion, conservative interactions, and, most importantly, decision-making elements. For simplicity, we focus on one-dimensional (1D) motion along the $x$-direction, though generalization to 2D is straightforward. To lowest order, the equations have the form (see Methods)

\begin{align}
    \partial_t \rt=& \nabla\cdot \left[{D}^{\text{T}}(\rt)\nabla\rt+\ktt\rt\nabla\rh \right]
    \label{eqn:T_field_1}\\
    \partial_t \rh= &\nabla\cdot\left[{D}^{\text{H}}(\rh)\nabla\rh- v_1(x)\rh\rt-v_2(x)\rh\nabla\rt \right]
     \label{eqn:H_field_1}
\end{align}

The intraspecies coupling in both equations is described by the renormalized diffusivities ${D}^{\text{A}}(\rho^{\text{A}})$ which arise from noise and short-range agent-agent repulsion (see Methods). Additionally, the cross-coupling term with coefficient $\tilde{k}^{\text{T}} > 0$ in Eq. (\ref{eqn:T_field_1}) combines the effects of both the reciprocal short-range repulsion and the nonreciprocal long-range repulsion exerted by the herders on the targets. This coupling results in the  gradient of $\rho^{\text{H}}$ generating a current for $\rho^{\text{T}}$.

The {\em decision-making} capability of the herders is captured by the last two terms in Eq.~(\ref{eqn:H_field_1}) that involve the space-dependent functions $v_1(x;\gamma,\delta)$ and $v_2(x;\gamma,\delta)$.
Examples of these functions are plotted in Fig.~\ref{fig:pde}(a), while their analytical expressions are derived in the Methods.
The role of the corresponding terms is sketched in Fig.~\ref{fig:pde} (e-g). The function $v_2$, which multiplies the term $\rh\nabla\rt$, combines two contributions: the short-range reciprocal repulsion between herders and targets and the long-range attraction exerted on a herder by targets at position $x$. In our parameter regime, $v_2$ remains strictly positive, causing herders to move preferentially toward regions of higher target concentration (as shown moving rightward in Fig.~\ref{fig:pde}(f)).

In contrast, function $v_1$, which multiplies the bilinear term $\rh\rt$, changes its sign at $x = 0$, the center of the goal region (indicated by a fence in Fig.~\ref{fig:pde}(g)). In particular, $v_1$ is negative for $x < 0$ and positive for $x > 0$. This change of sign reflects how the direction of chasing depends on the target's location relative to the goal. We may interpret $v_1$ as a task-dependent speed of herders when a non-zero concentration of targets $\rt$ is observed. The consequence is illustrated in Fig.~\ref{fig:pde}(g). Despite observing a symmetric (homogeneous) distribution of targets, the herder preferentially chases the rightmost target due to its greater distance from the fence, representing the goal region. While our agent-based model yields explicit expressions for $v_1$ and $v_2$ (see Methods), we expect the overall dynamics to remain robust under small changes
%perturbations
of these functions, provided their essential properties -- particularly their signs and symmetries -- are preserved.

Mathematically, the structure of the decision-making terms in Eq.~(\ref{eqn:H_field_1}), particularly the coupling term $v_1\rt\rh$, emerges from the core properties of our agent-based model. These reflect how herders select which target to pursue (controlled by $\gamma$) and how they choose their chasing position (determined by $\delta$). In both decisions, herders consider the positions of the targets relative to their own positions and, most importantly, to the location of the goal region. The equations are derived by expanding the {\em soft-max} function for $\textbf{T}_i^*$ in the small $\gamma$ regime, where $\gamma \lesssim 1/\xi$ (see Methods).

In the absence of decision-making ($\gamma=\delta=0$), 
Eq.~(\ref{eqn:T_field_1}) remains unchanged, while Eq.~(\ref{eqn:H_field_1}) reduces to
\begin{align}
    \partial_t \rh%|(\gamma=\delta=0)
    = &\nabla\cdot\left[D^{\text{H}}(\rh)\nabla\rh- v_2^0\rh\nabla\rt \right]
     \label{eqn:H_field_1_0}
\end{align}
where $v_1(x)|_{\gamma,\delta=0}=0$, $v_2^0>0$, and the negative sign in front of the last term reflects the attractive nature of the force acting on a herder by a target (see Methods and Supplementary Information Section II.A for further details on the derivation).
%
%From a conceptual point of view, equations~(\ref{eqn:T_field_1}) and (\ref{eqn:H_field_1}) significantly differ from conventional field equations for binary mixtures, such as, e.g., Cahn-Hillard (CH) dynamics for phase-separating Brownian systems. 
%
Even in this simplest case, 
%
%where $v_1(x)=0$ and $v_2(x)=v_2^0$ is  a positive constant (a limit case of our equations as shown below), 
%
the PDE system
(\ref{eqn:T_field_1})-(\ref{eqn:H_field_1_0}) is
{\em nonreciprocal} in the sense that the cross-couplings between targets and herders have different signs (or values).

While similar scalar nonreciprocal theories, such as the nonreciprocal Cahn-Hilliard models, have been studied extensively \cite{you2020nonreciprocity,saha2020scalar,brauns2024nonreciprocal} in relation to parity-time symmetry breaking and traveling states, our model exhibits a distinctive feature: for nonzero $\gamma$ and $\delta$, the current experienced by herders emerges from the control-oriented, three-body coupling that depends on the goal region's position. This three-body coupling gives rise to the final terms in Eq.~(\ref{eqn:H_field_1}).
These terms have not only space-dependent "prefactors", they also involve the bilinear nonreciprocal term $v_1(x)\rt\rh$, which is absent in simpler theories without goal-oriented decision-making.
%This selective \textcolor{red}{three-body} coupling embodies the herders' decision-making capability. 
The asymmetry between these terms and those in the simpler target dynamics in Eq.(\ref{eqn:T_field_1}) constitutes a fundamental, source of nonreciprocity in the system, which remains unexplored at the field level.
% OLD TEXT OF THE ABOVE BLUE TEXT
%%%%%%%%%%%%%%%%%%%%%%%%%%%
%While Cahn-Hillard models with similar nonreciprocity have been studied extensively \cite{you2020nonreciprocity,saha2020scalar} in relation to parity-time symmetry breaking and traveling states, our model possesses a unique feature: for nonzero $\gamma$ and $\delta$, the current experienced by herders arises from \al{the tendency of the herders to place at a {\em desired position} relative to {\em selected targets}, where both the selected targets and the desired positions depend on the position of the goal region}\st{selected targets, chosen based on their position relative to the goal region}.
% This three-body coupling leads to the last terms in Eq.~(\ref{eqn:H_field_1}). 
% These terms have not only space-dependent "prefactors", they also involve a bilinear \al{aren't all the couplings are bilinear?} combination of the densities absent in simpler theories without goal-oriented decision-making.
% %This selective \textcolor{red}{three-body} coupling embodies the herders' decision-making capability. 
% The asymmetry between these terms and those in the simpler target dynamics in Eq.(\ref{eqn:T_field_1}) constitutes a fundamental source of nonreciprocity in the system.
%%%%%%%%%%%%%%%%%%%%%%%%%%%%%%%%%%

\subsection{Field theory captures agent-based behaviour}
\begin{figure}[htb]
\vspace{.5cm}
\centering
        \begin{overpic}[width=.49\linewidth]{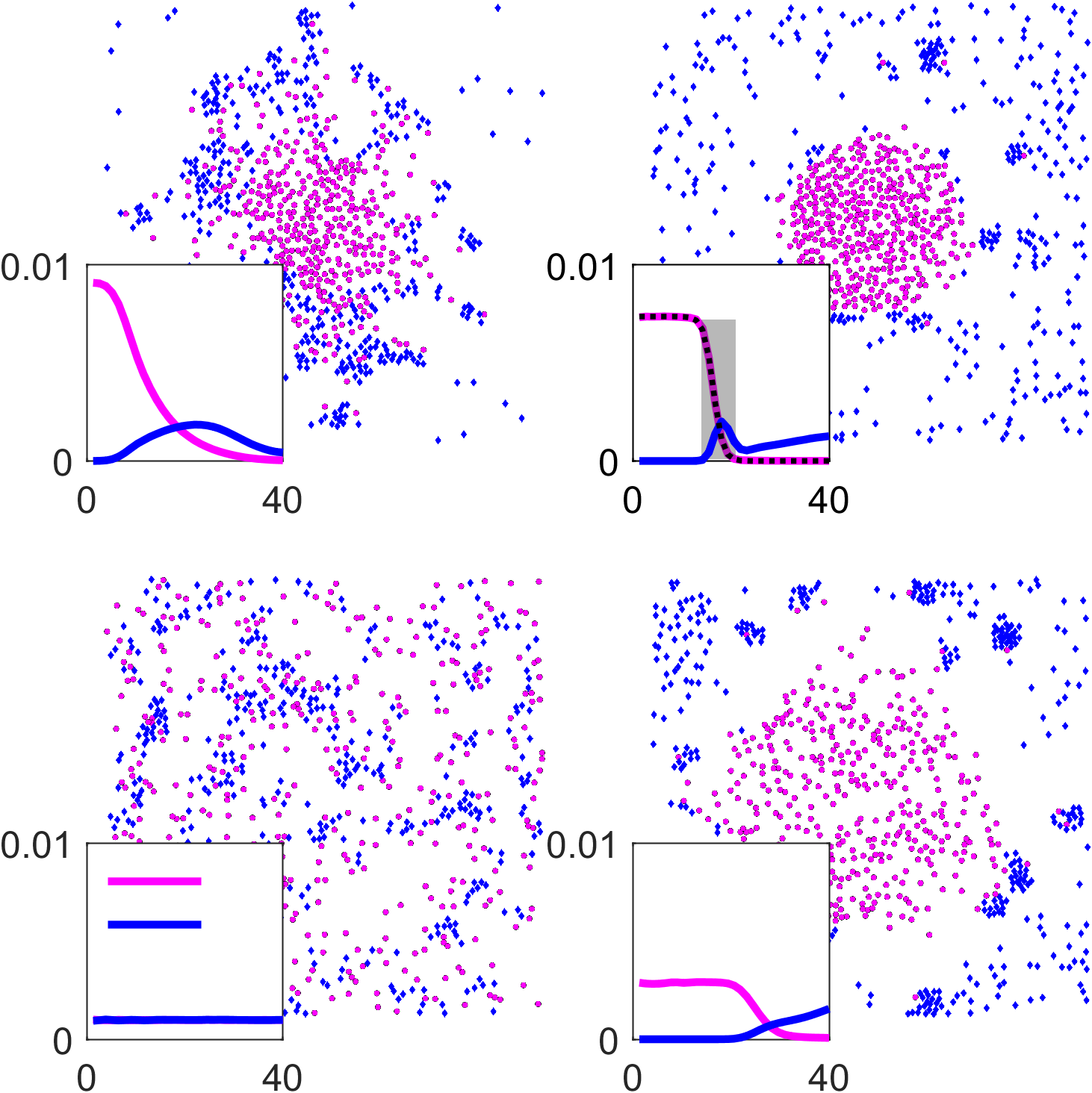}
        \put(6,102){\makebox(0,0){(a)}}
        \put(57,102){\makebox(0,0){(b)}}
        \put(6,50){\makebox(0,0){(c)}}
        \put(57,50){\makebox(0,0){(d)}}   
        \put(37,102){\makebox(0,0){$\gamma=0,\,\delta=\lambda/2$}}
        \put(83,102){\makebox(0,0){$\gamma\gg 1/\xi,\,\delta=\lambda/2$}}
        \put(38,50){\makebox(0,0){$\gamma=0,\,\delta=0$}}
        \put(85,50){\makebox(0,0){$\gamma\gg 1/\xi,\,\delta=0$}}      
        \put(15,55){\makebox(0,0){$r/\sigma$}}
        \put(65,55){\makebox(0,0){$r/\sigma$}}
        \put(15,2){\makebox(0,0){$r/\sigma$}}
        \put(65,2){\makebox(0,0){$r/\sigma$}}        
         \put(21,20){\makebox(0,0){$\hat{\rho}^\text{T}$}}
        \put(21,16){\makebox(0,0){$\hat{\rho}^\text{H}$}}   
        \put(65.5,73){\makebox(0,0){$\Delta$}}
            \begin{tikzpicture}[overlay]
                    \draw[->,thick] (1.8,-.25) -- (7.5,-.25);  % Simple arrow from (1,1) to (5,4)
                    \node at (4.,-.5) [] {$\gamma$};  % Label above the arrow
                    \draw[->,thick] (-.235,1.5) -- (-.235,7.5);
                    \node at (-.4,4) [] {$\delta$};
    \end{tikzpicture} 
        \end{overpic}
\vspace{.8cm}

    \caption{
    Long-time configurations and density profiles of the agent-based system under varying control parameters. (a-d) System states for different combinations of selectivity ($\gamma$) and trajectory planning ($\delta$) parameters, with insets showing time-averaged and angle-averaged density histograms of targets $\hrt(r)$ and herders $\hrh(r)$ as a function of the radial distance from the origin, $r$. At $\gamma=\delta=0$ we observe a disordered state with (on average) homogeneous density distribution (c). The other panels show that, as soon as $\delta$ or $\gamma$ are non zero,  inhomogeneous configurations are reached, 
    where herders tend to surround targets. The inset of panel (b) shows the fit of the target profile to a hyperbolic tangent (black dashed line) and indicates with a shaded grey rectangle the target-herder interface of width $\Delta$, see Methods for further details.}
    \label{fig:radial_snapshots}
\end{figure}
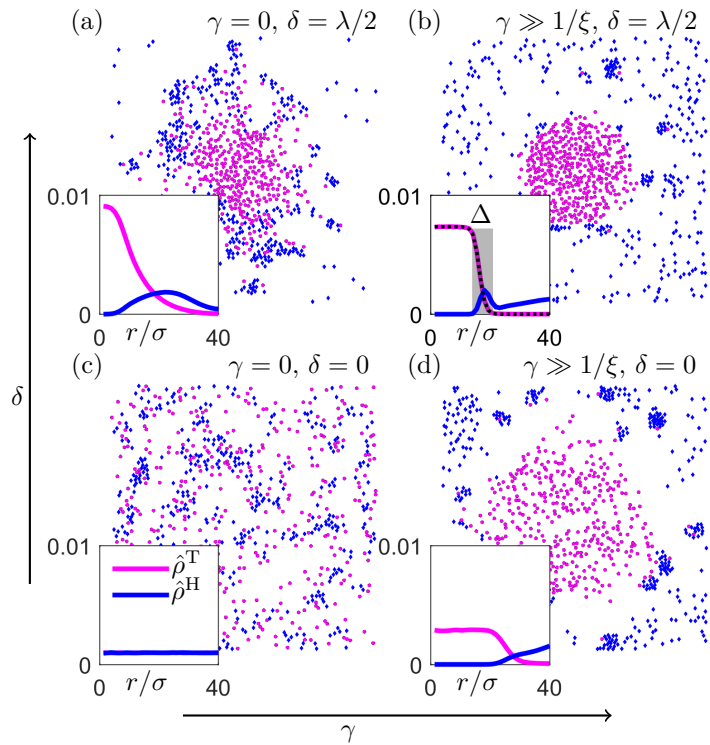

Our field equations capture the complex spatial patterns and dynamics observed in agent-based simulations, validating the theoretical framework. To provide a microscopic perspective of the emerging shepherding dynamics, we present simulation snapshots from the underlying agent-based model (see Methods) in Fig.~\ref{fig:radial_snapshots}. We explore several representative values of the two control parameters: $\gamma$, which determines the selectivity, and $\delta$, which controls the trajectory planning, focusing on cases where the number of herders equals the number of targets ($N_H=N_T$) (for other cases, see Supplementary Information Section I.C). The insets of Fig.~\ref{fig:radial_snapshots} show the corresponding time-averaged density profiles relative to the origin, $\hrt(r)$ and $\hrh(r)$, where $r$ is the radial distance, obtained by averaging over the polar angle $\theta$.

\begin{figure}[htb]
\vspace{.5cm}
        \begin{overpic}[width=.45\textwidth]{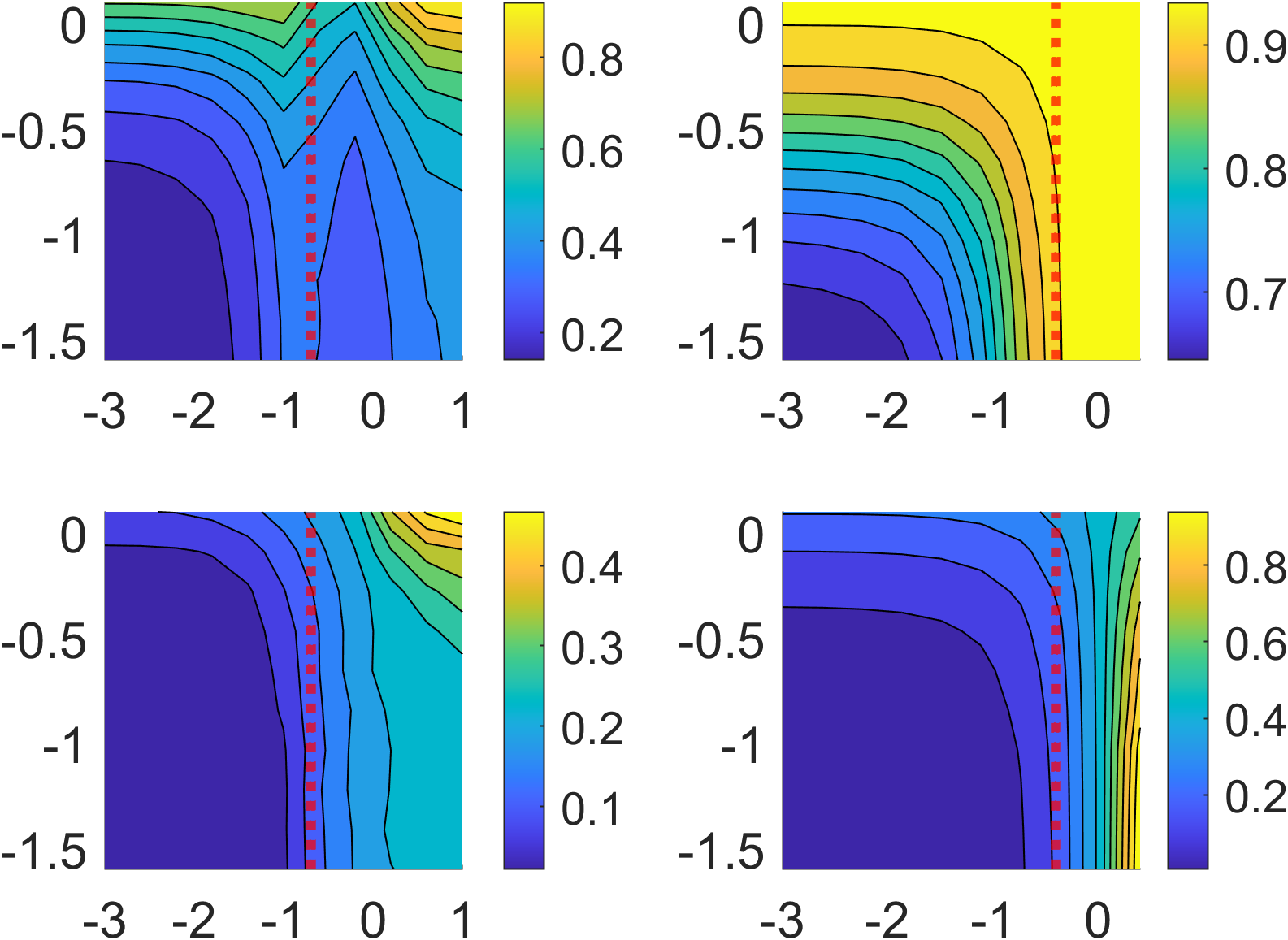}
                    \put(5,75){\makebox(0,0){(a)}}
                    \put(56,75){\makebox(0,0){(c)}}
                    \put(5,35){\makebox(0,0){(b)}}
                    \put(56,35){\makebox(0,0){(d)}}
                    \put(22,75){\makebox(0,0){$\chi$}}
                    \put(73.5,75){\makebox(0,0){$\chi$}}
                    \put(22,35){\makebox(0,0){$\sigma/\Delta$}}
                    \put(73.5,35){\makebox(0,0){$\sigma/\Delta$}}
                    \put(-3,58){\makebox(0,0){\rotatebox{90}{$log_{10}(\delta)$}}}
                    \put(-3,22){\makebox(0,0){\rotatebox{90}{$log_{10}(\delta)$}}}
                    \put(20,-3){\makebox(0,0){$log_{10}(\gamma)$}}
                    \put(73.5,-3){\makebox(0,0){$log_{10}(\gamma)$}}
                    \put(22,80){\makebox(0,0){\textbf{Agent-based}}}
                    \put(73.5,80){\makebox(0,0){\textbf{Field theory}}}
    \end{overpic}
\vspace{.5cm}
    \caption{Comparison of the values of $\chi$ (fraction of contained targets) and $\sigma/\Delta$ (sharpness of the herder-target interface) in the ($\gamma, \delta$) plane for (a-b) the agent-based model  and (c-d) the field theory. For small values of $\gamma$, where the linearization made to derive the field equations is justified, the level curves show that the results of the agent-based model and of the field theories are consistent. Some differences appear around $\gamma\sim \sigma/\xi$ (red dashed line) where the non-linearity of the selection rule
    (\ref{eqn:selection_rule}) becomes non negligible, hence where the linear approximation performed to derive the field equations starts to fail. A notable difference is that the combined maximum value of the two order parameters is reached for large $\gamma$ and $\delta$ for the agent-based model, and for large $\gamma$ and small $\delta$ on a field level.}
    \label{fig:phase_diagram}
\end{figure}

In the absence of decision-making ($\gamma=\delta=0$) the system reaches a disordered state [Fig.~\ref{fig:radial_snapshots}(c)]
with homogeneous densities characterized by constant profiles [Fig.~\ref{fig:phase_diagram}(c)]. This behaviour is also observed at the continuum level (see Fig. \ref{fig:pde}(c)). Note that even in this uncontrolled case, the underlying equations of motion remain nonreciprocal. However, for the system parameters considered here, this nonreciprocity does not lead to overall symmetry breaking (see Supplementary Information Section II.B), but only induces microstructural changes (clustering) on the agent-based level, which are not the focus here and will be the subject of future investigation.

The situation changes when one or both of the control parameters $\gamma$, $\delta$ are nonzero: the system develops inhomogeneities, as illustrated by the density profiles in Figs. 3(a), (b), (d). This behavior reflects a {\em radial symmetry-breaking} (relative to the case $\gamma=\delta=0$), 
where the target density $\hat{\rho}^{\text{T}}(r)$ reaches its maximum at the center and continuously decreases to zero radially outward. The most pronounced effect, which is key to obtain in real life scenarios such as in swarm robotics applications, occurs when $\gamma$ is large compared to the sensing radius and $\delta$ is non-zero, enabling herders to effectively guide their targets toward the desired direction. We then observe a concentrated disk of targets around the center, bounded by a sharp agglomeration of herders [cf. Fig.~\ref{fig:phase_diagram}(b)].

Interestingly, even when only one control parameter is nonzero, similar but less pronounced inhomogeneities emerge [Figs.~\ref{fig:phase_diagram}(a, d)].  When $\gamma = 0$ but $\delta > 0$, herders are attracted to the local center of mass of targets (rather than to a specific target), but approach them with an optimized distance strategy; this leads the herders to position themselves at the back of the barycenter of the observed targets, eventually creating an accumulation of herders around the targets. Conversely, when $\delta = 0$ but $\gamma > 0$, herders approach targets from random directions but intelligently select optimal targets; this leads the herders to move towards the observed target with the largest distance from the origin, eventually generating a spatial inhomogeneity.

%%% OLD %%%
% The situation changes when one or both of the control parameters $\gamma$, $\delta$ are nonzero: the system develops inhomogeneities, as illustrated by the density profiles in Figs.~\ref{fig:phase_diagram}(a), (b), (d). This behavior reflects a {\em radial symmetry-breaking} (relative to the case $\gamma=\delta=0$), %\al{can we talk of radial symmetry in a squared box/ 2d torus?} 
% where the target density $\hrt(r)$ reaches its maximum at the center and continuously decreases to zero radially outward. The most pronounced effect occurs when $\gamma$ is large compared to the sensing radius and $\delta$ is non-zero, enabling herders to effectively guide their targets toward the desired direction. We then observe a concentrated disk of targets around the center, bounded by a sharp agglomeration of herders [cf. Fig.~\ref{fig:phase_diagram}(b)]. When only one control parameter is nonzero, similar but less pronounced inhomogeneities emerge [Figs.~\ref{fig:phase_diagram}(a, d)].
%%%

The continuum theory captures the transition from homogeneous to spatially structured states when decision-making is present.
Two exemplary long-time sets of density profiles (from the numerical solution of Eq. (\ref{eqn:T_field_1}) and (\ref{eqn:H_field_1})) are shown in Figs.~\ref{fig:pde}(c-d), revealing the emergence of containment as we observed in their agent-based counterparts, despite differences in geometry (1D versus 2D) and specific values of $\gamma$ and $\delta$. Agent-based simulations in Supplementary Information Section I.D, where we consider a rectangular rather than circular goal region, further substantiate our claim that 1D field theories can capture the emergence of containment of the 2D agent-based dynamics. However, it is important to note that there are significant quantitative differences (for details, see Sec.~2.C).
%%% OLD
% A key achievement of our work is that the continuum theory we propose here successfully reproduces the essential features of agent-based density distributions. Two exemplary long-time sets of density profiles (from the numerical solution of (\ref{eqn:T_field_1}) and (\ref{eqn:H_field_1})) are shown in Figs.~\ref{fig:pde}(c-d), revealing a striking similarity with their agent-based counterparts,  despite differences in geometry (1D versus 2D) and specific values of $\gamma$ and $\delta$. Agent-based simulations in Supplementary Information Section I.D, where we consider a rectangular rather than circular goal region, further substantiate our claim that 1D field theories can capture the emergence of containment of the 2D agent-based dynamics. 
%%%
Indeed,  in the absence of decision-making [Fig.~\ref{fig:pde}(c)], the field theory produces a stable homogeneous state  (note that temporal instabilities are absent at the parameters considered, see Supplementary Information Section II.B). With decision-making,
the spatiotemporal evolution of the density difference $\rt(x,t)-\rh(x,t)$ in Fig.~\ref{fig:pde}(b) clearly demonstrates the emergence of inhomogeneities and finally the saturation into a steady state with density profiles shown in Fig.~\ref{fig:pde}(d). In this state, the targets are concentrated within the goal region around $x=0$, confined by herders accumulating at larger distances, creating a sharp interface between these two spatial domains. This configuration matches the expected outcome of successful shepherding, see Fig.~\ref{fig:task}.

Importantly, the appearance of the inhomogeneous steady state is a direct consequence of the nonreciprocal coupling arising from decision-making in our field equations. A stable homogeneous steady state solution exists only in the absence of decision-making (i.e., $\gamma=\delta=0$, $v_1=0$) (see Supplementary Information Section II.B). 
This solution ceases to be a steady state solution for {\em any} nonzero value of $\gamma$, $\delta$. Indeed, evaluated at the homogeneous solutions $\rho^{A}=\rho^A_0(t)$, Eq.~(\ref{eqn:H_field_1}) reads
\begin{align}
    \partial_t\rho^{\text{H}}_0(t)=-\frac{d}{dx}v_1(x;\gamma,\delta)\rho^{\text{T}}_0\rho^{\text{H}}_0.
    \label{eqn:steady_state}
\end{align}
Thus, when $v_1$ becomes space-dependent (which occurs for nonzero $\gamma$, $\delta$), no homogeneous steady state exists. This holds at any order in the gradient expansion of the densities (see Supplementary Information Section II.A).

\subsection{Phase Diagrams of Shepherding States}
\label{sec:quantitative}
Our analysis clearly reveals the emergence of inhomogeneous steady states whose details depend on two decision-making parameters. To quantify similarities and differences within these steady states,
%
%Our analysis reveals distinct phases of shepherding behavior and critical transitions controlled by some crucial decision-making parameters. To quantify the similarities and the differences between different inhomogeneous steady states 
 we analyze two key measures. First, we calculate the fraction $\chi$ of targets confined within 
 a circle of radius $R$ around the origin,
 a metric commonly used to evaluate shepherding performance \cite{lama2024shepherding,auletta2022herding}. Here, $R$ corresponds to the position of the interface shown in Fig.~\ref{fig:radial_snapshots}(b) (Fig.~\ref{fig:pde}(d) for the continuum case). Second, we evaluate the
 width of the interfacial region, $\Delta$, where the density $\hrt$ transitions from a non-zero value near the origin to zero at large distances. 
 For details on the definition and calculation of $R$ and $\Delta$ in both agent-based and continuum descriptions, see Methods.

 %%%%%%%%%% OLD
% Results for both the levels of description for $\chi$ and the inverse width (``sharpness'') $\sigma/\Delta$  as functions of the control parameters $\gamma$ and $\delta$ are shown in 
%%%%%%%%%%%
Results for both the levels of description for $\chi$ and the inverse width (``sharpness'') $\sigma/\Delta$ (where $\sigma$ is the repulsion distance representing the agents' physical size)  as functions of the control parameters $\gamma$ and $\delta$ are shown in Fig.~\ref{fig:phase_diagram}.
 Starting from the agent-based results [Figs.~\ref{fig:phase_diagram} (a-b)], we see that both quantities approach their minimum value in the limits  $\gamma\to 0$, $\delta\to 0$, as expected. The roles of $\gamma$ and $\delta$ are comparable for relatively small values, particularly when $\gamma\lesssim 1/\xi$ (red dashed line). Only beyond this value the nonlinearity of the selection rule given in Eq.~(\ref{eqn:selection_rule}) becomes relevant. The level-sets of $\chi$, represented by $\delta(\gamma)$, show non-monotonic behavior in $\gamma$ around this value. The largest values of $\chi$ and $\sigma/\Delta$ occur when both $\gamma$ and $\delta$ are large, confirming their roles as order parameters for the shepherding state \cite{lama2024shepherding}.

For comparison, continuum results for $\chi$ and $\sigma/\Delta$ are shown in Figs.~\ref{fig:phase_diagram} (c-d). As in the agent-based case, $\chi$ and $\sigma/\Delta$ reach their minimum values at $\gamma\to 0$, $\delta\to 0$. As $\gamma$ and $\delta$ increase from small values, $\chi$ increases monotonically with both parameters. This differs from the agent-based case [Fig.~\ref{fig:phase_diagram}(a)], where the level sets of $\chi$, $\delta(\gamma)$, exhibit non-monotonic behavior around $\gamma=1/\xi$. This difference arises because the continuum equations employ a linearization with respect to $\gamma$ (see Methods), valid for $\gamma\lesssim1/\xi$, a constraint absent in the agent-based case.

Notably, the sharpness reaches its maximum values at {\em small} values of $\delta$, a counterintuitive result explained by the competing roles of $v_1$ and $v_2$ at the continuum level (see Supplementary Information Section II.C).
Hence, it is important to note that while our field theory predicts the absence of homogeneous steady states for nonzero $\gamma$ and or $\delta$, and thereby the key qualitative feature of the shepherding dynamics, it does not fully capture all aspects of the spatial distributions observed in agent-based simulations.  These discrepancies highlight limitations in our current framework that could be addressed in future refinements, possibly by incorporating higher-order terms in the expansion of the selection rule or modifying the mean-field assumptions.
%%%OLD
% Despite these differences in detailed dependencies, the field theoretical results confirm that the decision-making parameters $\gamma$ and $\delta$ fundamentally shape the characteristics of the shepherding state. This demonstrates the theory's ability to describe complex, self-organizing behaviors driven by control-oriented, nonreciprocal interactions.
%%%

\subsection{Design Principles for Decision-driven Collective behavior in Natural and Artificial Systems}
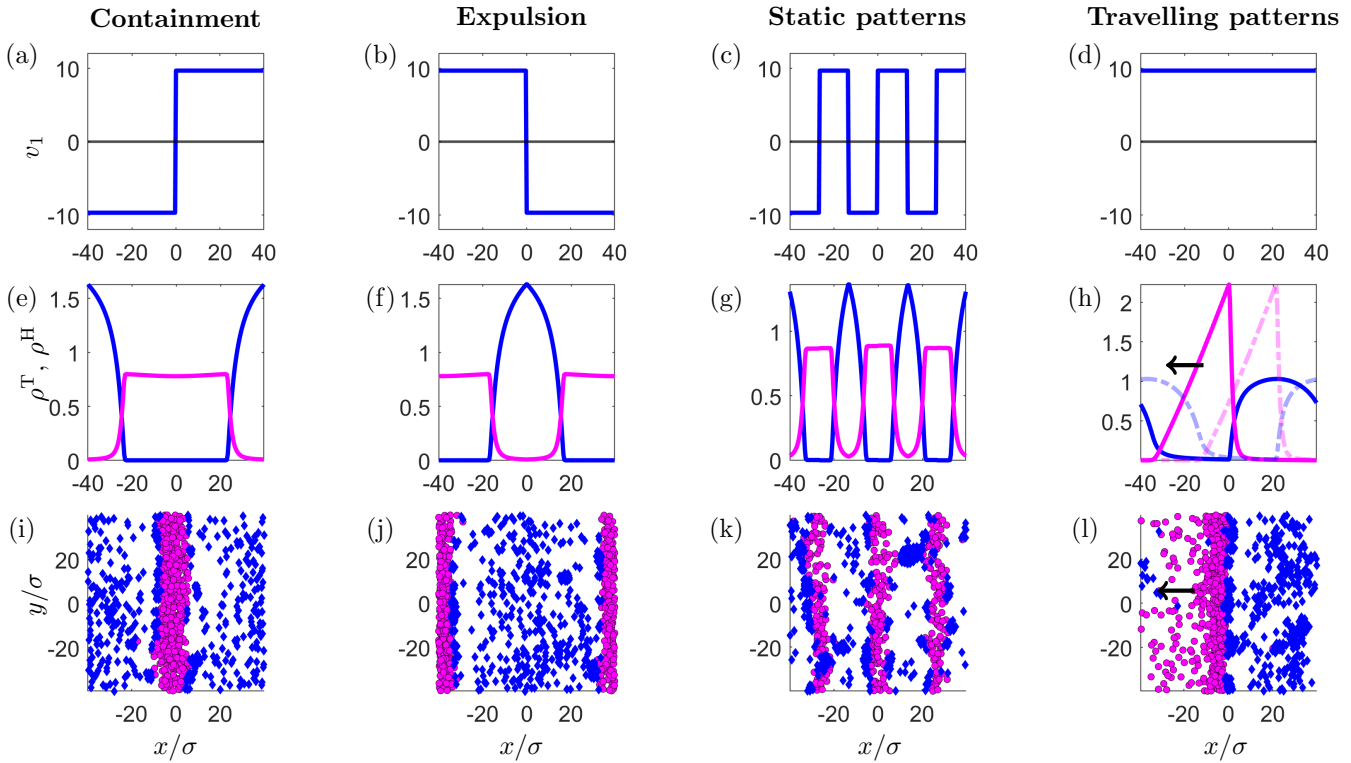
\begin{figure*}[htb]
\vspace{.5cm}
\begin{overpic}[width=.95\textwidth]{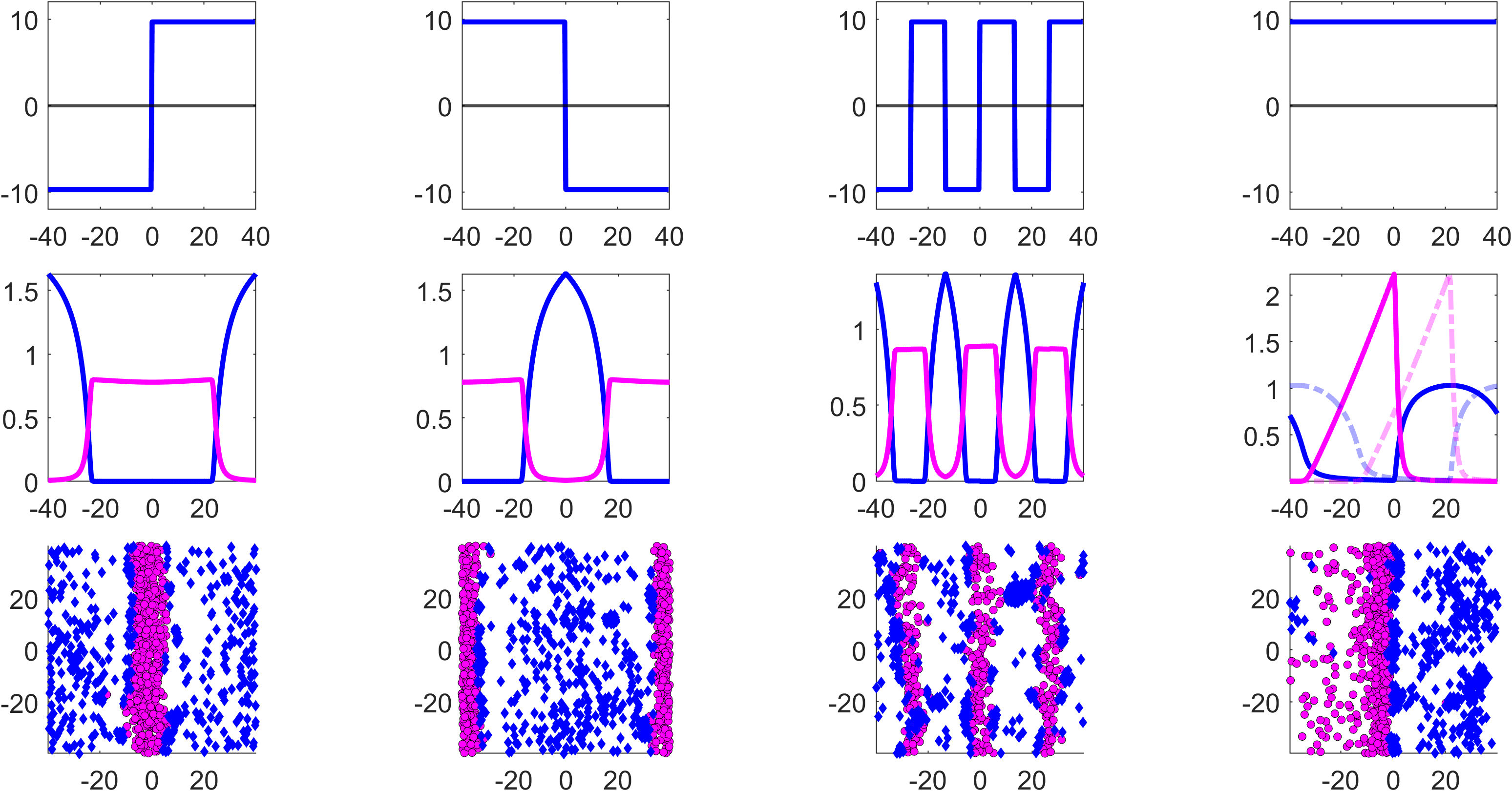}
    \put(-2,52){\makebox(0,0){(a)}}
    \put(26,52){\makebox(0,0){(b)}}
    \put(53,52){\makebox(0,0){(c)}}
    \put(81,52){\makebox(0,0){(d)}}
    \put(-2,33){\makebox(0,0){(e)}}
    \put(26,33){\makebox(0,0){(f)}}
    \put(53,33){\makebox(0,0){(g)}}
    \put(81,33){\makebox(0,0){(h)}}
    \put(-2,15){\makebox(0,0){(i)}}
    \put(26,15){\makebox(0,0){(j)}}
    \put(53,15){\makebox(0,0){(k)}}
    \put(81,15){\makebox(0,0){(l)}}
    \put(10,55){\makebox(0,0){\textbf{Containment}}}
    \put(37,55){\makebox(0,0){\textbf{Expulsion}}}
    \put(64,55){\makebox(0,0){\textbf{Static patterns}}}
    \put(91,55){\makebox(0,0){\textbf{Travelling patterns}}}
    % \put(15,26){\makebox(0,0){$x/\sigma$}}
    %  \put(40,26){\makebox(0,0){$x/\sigma$}}
    % \put(64,26){\makebox(0,0){$x/\sigma$}}
    % \put(88,26){\makebox(0,0){$x/\sigma$}}
    \put(10,-2){\makebox(0,0){$x/\sigma$}}
    \put(37,-2){\makebox(0,0){$x/\sigma$}}
    \put(65,-2){\makebox(0,0){$x/\sigma$}}
    \put(92,-2){\makebox(0,0){$x/\sigma$}}
    \put(-1,28){\makebox(0,0){\rotatebox{90}{$\rt$, $\rh$}}}
    \put(-1,10){\makebox(0,0){\rotatebox{90}{$y/\sigma$}}}
    \put(-1.,45){\makebox(0,0){\rotatebox{90}{$v_1$}}}
    \begin{tikzpicture}[overlay]
        \draw[<-,ultra thick] (14.75,1.75) -- (15.25,1.75); %rtl
    \end{tikzpicture}
    \begin{tikzpicture}[overlay]
        \draw[<-,ultra thick] (14.75,4.75) -- (15.25,4.75); 
    \end{tikzpicture}
    \end{overpic}
\vspace{.5cm}
\caption{Diverse collective state on the field level and at the agent-based level upon varying the control input and, thereby, the decision-making rules. (a-d) Possible choices of the function $v_1(x)$, which we set as a square wave with different periods (a-c) or a constant (d). (e-h) the corresponding steady state behaviours of the field equations. (i-l) The corresponding steady state behaviours of the agent-based model; see Supplementary Information Section III.B for details on the agent-based and field equations. By suitable choosing $v_1(x)$, starting from a homogeneous state, we can achieve (i,e) \textit{confinement} of the targets, as discussed above, (f,j) \textit{expulsion} of the targets, e.g. the herders expel the targets out from a region around the origin, (g,k) \textit{static patterns}, and (h,l) \textit{traveling patterns} (see Supplementary Information Video 3 and 4.); in (h) we also plot in shaded dashed lines the densities at a previous time to show that the pattern is moving coherently in time. }
\label{fig:relevance}
\end{figure*}

%\subsection{Emergent Collective States and Control through Field Theory}
%\subsection{Design Principles for Natural and Artificial Shepherding Systems}
Our field theory framework reveals fundamental design principles for the emergence of decision-driven collective behaviour in multi-agent systems that extend beyond the basic containment task that we focused on so far. Through appropriate choice of the functions $v_1$ and $v_2$ in Eq.~\eqref{eqn:H_field_1}, we can generate a variety of controlled collective states that could serve as building blocks to capture behaviour observed in  both natural and artificial  engineered systems. These principles emerge more clearly when we examine how different choices of these coupling functions lead to distinct collective behaviours at long times, as illustrated in Fig.~\ref{fig:relevance}.
%
%
%, as illustrated in Fig.~\ref{fig:relevance}, where we show the long-time behaviour of the field equations \eqref{eqn:H_field}, \eqref{eqn:T_field}, where in Eq. \eqref{eqn:H_field} the function $v_2$ is set to a constant $v_2=\tilde{v}_2>0$.

The function $v_1(x)$, which multiplies the product of density fields $\rt\rh$ in Eq.~(\ref{eqn:H_field_1}), plays a particularly crucial role as it encodes the task-dependent speed of herders. For simplicity, we set $v_1(x)$ to a square wave with different periods or phase shifts, while $v_2(x)$ is set to a positive constant (that, in this setting, represents an effective attraction of herders by targets). By choosing different spatial profiles for $v_1(x)$, we can generate (see Supplementary Videos):
\begin{itemize}
\item {\em Containment}: When $v_1(x)$ changes sign from positive to negative at $x=0$, the targets remain confined to the goal area centered around the origin [Fig.~\ref{fig:relevance}(e)];
\item {\em Expulsion}: Reversing the sign pattern, targets are expelled  from a predefined region [Fig.~\ref{fig:relevance}(f)];
\item {\em Static patterns}: When $v_1(x)$ alternates between positive and negative values periodically in space, stable, stationary stripe-like arrangements of targets are induced,  
[Fig.~\ref{fig:relevance}(g)];
\item {\em Traveling patterns}: When $v_1(x)$ is set to a constant, non-zero value throughout space  $v_1(x)=\tilde{v}_1$, targets and herders create traveling waves that propagate through the system [Fig.~\ref{fig:relevance}(h)].
\end{itemize}

These examples indicate that already simple manipulations of the control coupling functions that encode basic rules for how agents make decisions,
can induce a range of behaviors. For example, changes of sign of $v_1(x)$ and the location of the corresponding zeros can be used to realize desired static patterns without changing other coupling parameters. Such patterns are of interest, e.g., in swarm robotics \cite{giusti2023distributed}  and in engineered bacterial populations \cite{liu2011sequential,fu2012stripe}. 
Moreover, already the simplest choice for $v_1(x)$, namely a constant value, generates a traveling pattern (see Supplementary Information Section III.C for further details). 
%on the choice of $v_1(x)$ and the emergence of travelling patterns). %akin to those observed in nonreciprocal scalar field theories such as the nonreciprocal Cahn-Hilliard model \cite{you2020nonreciprocity,saha2020scalar}.
We note that stripe-like and travelling patterns also emerge from other PDEs well established in the physics literature, such as reaction-diffusion systems \cite{sherratt2008periodic} and variants of the nonreciprocal Cahn-Hilliard model studied recently \cite{you2020nonreciprocity,saha2020scalar,brauns2024nonreciprocal}.
The key insight presented here is that, in decision-making systems, such diverse behaviors can emerge solely from varying decision-making rules of the herders. For example, we can generate travelling patterns
in a parameter regime where such patterns are absent without decision-making (see Supplementary Information Section III.C). This perspective 
is further substantiated by agent-based simulation results (in the case of a rectangular goal region) shown in the bottom panel
of Fig.~\ref{fig:relevance}. To generate these results we did not change the microscopic interactions between the two types of agents, but only adapted the decision-making rules of the herders (see Supplementary Information Section III.B), using the same two-step strategy as in the shepherding problem:  
a proper selection rule, and a proper trajectory planning that finally determines $v_1(x)$ (see Supplementary Information Section III.A for further details).
%
%behaviours at the agent-based level: in Fig. \ref{fig:relevance}(i-l) we show typical snapshots of long-time behaviour of agent-based simulations, whose  detailed microscopic equations are reported in the SM. The key point to be noted is that  these behaviours where all obtained  at the agent level by using the same paradigm of diving the control input for the herders in two key ingredients: a proper selection rule, and a proper trajectory planning. This is the exact same paradigm we used to derive the function $v_1$ in the containment shepherding task.}
%
 The fact that we can create such different collective behaviors by only modifying $v_1(x)$ suggests a unifying principle that may be common to both natural and engineered  systems. Most notably, these behaviors emerge naturally from our field equations without requiring changes to their fundamental structure, indicating that our framework captures essential features of decision-driven collective behavior that extend beyond the specific task we considered.

%\textcolor{red}{Questions: do we know of any real-world example where one would like to establish, e.g. a stripe-pattern by decision-making processes?}
%\MdB{We could mention pattern formation in swarm robotics and bacterial populations.}
%\al{geometric pattern formation \cite{giusti2023distributed} in swarm robotics, formation of stripe patterns in engineered bacterial populations \cite{liu2011sequential,fu2012stripe}.}

%\al{To further substantiate our claim, we also show that we can reproduce such range of behaviours at the agent-based level: in Fig. \ref{fig:relevance}(i-l) we show typical snapshots of long-time behaviour of agent-based simulations, whose  detailed microscopic equations are reported in the SM. The key point to be noted is that  these behaviours were all obtained  at the agent level by using the same paradigm of diving the control input for the herders in two key ingredients: a proper selection rule, and a proper trajectory planning. This is the exact same paradigm we used to derive the function $v_1$ in the containment shepherding task.}

\section{Discussion}
\label{sec:discussion}
Our analysis uncovers a fundamental connection between microscopic decision-making and collective behavior: agents' decision-making capabilities inherently generate both nonreciprocity and, in presence of a pre-defined goal, entirely new forms of coupling in their continuum description.
%%%OLD
% agents' decision-making capabilities inherently generate both nonreciprocity and entirely new forms of coupling in their continuum description.
%%%OLD
This led us to discover a novel class of nonreciprocal interactions that has remained unexplored in the physics literature \cite{you2020nonreciprocity,saha2020scalar}. 

 %While previous field theories have considered non-reciprocal couplings, they typically lack strong connections to agent-based models that would justify their form. Our result bridge this gap by showing how non-reciprocal couplings emerge naturally from a widely studied class of models in  control theory and robotics; specifically, from herders' task-based decision-making. 
 %
The coupling we derive, which originates from feedback control at the microscopic level (\ref{eq:feedback_control}) and involves a goal that, in essence, breaks translational invariance, manifests as shepherding
 %%%OLD
 % The coupling we derive, which originates from feedback control at the microscopic level (\ref{eq:feedback_control}) and involves a predefined goal, manifests as shepherding
 %%%OLD
 behavior at the continuum scale [Fig.~\ref{fig:pde}(b)].
 This breakthrough establishes
 %, for the first time, 
 a rigorous mathematical framework connecting individual decision-making to collective behavior, unifying concepts from physics and control theory. 
In the field of shepherding, that we have chosen as a paradigmatic example, our continuum framework can be systematically extended and refined: For example, it would be very easy to incorporate other types of microscopic interactions such as, e.g., cohesion or alignment, that would lead to additional couplings  of the form $\nabla\cdot[\rho^A\nabla\rho^B]$ familiar from Cahn-Hillard-like theories.  
Further, applications in higher dimensions could reveal additional instabilities. Indeed, given recent experiments in human coordination \cite{nalepka2017herd}, we expect a polar instability induced by oscillatory motion of herders around targets. Another conceptually interesting question concerns changes imposed by considering nonlinear terms in the expansion of the selection rule.

The mathematical framework we developed here, where agents select whom to influence and how based on some control goal, can have implications beyond the specific shepherding task considered. Similar decision-making processes might play a role in biological systems, from marine predators \cite{haque2011biologically} and killer whales \cite{domenici2000killer} containing fish schools for hunting, to terrestrial cases like chimpanzee packs defending territories \cite{lemoine2020group} and ant colonies coordinating excavation \cite{prasath2022dynamics}. Similar principles might be present in social systems as well, particularly in opinion dynamics, \cite{Rainer2002-RAIODA} where influencers shape followers' opinion consensus by targeting certain individuals \cite{masuda2015opinion}, and extend to human coordination \cite{nalepka2019pnas,nalepka2017herd}, or engineered systems like swarm robotics \cite{lama2024shepherding,long2020comprehensive}, where artificial agents can be designed according to this paradigm. However, experimental validation would be needed to determine whether these natural systems actually employ similar mechanisms, an interesting open problem left for future study. 

From a more conceptual perspective,
our framework addresses an open challenge at the frontier between control theory and complex systems: steering collective behavior in large-scale multi-agent systems to solve distributed control tasks. While advances have been made in controlling complex networks \cite{d2023controlling}, distributed control of mobile agents faces unique challenges. Recent approaches in the field of active matter span optimal control \cite{norton2020optimal,davis2024active}, interaction design \cite{lavergne2019group}, feedback mechanisms \cite{khadka2018active}, and machine learning \cite{loffler2023collective,nasiri2022reinforcement}. Technically, at the agent level, two fundamental obstacles persist: the scaling of the number of dynamical equations with agent numbers and the time-varying nature of interaction matrices \cite{levis2020flocking,levis2017synchronization}. These challenges are particularly acute in shepherding, where the control of the targets is achieved  indirectly through herder dynamics \cite{licitra2019single}.

Here, we propose to tackle these challenges in complex, multi-agent control systems by deriving a
%through a
continuum macroscopic description where the functions $v_1(x)$ and $v_2(x)$ emerge as natural {\em control inputs} for shaping collective behavior. 
This aligns with growing interest from both physics \cite{doi:10.1073/pnas.2206994120} and control-theory communities to derive suitable partial differential equations that capture collective dynamics while incorporating precise control goals \cite{maffettone2022continuification}. The resulting PDEs provide insights that hold across different system sizes and are insensitive to specific microscopic initial conditions - which are often impossible to prescribe in real-world applications anyway.

The generality of our framework, is necessarily bounded by the specific hypotheses made in our microscopic model and derivation. Clearly, there are a number of problems in which decision-making occurs without broken translational invariance, such as in consensus finding \cite{liDistributed2011}, pattern formation in swarms \cite{giusti2023distributed}, and various socio- and economic problems that have been very recently analyzed through the lens of a field theory \cite{seara2025sociohydrodynamicsdatadrivenmodellingsocial,zakine2024socioeconomic}. Still, our framework, which incorporates decision-making on the continuum level through an appropriately defined feedback current, may have relevance also for these more general decision-making problems.

In the present work we have provided some first examples of how deliberate manipulation of the control input $v_1$ and $v_2$ leads to a variety of collective behaviors. Future research could focus on application of machine learning techniques to actually {\em learn} the functions $v_1$ and $v_2$ directly from experimental data of natural or artificial systems exhibiting decision-driven collective behavior. %for a given complex system with decison-making, based on patterns of emergent behavior. 
Similar strategies have been recently used to predict suitable PDEs for active and biological matter \cite{supekar2023learning}. A related intriguing question is to which extent such learned functions could provide new insight into microscopic decision rules and interactions. The framework could also be extended to incorporate more sophisticated decision-making mechanisms, such as those including memory effects or adaptive learning. Furthermore, investigating the interplay between noise, decision-making, and collective dynamics could reveal novel dynamical phases and transition mechanisms. Ultimately, such PDEs should predict not only steady-state behavior but also spatiotemporal patterns and even hierarchical self-organisation as it was recently found in biological communicating systems \cite{ziepke2022multi}.
The connection between nonreciprocity and decision-making uncovered here also suggests new ways to design and control collective behavior in synthetic systems, promising both deeper understanding of decision-driven collective phenomena and new possibilities for engineering complex systems with desired behaviors.

\clearpage
\section{Methods}

\subsection{Agent-based model}
\label{sec:agent_model}
The agent-based model underlying the field equations (\ref{eqn:T_field_1}) and
(\ref{eqn:H_field_1}) was chosen as a modified version of the minimal shepherding model considered in earlier work by some of the authors \cite{lama2024shepherding}. It involves all of the essential ingredients for shepherding dynamics and is, at the same time, particularly suitable for the derivation of field equations.

Specifically, we consider a binary mixture of mobile, disk-like agents of type T or H (with 2D Cartesian position
vectors $\textbf{T}_a$ and $\textbf{H}_i$, respectively)
moving in a square box of side length $L$ with periodic boundary conditions. We discuss here the case of a circular goal region around the origin; the case with a rectangular goal region, essentially representing a 1D problem, is discussed in Supplementary Information Section I.D.

The positions evolve in time according to overdamped Langevin dynamics, that is, both species are subject to Gaussian white noises representing the effects of (e.g., environmental or heterogeneity-induced) disorder on the coarse-grained scale (similar approaches were taken, e.g., 
in swarm robotics \cite{sartoretti2014decentralized,vasarhelyi2018optimized}, collective behaviours in biology \cite{ariel2015locust,cavagna2022marginal}, and in crowd dynamics systems \cite{heliovaara2012counterflow}). Further reasoning particularly for herder's noise is outlined in Supplementary Information Section I.A. We neglect inertial effects as it is often done in the robotics and control literature on shepherding \cite{sebastian2022adaptive,zhang2024outmost}. 
The targets' and herders' dynamics are given by 
\begin{align}
\dot{\textbf{T}}_a  =  k^{\text{SR}}\left(\sum_{b\neq a}^{N_{\text{T}}}\textbf{F}^{\text{rep}}_{\text{SR}}(\textbf{T}_a,\textbf{T}_b)
+\sum_{i=1}^{N_{\text{H}}}\textbf{F}^{\text{rep}}_{\text{SR}}(\textbf{T}_a,\textbf{H}_i)\right)
+k^T\sum_{i=1}^{N_{\text{H}}}\textbf{F}^{\text{rep}}_{\text{LR}}(\textbf{T}_a,\textbf{H}_i)
+\sqrt{2D}\mathcal{\bm{N}}_a 
\label{eq:target}
\end{align}
and 
\begin{align}
    \dot{\textbf{H}}_i=&k^{\text{SR}}\left(\sum_{j\neq i}^{N_{\text{H}}} \textbf{F}^{\text{rep}}_{\text{SR}}(\textbf{H}_i,\textbf{H}_j) + \sum_{a= 1}^{N_{\text{T}}} \textbf{F}^{\text{rep}}_{\text{SR}}(\textbf{H}_i,\textbf{T}_a)\right)
%     -& k^H (\textbf{H}_i-(\textbf{T}^*_i +\delta\,\widehat{\textbf{T}}^*_i))+\sqrt{2D}\mathcal{\bm{N}}_i
       + k^{\text{H}} \textbf{u}_i+\sqrt{2D}\mathcal{\bm{N}}_i
     \label{eq:herder}
\end{align}
respectively, where $\mathcal{\bm{N}}_{a,i}$ 
are zero-mean white (i.e., delta-correlated) noises with unit variances, and the diffusion constants $D$ are assumed to be equal. All agents interact through symmetric (reciprocal) short-range (SR) repulsive forces $\textbf{F}^{\text{rep}}_{\text{SR}}$ of amplitudes $k^{\text{SR}}>0$ and range $\sigma$, the latter representing the (uniform) particle diameter. In addition, the targets [see (\ref{eq:target})] are subject to a long-range (LR) repulsive force $\textbf{F}^{\text{rep}}_{\text{LR}}$ from the herders, with amplitude $k^{\text{T}}>0$ and range $\lambda>\sigma$. The LR repulsion reflects the tendency of a target to escape from an approaching herder.
For both, SR and LR types of repulsion, as is common in the Literature (e.g., \cite{dolai2018phase,kreienkamp2022clustering}), we assume a linearly decreasing dependence on the distance, such that, e.g.,
\begin{equation}
\textbf{F}^{\text{rep}}_{\text{LR}}(\textbf{T}_a,\textbf{H}_i)
=\frac{\textbf{d}_{ai}}{d_{ai}}(\lambda-d_{ai})\Theta(\lambda-d_{ai})
\label{eq:repulsion}
\end{equation}
where $d_{ai}=|\textbf{d}_{ai}|=|\textbf{T}_a-\textbf{H}_i|$, and $\Theta(x)=1$ if $x>0$, $0$ otherwise.
This linear ansatz is particularly useful for our later derivation of field equations \cite{you2020nonreciprocity}. For the SR repulsion we have the same functional form of the interaction, but the range $\lambda$ is replaced by the particle diameter $\sigma$.

Beyond SR repulsion, the equation of the herders (\ref{eq:herder}) contains the essential ingredient to realize shepherding, that is, the feedback control input $\textbf{u}_i$ defined in Eq.~(\ref{eq:feedback_control}). As described in Sec.~\ref{sec:intro}, the feedback term represents the (linear) attraction
of a herder towards a strategic position behind its selected target. Specifically, this position is given by $\textbf{T}^*_i+\delta\widehat{\textbf{ T}^*}_i$, where $\textbf{T}_i^*$ [see Eq.~(\ref{eqn:selection_rule})] is the position of the selected target (within a sensing radius $\xi$ and with selectivity $\gamma$), and $\delta\widehat{\textbf{T}^*}_i$ represents a displacement along the target's position vector ($\widehat{\textbf{ T}^*}_i$ being the unit vector along $\textbf{T}_i^*$). The parameter $\delta$ determines how far behind the target the herder should position itself to effectively steer the target's dynamics towards the origin. (As long as $0<\delta<\lambda$ the herder can successfully navigate the selected target towards the origin.)

Equations~(\ref{eq:target}), (\ref{eq:herder}), combined with Eqs.~(\ref{eqn:selection_rule}) and (\ref{eq:feedback_control}), constitute the fundamental components of our agent-based model. A distinctive feature of these equations is that the coupling between targets and herders cannot be derived from a Hamiltonian and exhibits strong nonreciprocity in two key aspects: first, there is an inherent asymmetry in the interaction forces -- targets experience repulsion from herders while herders are attracted to targets, which appears to violate Newton's third law (though this apparent contradiction is resolved when considering the coarse-grained nature of our description). Second, the selective nature of the herders' attraction introduces another form of nonreciprocity -- herders interact only with specific targets chosen according to their selection rule and guide them toward the goal, whereas targets respond uniformly to all herders without any such selectivity.

To achieve effective confinement of targets (see Fig.~\ref{fig:task}), we assume that the sensing radius of the herders exceeds that of target-herder repulsion ($\xi>\lambda$), consistent with earlier literature \cite{zhang2024outmost,auletta2022herding,lama2024shepherding}. Details of the simulations are provided in the subsequent paragraph \ref{sec:technical_agent}.

\subsection{Numerical calculations based on the agent-based model}
\label{sec:numerical_agent}

\subsubsection{Technical details}
\label{sec:technical_agent}
All simulations were carried out in MATLAB using an Euler-Maruyama integration scheme. The simulations are performed in a square periodic box of size $L\times L$ with $L=80\sigma$, where $\sigma=1$ defines the particle diameter. The diffusion constant is set to $D=1$, which characterizes the strength of noise in the system. Using the characteristic time scale $\tau=\sigma^2/D$, we employ a time step of $dt=1\cdot 10^{-3}\tau$. For initial conditions, we randomly and uniformly distribute $N_{\text{H}}=400$ herders and 
$N_{\text{T}}=400$ targets in the simulation box, resulting in equal initial densities of $\rho_0^{\text{T}}=\rho_0^{\text{H}}=0.0625/\sigma^2$ for both agent types (see Supplementary Information Section I.B  for further details). The system typically reaches stationary states after a transient period of $t_{\text{tran}}=200\tau$, and we run the simulations for a total duration of $t=300\tau$.

Results for different values of the ratio $N_{\text{H}}/N_{\text{T}}$ are reported in Supplementary Information Section I.C.

The agent-agent interactions are characterized by three distinct force contributions, each with specific parameters. For the short-range repulsive forces, which act between all agents regardless of type, we choose the particle diameter as $\sigma=1$ and amplitude $k^{\text{SR}}=100$. The long-range target-herder repulsion is characterized by an interaction radius $\lambda=2.5$ and amplitude $k^\text{T}=3$. The long-range herder attraction has a sensing radius $\xi=5$ and force amplitude $k^\text{H}=3$.
The snapshots in Fig.~\ref{fig:relevance}(i-l) are obtained with the same numerical values, and with $\gamma=5$ and $\delta=\lambda/2=1.25$; the dynamical equations are presented and discussed in Supplementary Information Section III.A.

\subsubsection{Density profiles and order parameters}
\label{sec:profiles_agent}
The radial density profiles $\hrt(r)$ and $\hrh(r)$ are computed by constructing histograms of the spatial distribution of targets and herders, respectively, as a function of their radial distance $r$ from the origin. These profiles are obtained by time-averaging over an interval of $t_{\text{avg}}=100\tau$ after the system has reached a steady state.

The radial densities are computed using an array of equispaced points from the origin to $L/2$ (half the box length), with a step size of $\Delta r=\xi/4$. The normalization of these density profiles follows the condition $\int_0^{L/2}\int_{0}^{2\pi}\hat{\rho}(r)r,drd\theta=\chi(L/2)$, where $\chi(L/2)$ represents the fraction of  targets contained within a circle of radius $L/2$ centered at the origin. 
The profiles $\hrt(r)$ and $\hrh(r)$ are obtained by averaging over the polar angle $\theta$, which is appropriate given our focus on the radial structure of the confined state. Final results are obtained by averaging over an ensemble of $48$ independent simulations.

We now define the two order parameters $\chi$ and $\sigma/\Delta$ considered in Fig.~\ref{fig:phase_diagram}.
First, to retrieve information on the width of the target-herder interface, $\Delta$, we fit the calculated density profile
of the targets to a hyperbolic profile $\rho_{f}(r)$. This choice is inspired by theoretical studies on the vapor-liquid interface in phase-separating fluids \cite{evans1979nature}. The explicit expression for $\rho_{f}(r)$ is given by
\begin{equation}
    \rho_{f}(r)=\rho_{f}(r; c,R,\Delta)=\frac{c}{2} \left( 1-\tanh \left(\frac{x-R}{\Delta}\right)  \right)
\end{equation}
where $c$, $R$, $\Delta$ are the fitting parameters representing, respectively, the maximum values of the profile, the position of the interface, and its width. We use the inverse of the width, $\sigma/\Delta$, as an estimate of the sharpness of the target-herder interface.

The confinement order parameter $\chi(R)$ measures the fraction of targets located within a distance $R$ from the origin.  In Fig.~\ref{fig:phase_diagram}, the reference radius $R$ is chosen as the fitting parameter $R$ of the density profile obtained at maximum values of $\gamma$ and $\delta$, where the spatial segregation between targets and herders is most distinct.
% In this case, the spatial separation between targets and herders is most pronounced.
The quantity $\chi$ is then calculated by averaging the instantaneous fraction of targets inside a circle of radius $R$, both over time ($t_{\text{avg}}=100\tau$) and over an ensemble of $48$ independent simulations. In Supplementary Information Section I.E we show that different choices of $R$ do not impact the qualitative behaviour of our results.
\subsection{Derivation of the field equations}
\label{sec:field_theory}
Starting from the agent-based stochastic equations (\ref{eq:target}), (\ref{eq:herder}) we now derive corresponding partial differential equations (PDEs) describing the spatial-temporal dynamics of the two density fields $\rho^{\text{A}}$, with A$=$H,T. The final structure of these equations is given in Eqs.~(\ref{eqn:T_field_1}) and (\ref{eqn:H_field_1}). As in Sec.~\ref{sec:key_result} we consider $1$-dimensional dynamics along the $x$-direction; the derivation can be easily generalized to higher dimensions.

Due to the conservation of the number of particles, both density fields
$\rho^{{\text{T}}({\text{H}})}(x,t)=\sum_{a(i)=1}^{N^{{\text{T}}({\text{H}})}}\delta\left(x-x_{i(a)}^{{\text{T}}({\text{H}})}(t)\right)$ obey a continuity equation, that is,
\begin{equation}
    \partial_t \rho^{{\text{T}}({\text{H}})}(x,t)=\nabla \cdot j^{{\text{T}}({\text{H}})}(x,t)+D\nabla^2 \rho^{{\text{T}}({\text{H}})}(x,t)
 \label{eq:continuity}   
\end{equation}
where $\nabla=\partial/\partial x$ and the currents $j^{{\text{T}}({\text{H}})}(x,t)$ take into account the various force terms in Eqs.~(\ref{eq:target}), (\ref{eq:herder}). 

The (piecewise linear) pair forces comprising the (SR or LR) repulsion in Eqs.~(\ref{eq:target}) and (\ref{eq:herder}) can be handled straightforwardly \cite{you2020nonreciprocity} (see 
Supplementary Information Section 2.A for details). Here we sketch the main idea. Each of the pair terms is of the form $F_i=\sum _{j=1}^M F_{ij}$, where $M$ is the number of interacting neighbors. Under a mean field assumption, $F_i$ gives rise to an average force %n the field level 
that, when
neglecting nontrivial spatio-temporal correlations, has the form
$\langle F(x,t) \rangle =\int F(x,y)\rho(y,t)\,dy$ where the integral extends over the interaction zone around $x$. For both the SR and LR forces, $F(x,y)$ represents a {\em translationally invariant} kernel, i.e., $F(x,y)=F(x-y)$. The resulting contribution to the current then has the form
$-\langle F \rangle(x,t)\rho(x,t)$. To compute the integrals in $\langle F\rangle$,
we perform a gradient expansion of $\rho(y,t)$ around $x$, i.e.,
$\rho(y)=\rho(x)+\nabla\rho(x)(y-x)+{\cal O}(\nabla\rho^2)$. Zeroth-order terms vanish due to translational invariance of the kernel. Keeping only linear terms in the gradients, we obtain the following currents stemming from repulsive forces
\begin{align}
    j_{\text{rep}}^{\text{T}}(x,t)&=\rho^{\text{T}}(x,t)\left(\alpha^{\text{SR}}\nabla\rho^{\text{tot}}(x,t)
+\alpha^{\text{LR}}
    \nabla\rho^{\text{H}}(x,t)\right)\nonumber\\
    j_{\text{rep}}^{\text{H}}(x,t)&=\rho^{\text{H}}(x,t)\alpha^{\text{SR}}\nabla\rho^{\text{tot}}(x,t)
    \label{eqn:jrep}
\end{align}
where $\rho^{\text{tot}}(x,t)=\rt(x,t)+\rh(x,t)$ is the combined density field, and
$\alpha^{\text{SR}}=k^{\text{SR}}\sigma^3/3$ and $\alpha^{\text{LR}}=k^T\lambda^3/3$
are positive constants related to the respective interaction radii and coupling constants in Eqs.~(\ref{eq:target}), (\ref{eq:herder}). 

With careful consideration, the third term on the right side of Eq.(\ref{eq:herder}), which describes herders' decision-making abilities, can also be handled using a mean-field-like approach and a density gradient expansion. This term depends on two key components defined in previous equations: the feedback control input $\textbf{u}_i$ [Eq.~(\ref{eq:feedback_control})] and the position of the selected target $\textbf{T}_i^*$ 
[Eq.~(\ref{eqn:selection_rule})]. Unlike typical classical systems where forces have a pairwise character, the resulting ``force'' here represents a three-body coupling between the herder's position, the selected target's position, and the origin. This leads to a mean-field force $\langle F(x,t) \rangle =\int F(x,y)\rho(y,t)\,dy$ with a kernel that lacks translational invariance and depends nonlinearly on both coordinates $x$ and $y$. To proceed, we (i) linearize the (exponential) weight function appearing in the numerator of Eq.~(\ref{eq:feedback_control}) (valid for $\gamma\lesssim 1/\xi$), and (ii) neglect the denominator. The kernel can then be expressed as
\begin{equation}
    F(x,y)=-k^{\text{H}}(1+\gamma(|y|-|x|))(x-(y+\text{sign}(y)\delta))
\end{equation}
This simplified form where the selection rule has been linearized in $x$ and $y$, still lacks translational invariance. A notable consequence is that even the zeroth-order term in the gradient expansion of $\rt(y)$ around $\rt(x)$ contributes to the average force. Combining this with the first-order (linear) gradient term, we obtain the following expression for finite values of the decision-making parameters $\gamma$ and $\delta$:
\begin{align}
    j_{\text{dm}}^{{\text{H}}}(x,t)=-g_1(x;\gamma,\delta)\rh(x,t)\rt(x,t)
    -g_2(x;\gamma,\delta)\rh(x,t)\nabla\rt(x,t).
    \label{eq:jatt}
\end{align}
Here, $g_{1/2}(x;\gamma,\delta)$ are space-dependent functions whose explicit expression is given by 

\begin{align}
    &g_1(\gamma,\delta,x)= \text{sign}(x) k^{\text{H}}\left\{
    \begin{aligned}
        &2\delta\xi+\frac{2}{3}\gamma\xi^3,\quad & \xi\leq|x|\leq L/2-\xi\\  
        & \left[\delta\left(1-2\gamma |x|\right)(|x|-\xi)+\delta(|x|+\xi) +\gamma x (\xi^2-x^2)+\frac{2}{3}\gamma|x|^3\right], \quad &|x|<\xi\\
    & \left[\delta\left(1-2\gamma (L/2-|x|)\right)((L/2-|x|)-\xi)+\delta((L/2-|x|)+\xi)\right.\\
        &\left. +\gamma (L/2-|x|) (\xi^2-(L/2-|x|)^2)+\frac{2}{3}\gamma(L/2-|x|)^3\right], \quad &|x|>L/2-\xi
    \end{aligned} \right .\\
    &g_2(\gamma,\delta,x)= k^{\text{H}} \left\{
    \begin{aligned}
        &\frac{2}{3}(\delta\gamma+1)\xi^3, \quad & \xi\leq|x|\leq L/2-\xi\\  
        &  \left[\delta(1-\gamma|x|)(\xi^2-x^2) +\frac{2\xi^3}{3}(\delta\gamma+1)-\frac{2}{3}(\gamma |x|)(\xi^3-|x|^3)+(\frac{\gamma}{2})(\xi^4-x^4)\right],    &|x|<\xi\\
        &  \left[\delta(1-\gamma (L/2-|x|))(\xi^2-(L/2-|x|)^2) +\frac{2\xi^3}{3}(\delta\gamma+1)\right.\\
        &\left.-\frac{2}{3}(\gamma (L/2-|x|))(\xi^3-(L/2-|x|)^3)+(\frac{\gamma}{2})(\xi^4-(L/2-|x|)^4)\right],   &\quad |x|>L/2-\xi
    \end{aligned} \right.
    \label{eqn:v1v2}
\end{align}

Note that the functions $v_1$ and $v_2$ in \eqref{eqn:H_field_1} are scaled versions of $g_1$ and $g_2$; furthermore, to obtain $v_2$,  we have to subtract from $g_2$ the (constant) contribution arising from the SR repulsion between herders and targets $\alpha^{\text{SR}}$ (see below). 
For further details on the derivation of $j_{\text{dm}}^{{\text{H}}}(x,t)$ see Supplementary Information Section II.A. Figure~\ref{fig:pde}(a) illustrates the behavior of $v_1$ and $v_2$. The consistently positive values of $v_2$ across all positions demonstrate the attractive force that targets exert on the herders. %\st{The maximum of v 2 occurs 
%at x=0, that is, the center of the goal region.}
In contrast to $v_2$, function $v_1$ changes its sign at $x=0$, with positive (negative) values at positive (negative) values of $x$. The resulting current acting on the herders thus depends on {\em where} these targets are relative to the goal region, consistent with our decision-making strategy. We stress that such a term is absent in more conventional nonreciprocal field theories of mixtures, e.g., the Cahn-Hillard model \cite{you2020nonreciprocity}.

In the special case $\gamma=\delta=0$ (no target selection and no trajectory planning) these functions reduce to
the constants $g_1=0$ and $g_2=g_2^0$ with $g_2^0=2k^H\xi^3/3>0$, yielding the current
\begin{equation}
    j_{\text{dm}}^{{\text{H}}}(x,t)|_{\gamma=\delta=0}=-g_2^0\rho^{{\text{H}}}(x,t)\nabla\rho^{\text{T}}(x,t)
    \label{eq:jatt_0}
\end{equation}
Note that, even in this "uncontrolled" case, where herders are unable to make decisions, the resulting field equations (\ref{eq:continuity}) for the two density field are nonreciprocal due to the different signs and values of the cross coupling terms, as are the corresponding Langevin equations (\ref{eq:target}), (\ref{eq:herder}) in this limit.

Finally, inserting the currents~(\ref{eqn:jrep}) and (\ref{eq:jatt}) into the continuity equations
(\ref{eq:continuity}) we obtain the PDEs 
\begin{align}
    \partial_t \rt&= \nabla\cdot\left[\rho^{\text{T}}\left(\alpha^{\text{SR}}\nabla\rho^{\text{tot}}
+\alpha^{\text{LR}}\nabla\rho^{\text{H}}\right)\right]+D\nabla^2 \rt
    \label{eqn:T_field}
    \end{align}
and
\begin{align}
    \partial_t \rh= \nabla\cdot\left[\rho^{{\text{H}}}\alpha^{\text{SR}}\nabla\rho^{\text{tot}}
    -g_1(x;\gamma,\delta)\rt\rh-g_2(x;\gamma,\delta)\rh\nabla\rt\right]+D\nabla^2 \rh   
     \label{eqn:H_field}
\end{align}
where we have omitted the arguments of the density fields.
We non-dimensionalize these equations by choosing the Brownian time, $\tau\equiv\sigma^2/D$, as characteristic time scale and the particle diameter, $\sigma$, as characteristic length scale.
Densities are scaled with the overall density 
$\rho_0=(N_\text{T}+N_\text{H})/L$. 
We further introduce dimensionless and renormalized diffusion coefficients 
$D^{\text{A}}(\rho^{\text{A}})=D\frac{\tau}{\sigma^2}+\alpha^{\text{SR}}\frac{\tau\rho_0}{\sigma^2}\rho^{\text{A}}$ with ${\text{A}}=\{{\text{T}},{\text{H}}\}$, which describe both, the effect of the noise and the intra-species short-range (SR) repulsion. With $\ktt\equiv(\alpha^{\text{LR}}+\alpha^{\text{SR}})\frac{\tau\rho_0}{\sigma^2}$, $v_1\equiv g_1 \frac{\tau\rho_0}{\sigma}$, and $v_2=(g_2-\alpha^{\text{SR}})\frac{\tau\rho_0}{\sigma^2}$ (and $v_2^0=(g_2^0-\alpha^{\text{SR}})\frac{\tau\rho_0}{\sigma^2}$), we then arrive at the field equations given in (\ref{eqn:T_field_1}) and (\ref{eqn:H_field_1}).

\subsection{Numerical calculations in continuum}
\label{sec:numerical_continuum}
All simulations were carried out in MATLAB. The partial differential equations are integrated using a pseudospectral code combined
with an operator splitting technique \cite{canuto2006spectral}, which allows to accurately treat the linear part of the spatial operator (the linear diffusion). The non-linear parts of the spatial operator (non reciprocal interactions and reciprocal SR repulsion) are treated as source terms: at every time-step, they are evaluated in real space using the values from the previous step. Then they are transformed into the Fourier space, where they are considered as source terms and integrated using a fourth-order Runge-Kutta time integration scheme.

In the actual calculations, the periodically repeated, one-dimensional 
segment $[-L/2,L/2]\subset \R$ with $L=80\sigma$ is divided into $N=201$ equispaced grid points. The time step is chosen as $dt=1\cdot 10^{-4}\tau$. The simulations are run for a total duration of $t=150\tau$, which is sufficient to reach a steady state. For Fig. \ref{fig:relevance}(h), $dt=1\cdot 10^{-3}\tau$, $t=750\tau$. The initial conditions are assumed to be slightly perturbed disordered states with $\rt_0=\rh_0=\rho_0/2=0.25$.

In our continuum model, most parameters can be directly adopted from the considered agent-based simulation parameters.
Exceptions are, first, the sensing radius $\xi$ which we set to $\xi=2.5\sigma$ to mitigate the effects of neglecting the denominator in Eq.~(\ref{eqn:selection_rule}). Second, the diffusion constant $D$ is set to $D=5$ to increase numerical stability. Third, the amplitude of the short-range repulsion is set to $k^{\text{SR}}=75$ so that $v_2^0>0$.
To reproduce Fig.~\ref{fig:pde}(b, d) we set $\gamma=2.5$, $\delta=\lambda/2$.
To reproduce Fig.~\ref{fig:relevance}(i-l), we set $v_2(x)=\tilde v_2=const$, while $v_1(x)$ is described by a square wave with different periods and amplitude $\tilde v_1$ (details are given in the Supplementary Information Section III.A). Specifically, we set $\tilde{v}_1=k^{\text{H}}(2\delta\xi+(2/3)\gamma(\xi^3))(\sigma/D)\rho_0$ and $\tilde v_2=  [k^{\text{H}} (2/3) (\delta\gamma+1)(\xi^3)-\alpha^{\text{SR}}](\rho_0/D)$, which would be the (constant) absolute values of $v_1(x)$ and $v_2(x)$ in the range $\xi\leq|x|\le\ L/2-\xi$; within these expressions, we set $\gamma=2.5$, $\delta=\lambda/2$. Only to reproduce Fig.~\ref{fig:relevance}(h) we use $k^{\text{T}}=20$, while for remaining part of the manuscript, $k^{\text{T}}=3$ .

The order parameters are calculated in analogy to the corresponding agent-based calculations; see Sec.~\ref{sec:profiles_agent}. For the quantity $\chi(R)$, we use the definition $\chi(R)=\frac{1}{\mathcal{C}}\int_{-R^*}^{R^*} \rt(x)\,dx$, with $\mathcal{C}$ being a proper normalization factor.

%\section{Numerical Simulations}
%All simulations were carried out in MATLAB. [AGGIUNGERE DETTAGLI SE VUOI SU DISCRETIZATION STEP, ROUTINE DI MATLAB UTILIZZATE, SCELTA CONDIZIONI INIZIALI ECC.]

\section*{Data Availability}
The authors declare that the data supporting the findings of this study are available within the paper and its Supplementary Information files or from the corresponding author on reasonable request. %The source data underlying all figures in the main text and supplementary information are provided as a Source Data file.

\section*{Code Availability}
The code used for agent-based simulations and continuum field theory calculations is available at {\tt bit.ly/43BdAQi} or from the corresponding authors on reasonable request.

\section*{Acknowledgements}
MdB and AL wish to acknowledge support from the MUR (Italian Ministry of University and Research) Research project ``Machine-learning based control of complex multi-agent systems for search and rescue operations in natural disasters'' (MENTOR) -- [PRIN 2022 -- CUP: E53D23001160006  -- SETTORE ERC PE7] and funding from the Scuola Superiore Meridionale in Naples, Italy which supported AL visit to TUB.

\section{Author contributions}
SK, MdB and AL jointly conceived and designed the research; AL wrote and checked the numerical code, carried out the derivations and data analysis with inputs from SK and MdB. All authors contributed to the analysis and interpretation of the results and were involved with the writing of the manuscript. 

\section*{Competing Interests}
The authors declare no competing interests.

\clearpage

%\textbf{About the amplitudes(do we need this)}:
%OLD: we define their amplitudes $\tilde{v}_1$ and $\tilde{v}_2$ as the (absolute) value they assume in all the domain beside the regions of size $\xi$ around the origin ($x=0$) and the boundaries of the domain ($x=\pm L/2$)
%\begin{align}
%    &\tilde{v}_1(\gamma,\delta)=k^H(2\delta\xi+\frac{2}{3}\gamma(\xi^3))\rho_0\frac{\sigma}{D}\nonumber \\
%    &\tilde{v}_2(\gamma,\delta)=k^H\frac{2\xi^3}{3}(\delta\gamma+1)\frac{\rho_0\tau}{\sigma}
%    \label{eqn:v1v2_amplitude}
%\end{align}
% and we can observe that, if we consider  the case where herders do not possess control information ($\gamma=0$, $\delta=0$), we have
%\begin{align}
%    &\tilde{v}_1(0,0)=0\\
%    &\tilde{v}_2(0,0)=k^H\frac{2\xi^3}{3}\frac{\rho_0\tau}{\sigma}
%    \label{eqn:v1v2_g0d0}
%\end{align}

\bibliography{biblio}

\end{document}

\renewcommand{\theequation}{S\arabic{equation}}
\renewcommand{\thefigure}{S\arabic{figure}}

\preprint{APS/123-QED}
\definecolor{myspecialcolor}{RGB}{230, 230, 230} % Definisci un colore personalizzato

\title{ \Large Supplemental material for  \\
``Nonreciprocal field theory for decision-making in multi-agent control systems''}

\author{Andrea Lama}
\email{andrea.lama-ssm@unina.it}
\affiliation{Modeling and Engineering Risk and Complexity, Scuola Superiore Meridionale, Naples, 80125, Italy}
\author{Mario di Bernardo}
\email{mario.dibernardo@unina.it}
\affiliation{Department of Electrical Engineering and ICT, 
University of Naples Federico II, Naples, 80125, Italy}
\author{Sabine.~H.~L.~Klapp}
\email{sabine.klapp@tu-berlin.de}
\affiliation{Institute for Theoretical Physics, Sec. EW 7-1,
Technical University Berlin, Hardenbergstrasse 36, 10623 Berlin, Germany}
\date{\today}%

%\onecolumngrid


\maketitle

In this Supplemental Information we provide, first, background information and additional results for our agent-based model. Second, we discuss several aspects of the continuum model regarding, in particular, the derivation and the linear stability analysis. Third, we discuss relations between continuum and agent-based results used in the context of the more general design rules considered in Fig.~5 of the main text. Finally, we present a list of the supplementary videos.
A list of references list is given at the end of this Supplemental Information.

\tableofcontents

\clearpage

\section{Agent-based model}
%\section{Supplemental information on the agent-based model}
\subsection{Comparison with other models}
%\subsection{The agent-based model compared with other shepherding models}
\label{sec:abm_comparison}
 In this section we discuss our agent-based model (see Methods) in the light of earlier models describing shepherding dynamics, particularly the one considered by some of the authors in~\cite{lama2024shepherding}. The latter model, along with other established ones \cite{nalepka2017herd,zhang2024outmost,auletta2022herding}, includes elements that are plausible on the microscopic level but make the derivation of a continuum theory unnecessarily complicated. It is for this reason that we have chosen here to simplify the model. The specific modifications compared to \cite{lama2024shepherding} are as follows:

First, and most importantly, we have removed a cooperation strategy between herders that prevents two or more of them to chase the same target. We recall [see Eq.~(1) of the main text] that a herder $i$ at position $\textbf{H}_i$ only considers targets $a$ at positions $\textbf{T}_a$ within its sensing radius $\xi$ (i.e., $|\textbf{H}_i-\textbf{T}_a|\leq\xi$). Ref.~\cite{lama2024shepherding} additionally involves a proximity-based rule: even if another target $b$ at $\textbf{T}_b$ is within herder $i$'s sensing region, it will not be followed by the herder $i$ if there is another herder $j$ (inside herder $i$'s sensing region) that is closer to $b$ (i.e., if $|\textbf{H}_j-\textbf{T}_b|<|\textbf{H}_i-\textbf{T}_b|$). This mechanism promotes the desired pattern, namely a spreading of herders around the circular arrangement of targets, but is hard to translate to the continuum level.
Upon removing this cooperative element, multiple herders can cluster around the same target that locally maximizes its distance from the origin. To qualitatively reproduce the herder spreading behavior, we introduce noise into the herders' dynamics -- an element often absent in earlier works \cite{lama2024shepherding,auletta2022herding,nalepka2017herd,zhang2024outmost}. For simplicity, we apply the same noise term used for target dynamics (see Eqs.~(7-8) in the Methods). This noise  mitigates the tendency of herders to cluster while chasing the same target; this partially recreates the desired spreading behavior. At the same time, the noise can be easily translated to the continuum by a diffusive term. We note that 
the introduction of noise within herder dynamics 
is not uncommon; it is also used, e.g., to model environmental disturbances in 
collective animal behavior \cite{ariel2015locust,cavagna2022marginal} or measurement errors in swarm robotics \cite{sartoretti2014decentralized,vasarhelyi2018optimized}.

As a second modification compared to \cite{lama2024shepherding}, we have eliminated the saturation of the herder speed, $\dot{\textbf{H}}_i$, that was implemented in \cite{lama2024shepherding} to prevent unrealistically high values when varying the herders' sensing radius $\xi$ over more than an order of magnitude. In the present work, $\xi$ was fixed to a moderate value, making the saturation redundant.

Finally, we here employ a finite simulation box with periodic boundary conditions, consistent with the large body of literature in active and nonreciprocal systems (see, e.g., \cite{you2020nonreciprocity}).
Given the presence of a ``special'' point being the center of the goal region, the resulting system can be visualized as a torus with one special point (i.e., the origin).

\subsection{Guidelines for parameter choices}
Equations (7-8) of the Methods contain multiple parameters that we have fixed based on the following considerations.

The domain size $L$ and number of targets $N_\text{T}$ were chosen to ensure the average target density $(N_\text{T})/L^2$
%$(N_\text{T}+N_\text{H})/L^2$
remains sufficiently high to prevent issues arising from herders' limited sensing capabilities. This is achieved by maintaining $N_\text{T}$ above the critical threshold $N_\text{T}^{\text{low}}(L,\xi)$ described in \cite{lama2024shepherding}.

Following standard practices in the control and robotics literature \cite{long2020comprehensive}, we set the herder sensing radius $\xi$ larger than the target repulsion range $\lambda$ (i.e., $\xi>\lambda$). Additionally, we ensure that target-herder nonreciprocal interactions dominate over noise effects by maintaining $k^T\lambda^2\gg D$ and $k^H\xi^2\gg D$.
The soft, reciprocal, short-range repulsion is configured to dominate other interactions ($k^{\text{rep}}\gg k^\text{H},k^\text{T}$) while maintaining the hierarchy of ranges, $\sigma<\lambda<\xi$. The concrete values for these parameters are given in the Methods.

\subsection{Impact of the number of herders}
%\subsection{Simulations with different $N^{\text{H}}$ value}

\begin{figure}[htb]
\vspace{1cm}

\centering
\vspace{.3cm}
        \begin{overpic}[width=.45\linewidth]{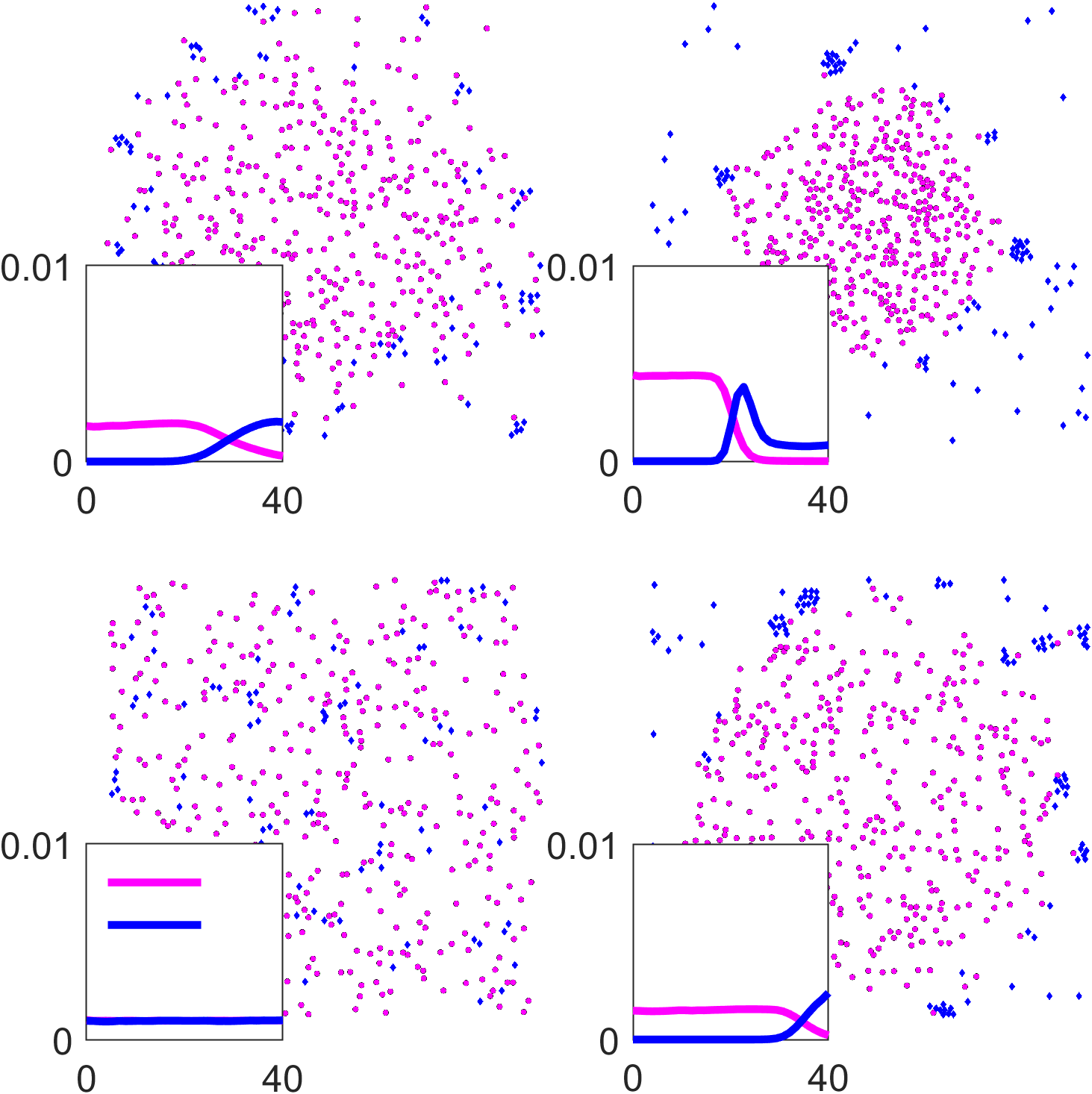}
        \put(3,102){\makebox(0,0){(a)}}
        \put(56,102){\makebox(0,0){(b)}}
        \put(5,50){\makebox(0,0){(c)}}
        \put(56,50){\makebox(0,0){(d)}}
        
        \put(37,102){\makebox(0,0){$\gamma=0,\,\delta=\lambda/2$}}
        \put(83,102){\makebox(0,0){$\gamma\gg 1/\xi,\,\delta=\lambda/2$}}
        \put(38,50){\makebox(0,0){$\gamma=0,\,\delta=0$}}
        \put(85,50){\makebox(0,0){$\gamma\gg 1/\xi,\,\delta=0$}}
        
        \put(15,55){\makebox(0,0){$r/\sigma$}}
        \put(65,55){\makebox(0,0){$r/\sigma$}}
        \put(15,2){\makebox(0,0){$r/\sigma$}}
        \put(65,2){\makebox(0,0){$r/\sigma$}}  
        
         \put(21,20){\makebox(0,0){$\hat{\rho}^\text{T}$}}
        \put(21,16){\makebox(0,0){$\hat{\rho}^\text{H}$}}

            \begin{tikzpicture}[overlay]
                    \draw[->,thick] (1.8,-.25) -- (7.5,-.25);  % Simple arrow from (1,1) to (5,4)
                    \node at (4.,-.5) [] {$\gamma$};  % Label above the arrow
                    \draw[->,thick] (-.235,1.5) -- (-.235,7.5);
                    \node at (-.4,4) [] {$\delta$};
    \end{tikzpicture}
        \end{overpic}
                \hspace{1.5cm}
        \begin{overpic}[width=.45\linewidth]{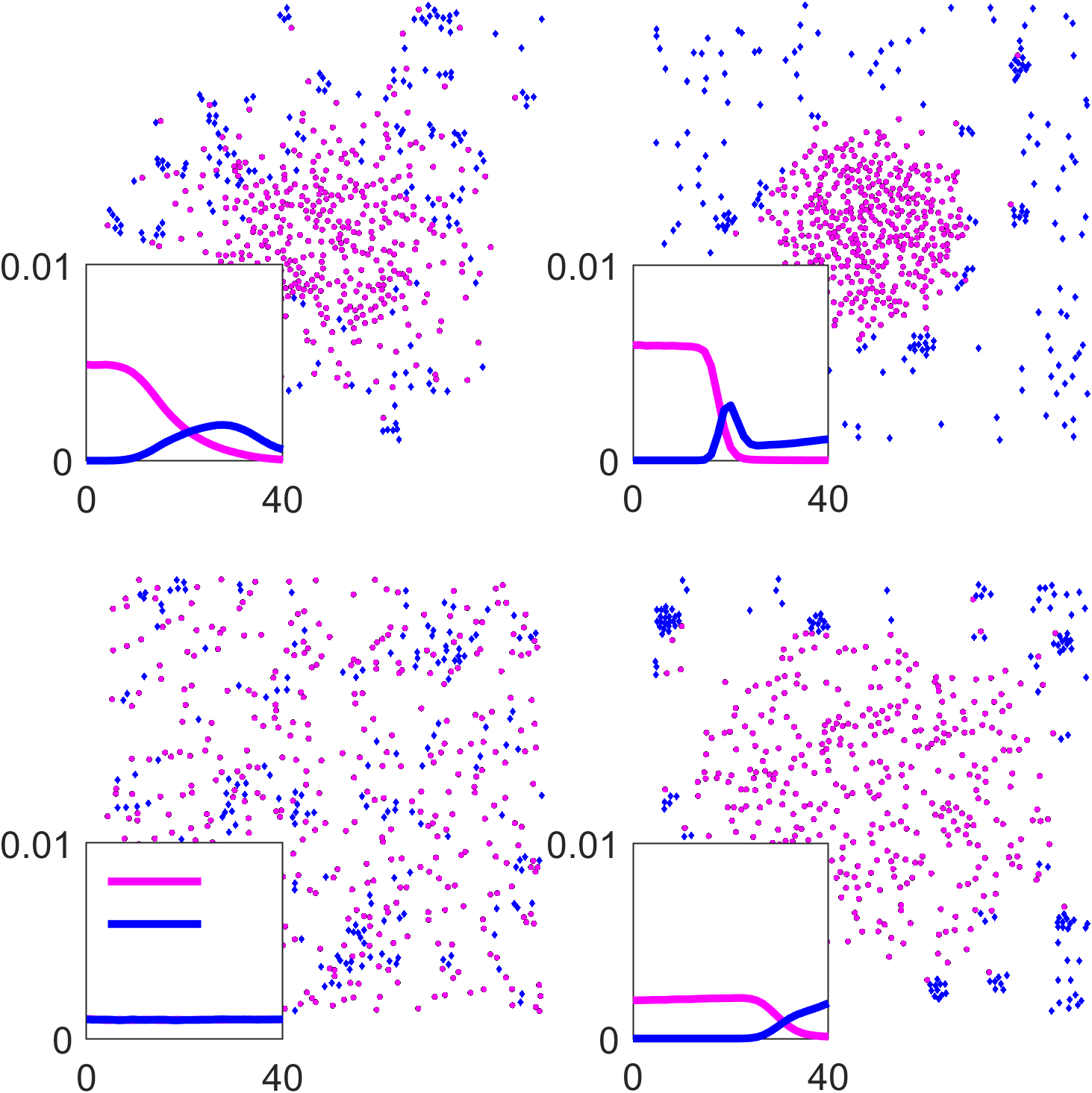}
        \put(3,102){\makebox(0,0){(e)}}
        \put(56,102){\makebox(0,0){(f)}}
        \put(3,50){\makebox(0,0){(g)}}
        \put(56,50){\makebox(0,0){(h)}}
        
        \put(37,102){\makebox(0,0){$\gamma=0,\,\delta=\lambda/2$}}
        \put(83,102){\makebox(0,0){$\gamma\gg 1/\xi,\,\delta=\lambda/2$}}
        \put(38,50){\makebox(0,0){$\gamma=0,\,\delta=0$}}
        \put(85,50){\makebox(0,0){$\gamma\gg 1/\xi,\,\delta=0$}}
        
        \put(15,55){\makebox(0,0){$r/\sigma$}}
        \put(65,55){\makebox(0,0){$r/\sigma$}}
        \put(15,2){\makebox(0,0){$r/\sigma$}}
        \put(65,2){\makebox(0,0){$r/\sigma$}}  
        
         \put(21,20){\makebox(0,0){$\hat{\rho}^\text{T}$}}
        \put(21,16){\makebox(0,0){$\hat{\rho}^\text{H}$}}

            \begin{tikzpicture}[overlay]
                    \draw[->,thick] (1.8,-.25) -- (7.5,-.25);  % Simple arrow from (1,1) to (5,4)
                    \node at (4.,-.5) [] {$\gamma$};  % Label above the arrow
                    \draw[->,thick] (-.235,1.5) -- (-.235,7.5);
                    \node at (-.4,4) [] {$\delta$};
    \end{tikzpicture}
        \end{overpic}
        \vspace{.3cm}
    \caption{Snapshots of typical long time configurations for representative values of the control parameters $\gamma$ and $\delta$ for $N_\text{H}=100$ (a-d) and $N_\text{H}=200$ (e-h); in both cases $N_\text{T}=400$ as for the main text. The insets show the time-averaged and angle-averaged densities of agents, $\hrt(r)$ and $\hrh(r)$, as a function of the distance $r$ from the origin for $r\leq L/2$. The simulations show no significant difference with respect to those presented in the main text.}
    \label{fig:radial_snapshots}
\end{figure}

The numerical results in the main text have been obtained by choosing equal numbers of targets and herders, that is,
$N_\text{T}=N_\text{H}=400$. Here we present results for systems where $N_\text{H}<N_\textbf{T}$, which is the common choice in the control literature on shepherding \cite{auletta2022herding,lama2024shepherding,van2023steering}.
%seems realistic in many real-world scenarios.
Figure~\ref{fig:radial_snapshots} shows representative long-time configurations for different values of the control parameters $\gamma$ and $\delta$ and $N_\text{H}=100$ (panel (a)-(d)) and $N_\text{H}=200$ (panel (e-h)), while 
$N_\text{T}=400$ is fixed. Visual inspection of the respective snapshots suggests that the qualitative features of the dynamics 
observed for $N_\text{H}=400$ (see Fig.~3 of the main text) remain. 
This robustness is further confirmed by the radial density profiles $\hrt(r)$ and $\hrh(r)$ shown in the insets, which exhibit spatial structures similar to those observed in our main analysis.

%To further validate our findings, we explored additional parameter regimes by varying both the control parameters $\gamma$ and $\delta$ and the number of herders $N_H$. Figure \ref{fig:radial_snapshots} shows representative long-time configurations for different values of these parameters, maintaining the number of targets fixed at $N_T=400$ as in the Main Manuscript. The spatial distributions obtained with $N_H=100$ (panels a-d) and $N_H=200$ (panels e-h) demonstrate that the qualitative features of the dynamics remain consistent across different herder populations. This robustness is further confirmed by the radial density profiles $\hrt(r)$ and $\hrh(r)$ shown in the insets, which exhibit patterns similar to those observed in our primary analysis.

\subsection{Impact of the geometry of the goal region}
\label{sec:rec_geom}
%al{Title modified}
%\subsection{Impact of the geometry of the goal region}

%\subsection{The shepherding task in a "rectangular" geometry}

The agent-based results in Fig. 3 of the main text are based on the assumption
of a circular goal region centered around the origin, consistent with the established shepherding literature \cite{auletta2022herding,lama2024shepherding}. Importantly, this circular geometry influences both steps of the decision-making process of the herders, namely (i), the selection rule for
$\textbf{T}_i^{*}$ [Eq.~(1) of the main text] that is based on the distance of targets from the origin, and (ii), the specification of the herder's positioning relative to $\textbf{T}_i^*$ that is crucial for the trajectory planning.  

To explore the possible role of geometry on the shepherding dynamics we here present, as an alternative, agent-based results for a rectangular geometry. Concretely, the simulations are still performed in two dimensions (in a square box of side length $L$ with periodic boundary conditions), but the goal region is now defined as a stripe centered around the origin at $x=y=0$ and extending along the 
$y$-direction. To realize this goal, each herder $i$ now selects its target 
$\textbf{T}_i^*$ based on the maximum distance of each target $a$ from the $y$-axis, that is, based on the $x$-component $\textbf{T}_{a,x}$.
The selection rule then becomes
\begin{equation}
\textbf{T}_i^* = \frac{\sum_{a\in N_{i,\xi}}e^{\gamma (|\textbf{T}_{a,x}|-|\textbf{H}_{i,x}|)} \textbf{T}_{a}}{\sum_{a\in N_{i,\xi}}e^{\gamma (|\textbf{T}_{a,x}|-|\textbf{H}_{i,x}|)}}.\label{eqn:selection_rule_rectangular}
\end{equation}
Furthermore, herders position themselves behind $\textbf{T}_i^*$ with shift vector
$\widehat{\textbf{T}}_{i}^{*}=
\sign(\textbf{T}_{i,x}^*)\hat{\textbf{x}}$, where 
$\hat{\textbf{x}}$ is the unit vector in $x$-direction. 
Accordingly, the control input $\textbf{u}_i$ for each herder becomes
\begin{equation}
    \textbf{u}_i=-\left( \textbf{H}_i-(\textbf{T}_i^*+\delta\,\sign(\textbf{T}^*_{i,x} )\hat{\textbf{x}}) \right)
\end{equation}

\begin{figure}[htb]
\vspace{1cm}

\centering
\vspace{.3cm}
        \begin{overpic}[width=.49\linewidth]{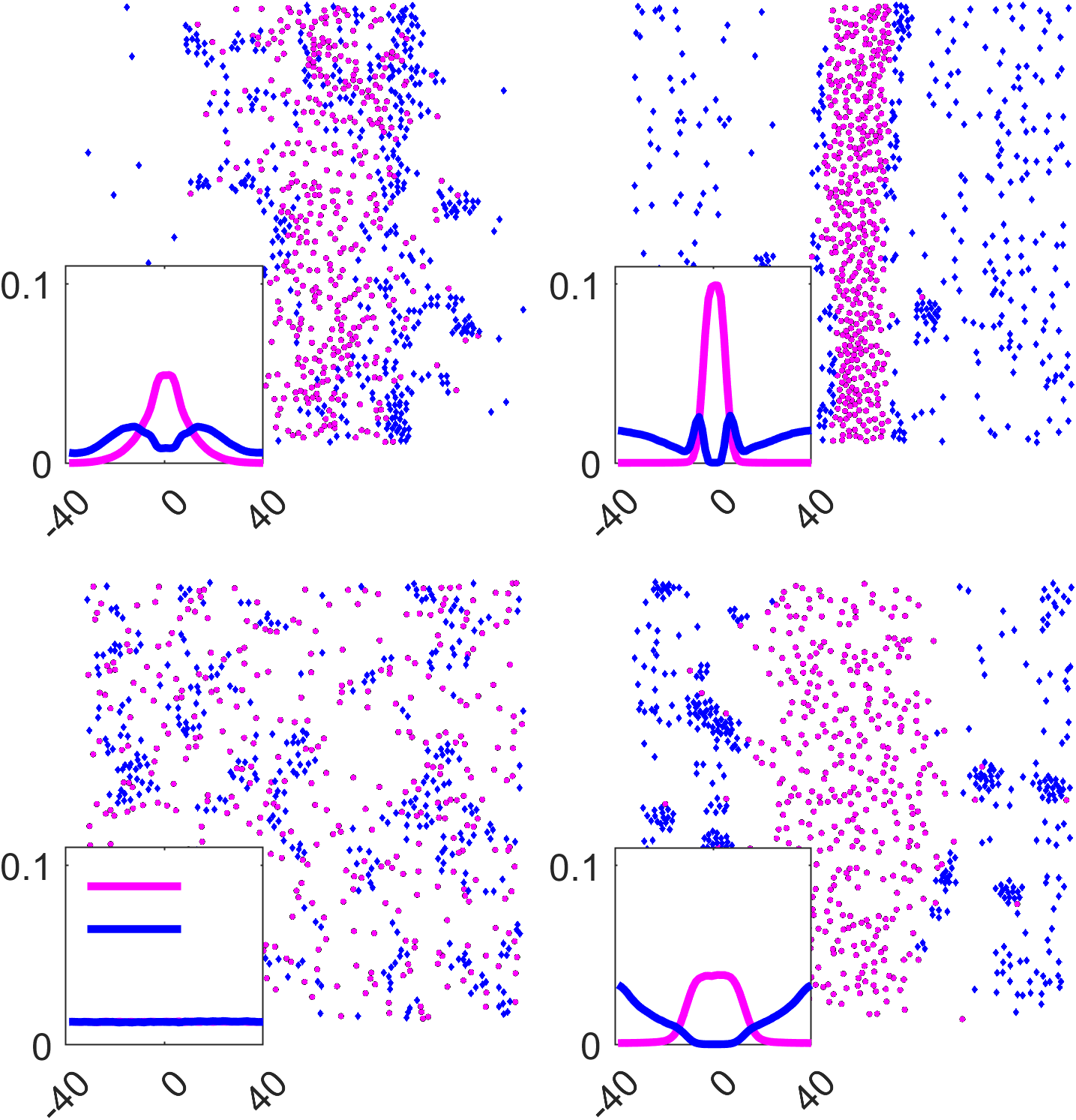}
        \put(3,102){\makebox(0,0){(a)}}
        \put(56,102){\makebox(0,0){(b)}}
        \put(3,50){\makebox(0,0){(c)}}
        \put(56,50){\makebox(0,0){(d)}}
        
        \put(37,102){\makebox(0,0){$\gamma=0,\,\delta=\lambda/2$}}
        \put(83,102){\makebox(0,0){$\gamma\gg 1/\xi,\,\delta=\lambda/2$}}
        \put(38,50){\makebox(0,0){$\gamma=0,\,\delta=0$}}
        \put(85,50){\makebox(0,0){$\gamma\gg 1/\xi,\,\delta=0$}}
        
        \put(20,56.5){\makebox(0,0){$r/\sigma$}}
        \put(69,56.5){\makebox(0,0){$r/\sigma$}}
        \put(20,4){\makebox(0,0){$r/\sigma$}}
        \put(69,4){\makebox(0,0){$r/\sigma$}}  
        
         \put(20,22){\makebox(0,0){$\hat{\rho}^T$}}
        \put(20,18){\makebox(0,0){$\hat{\rho}^H$}}

            \begin{tikzpicture}[overlay]
                    \draw[->,thick] (1.8,-.25) -- (7.5,-.25);  % Simple arrow from (1,1) to (5,4)
                    \node at (4.,-.5) [] {$\gamma$};  % Label above the arrow
                    \draw[->,thick] (-.235,1.5) -- (-.235,7.5);
                    \node at (-.4,4) [] {$\delta$};
    \end{tikzpicture}
        \end{overpic}
        \vspace{.3cm}
    \caption{Long-time configurations of the agent-based system for exemplary values of the two control parameters $\gamma$ and $\delta$; the insets represent the time-averaged histograms of targets $\hat{\rho}^T(x)$ and herders $\hat{\rho}^H(x)$ as a function of the relative position from the $y$-axis $x$, obtained from the $2$ dimensional histogram after integrating over the $y$ direction. At $\gamma=\delta=0$ we observe a disordered state with homogeneous density distribution (c). The other panels show that, as soon as $\delta$ or $\gamma$ are non zero,  inhomogeneous configurations are reached (consistent with Eq.~(6) of the main text), where we observe that herders tend to surround targets.}
    \label{fig:rectangular_snapshots}
\end{figure}

The remaining dynamical elements from Eq. (7-8) in the Methods and discussed in the main text are preserved. Figure \ref{fig:rectangular_snapshots} presents snapshots of the agent-based simulations for different values of control parameters $\gamma$ and $\delta$, along with the corresponding agent densities along the $x$ direction, $\hrt(x)$ and $\hrh(x)$.
The numerical density calculations follow a methodology similar to that used for the radial densities $\hrt(r)$ and $\hrh(r)$ in the circular case, with two key differences: first, we compute histograms using Cartesian coordinates $(x,y)$ rather than polar coordinates $(r,\theta)$, and second, we integrate over the $y$ axis instead of averaging over the polar angle $\theta$. Additionally, unlike the radial case where geometric constraints limited our analysis to agents within a circle of radius $L/2$, here we include all agents in our density calculations.
The density profiles exhibit striking similarities between the circular and rectangular geometries. Notably, the rectangular configuration's snapshots demonstrate that the system can be effectively reduced to a one-dimensional setting through $y$-axis averaging. This simplification establishes direct connections to our one-dimensional continuum framework and supports our assertion that one-dimensional field theories can capture the essential features of the shepherding problem.

\subsection{Robustness of results for the fraction of confined targets}
%\subsection{The effects on $\chi(R^*)$ of different $R^*$ values}

\begin{figure*}[htb]
\vspace{1cm}

        \begin{overpic}[width=.95\textwidth]{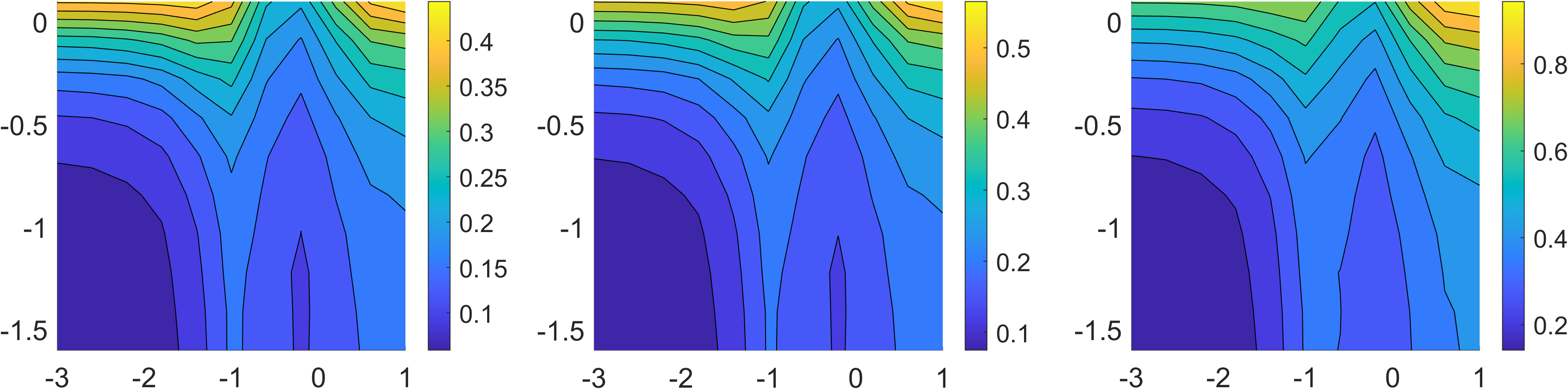}
                    \put(3,28){\makebox(0,0){(a)}}
                    \put(37,28){\makebox(0,0){(b)}}
                    \put(70,28){\makebox(0,0){(c)}}
                    \put(15,28){\makebox(0,0){$\chi(R^{\text{min}})$}}
                    \put(50,28){\makebox(0,0){$\chi(R^{\text{max}})$}}
                    \put(85,28){\makebox(0,0){$\chi(R)$}}
                    \put(85,-1){\makebox(0,0){$log_{10}(\gamma)$}}
                    \put(50,-1){\makebox(0,0){$log_{10}(\gamma)$}}
                    \put(15,-1){\makebox(0,0){$log_{10}(\gamma)$}}
    \put(-1,13){\makebox(0,0){\rotatebox{90}{$log_{10}(\delta)$}}}
                    
    \end{overpic}
    \vspace{.5cm}
    
    \caption{Fraction of confined targets $\chi(R)$ for different choices of the size of the goal region, $R$, in the steady state of the agent-based simulations. (a-b) The values  of $R$ are derived from the minimum and  maximum allowed densities for a solid phase of hard spheres in 2D \cite{schmidt1996freezing}. These two values are $R^\text{min}\sim \sqrt{N_{\text{T}}\sigma^2/(1.15\pi)}$ and $R^\text{max}\sim\sqrt{N_{\text{T}}\sigma^2/(0.9\pi)}$. (c) Results for the value of $R$ considered  in  the main text, which are the same of those shown in Fig. 4(a) of the main text. We can  see that there is no significant difference in the behaviour of $\chi(\gamma  ,\delta)$.}
    \label{fig:chi_threshold_comparison}
\end{figure*}

A key order parameter to characterize the inhomogeneous steady state resulting from the shepherding process is the fraction of confined targets, $\chi(R)$, within a circle of radius $R$ centered around the origin, see Fig. 4(a) of the main text.
In these agent-based calculations, $R$ has been chosen based on the fitted density profile of the targets in the case with the largest considered values for both $\gamma$ and $\delta$. Here we present data for two alternative choices of $R$. These are inspired by the fact that the targets in our model are represented by mutually repelling disks. It is well known that hard-disk systems, in thermal equilibrium, exhibit a first-order freezing transition from a fluid into a solid state \cite{schmidt1996freezing}. Considering the target distribution within the goal region as solid-like, we can deduce the radii related to the minimum and maximum area fraction in such a solid, yielding $R^{\text{min}}\sim \sqrt{N_{\text{T}}\sigma^2/(1.15\pi)}$ and 
 $R^{\text{max}}\sim\sqrt{N_{\text{T}}\sigma^2/(0.9\pi)}$. The resulting phase diagrams (obtained after time - and ensemble averaging in the steady state) are shown in Fig.~\ref{fig:chi_threshold_comparison}(a-b). Comparing with the original result we conclude that the precise choice of $R$ has no significant impact on the structure of the diagram.

%\clearpage

\section{Continuum theory}
\subsection{Derivation of coupling functions}
%\subsection{Derivation of the mean field coupling functions}
%
In this section we provide details on the derivation of the one-dimensional field equations~(3-4) of the main text for the targets and herder densities, $\rt(x,t)$ and $\rh(x,t)$, respectively (see the Methods for a summary). 
The field equations for these conserved fields take the general form \cite{you2020nonreciprocity}
%Using a mean field approximation, 
%the field equations take the general form %\cite{you2020nonreciprocity}:
%
\begin{equation}
    \partial_t \rho(x,t)=-\nabla  \left[\langle F \rangle(x,t)\rho(x,t)\right]+D\nabla^2 \rho(x,t)
    \label{eqn:general_PDE}
\end{equation}
where the average force $\langle F \rangle(x,t)$ at position $x$ and time $t$ is derived from the microscopic forces $F_{i(a)}$ acting on individual agents through a mean-field assumption. Specifically, 
we employ the following ``recipe'':
%we transition from the discrete force expression to its continuum counterpart:
%
\begin{equation}
F_i(t)=\sum_{j\in \mathcal{N}_i} F_{ji}(t)\to \langle F(x,t) \rangle =\int_{\mathbb{B}(x)} F(x,y)\rho(y,t)\,dy
\label{eq:recipe}
\end{equation}
where $\mathbb{B}(x)$ denotes the finite interaction range between the agents centered around $x$. As explained in the Methods, the function $F(x,y)$ is translationally invariant (i.e., $F(x,y)=F(x-y)$ for the short-range and long-range repulsion in Eqs. (7-8) while it is not for the ``force'' imposed by the feedback control input in the dynamics of the herders. In any case, direct computation of the integral over $y$ is not possible since the density $\rho(y,t)$ is not known a priori. Following standard procedures \cite{you2020nonreciprocity}, we employ in all cases a gradient expansion
%
%We must calculate the term $\langle F \rangle(x)$ for each interaction type present in the agent-based equations Eq. \al{ref}, with each interaction corresponding to a different kernel $F(x,y)$. However, direct computation of this integral is not possible since the density $\rho(y,t)$ is not known a priori. Following standard procedures \al{ref}, we employ a gradient expansion of the density:
%
\begin{equation}
\rho(y)\simeq\rho(x)+\nabla\rho(x)(y-x)+\mathcal{O}(\nabla^2\rho(x))
\label{eq:gradient}
\end{equation}
where we take into account (at most) the term linear in the gradient.
%
%\begin{center}
 %   \textbf{Repulsion}\\
%\end{center} 
%

We first apply the procedure of Eq.~(\ref{eq:recipe}) with 
 Eq.~(\ref{eq:gradient}) 
to the pairwise repulsive
forces $\textbf{F}^{\text{rep}}_{\text{SR(LR)}}$ that
share the same functional form [see Eq (9)] and, crucially, depend only on agents relative distances.
Therefore, the corresponding kernels 
satisfy the condition of translational invariance; $F(x,y)=F(x-y)$. We also note that, for each $x$, we 
integrate over a domain in $y$ that is symmetric around $x$, and that $F(x-y)$ is, in our case, an odd function in this domain.
As an important consequence we find that the zeroth-order term in the gradient expansion of Eq. (\ref{eq:gradient}) cancels out. Therefore, the leading non-zero contribution comes from the gradient term, yielding
%Moreover, due to translational invariance, we can evaluate this integral at any point $x$, yielding
%
\begin{equation}
    \langle F^{\text{rep}}_{\text{SR(LR)}} \rangle (x)=\int_{x-\sigma}^{x+\sigma} F^{\text{rep}}_{\text{SR(LR)}}(x-y)(y-x)\, dy\,\nabla\rho(x)
    \label{eq:fpair}
\end{equation}
where we have used that $\nabla\rho(x)$ is independent of $y$ and can therefore be pulled out of the integral.

From Eq.~(\ref{eq:fpair}) we obtain for the short-range (SR) average forces
%
%We begin with the simpler cases of Short-Range (SR) and Long-Range (LR) repulsive forces, which share the same functional form and, crucially, are both translationally invariant -- that is, their kernels satisfy $F(x,y)=F(x-y)$. This translational invariance has an important consequence: when we apply the gradient expansion, the first-order term vanishes due to symmetry considerations, i.e.
%
%\begin{equation}
%    \langle F^{\text{SR}} \rangle (x)=\int_{x-\sigma}^{x+\sigma} F^{\text{rep}}_{\text{SR}}(x-y)\left( \cancel{\rho(x)}+\nabla\rho(x)(y-x) \right)\, dy
%\end{equation}
%The first-order term vanishes because we integrate over a domain in $y$ that is symmetric around $x$. 
%

\begin{align}
    k^{\text{rep}}\langle F^{\text{SR}} \rangle (x)&=
    %k^{\text{rep}}\int_{x-\sigma}^{x+\sigma} F^{\text{rep}}_{\text{SR}}(x-y)(y-x)\nabla\rho(x)\, dy =\nonumber\\
    &=k^{\text{rep}}\int_{x-\sigma}^{x+\sigma} \sign(x-y)(\sigma-|x-y|)(y-x)\nabla\rho(x)\, dy  =-\alpha ^{\text{SR}}\nabla\rho(x)=-k^{\text{rep}}\frac{\sigma^3}{3}\nabla\rho(x)
\end{align}
Analogously, the contribution from the LR forces follows as
\begin{equation}
    k^{\text{T}}\langle F^{\text{LR}} \rangle (x)=k^{\text{T}}\int_{x-\lambda}^{x+\lambda} F^{\text{rep}}_{\text{LR}}(x-y)(y-x)\nabla\rho(x)\, dy =-\alpha ^{\text{LR}}\nabla\rho(x)=-k^{\text{T}}\frac{\lambda^3}{3}\nabla\rho(x)
\end{equation}

%\begin{center}
%    \textbf{Decision-making interaction}\\
%\end{center}
We now turn to the feedback control term that represents decision-making. As discussed in the Methods, the corresponding kernel describing how the herders select the target to chase and how to place themselves relative to the target is not translationally invariant (since the decision-making depends on agent positions relative to the goal region at $x=0$), that is, $F(x,y)$ also depends separately on $x$ and $y$. To simplify the calculations, We first take the control input from Eq. (2) of the main text in its 1D form, $u_i=-k^{\text{H}}(H_i-T_i^*-\delta \sign(T_i^*))$, and invert the order of the $\sign(\cdot)$ and the weighted average relative to herder $i$, denoted as $(\cdot)_i^*$. This yields $u_i=-k^{\text{H}}(H_i-T_i^*-\delta (\sign(T))_i^*)$. Notice that these two expressions are equivalent throughout most of the domain, specifically for $\xi\leq |H_i|\leq L/2-\xi$, with $L\gg\xi$. We then linearize the exponent and neglect the denominator in the selection rule (see Methods), yielding
%
% $F(x,y)$ depends separately and nonlinearly on $x$ and $y$. To deal with the latter point, we linearize the exponent and neglect the denominator in the selection rule (see Methods), yielding

\begin{equation}
    F^{\text{DM}}(x,y)=-k^{\text{H}}(1+\gamma(|y|-|x|))(x-(y+\sign(y)\delta))\,.
\label{eq:kernel_DM}
\end{equation}

As a consequence of the lacking translational invariance, the symmetry arguments used above, which lead to the vanishing of the zeroth order term in the gradient expansion, do not apply any more. Thus, the general form of this term is given by
\begin{equation}
    \langle F^{\text{DM}} \rangle (x)=\int_{x-\xi}^{x+\xi} F^{\text{DM}}(x,y)\left( {\rho(x)}+\nabla\rho(x)(y-x) \right)\, dy =g_1(x) \rho(x) + \nabla \rho(x) g_2(x)
    \label{eq:g1_g2}
\end{equation}
where $g_1(x)$ and $g_2(x)$ are functions depending explicitly on $x$ (rather than constants as in the case of translational invariance discussed before).
Clearly, $g_{1,2}(x)$ also depend on the parameters $\gamma$ and $\delta$, but we drop this here for notational ease.

To proceed, we differentiate the domain into three regions: (i) $x\in[-\xi,0]$ and $x\in[0,\xi]$, (ii) $x\in(\xi,L/2-\xi]$ and 
 $x\in[-L/2+\xi,-\xi)$, (iii) $x\in(L/2-\xi,L/2)$ and $x\in[-L/2,-L/2+\xi)$. 
%If one pays attention to this aspect, the calculations are straightforward. 
We here focus on the intervals $x\in[0,\xi]$ and $x\in(\xi,L/2-\xi ]$, the other cases follow analogously. 
%
%A second consequence of the lack of translational invariance, is that the resulting coupling functions will depend on $x$ as well, requiring careful treatment when computing the average forces at the continuum level.  Specifically we need to differentiate the domain into three regions: i) $x\in[-\xi,0]$and$x\in[0,\xi]$, ii) $x\in(\xi,L/2-\xi]$ ,iii) $x\in(L/2-\xi,L/2)$ and $x\in[-L/2,-L/2+\xi)$. If one pays attention to this aspect, the calculations are straightforward. We report the key steps  for $x\in[0,\xi]$ and $x\in(\xi,L/2-\xi$, the other cases being extremely similar. Finally, we recall that the interaction kernel, $F^{\text{DM}}(x,y)$, is given by
%
Starting with $g_1(x)$, we find for $0\leq x \leq \xi$

\begin{align}
    g_1(x)|_{x\in [0,\xi] }=&-k^{\text{H}}\int_{x-\xi}^{x+\xi}(1+\gamma(|y|-|x|))(x-(y+\sign(y)\delta))\,dy=\nonumber\\
    =& -k^{\text{H}}\left( \int_{x-\xi}^{0}   (1+\gamma(-y-x))(x-(y-\delta))\, dy   + \int_{0}^{x+\xi}   (1+\gamma(y-x))(x-(y+\delta))\, dy \right)=\nonumber\\ 
    =&k^{\text{H}} \left( \delta(1-2\gamma x)(x-\xi)+\delta (x+\xi) +\gamma x (\xi^2-x^2)+\frac{2}{3}\gamma x^3\right)
\end{align}
For the interval $\xi< x < L/2-\xi$, the calculations are simpler since $y$ keeps the same sign throughout the integration domain. We obtain:

\begin{align}
    g_1(x)|_{x\in (\xi,L/2-\xi] }=&-k^{\text{H}}\int_{x-\xi}^{x+\xi}(1+\gamma(|y|-|x|))(x-(y+\sign(y)\delta))\,dy=\nonumber\\
    =& -k^{\text{H}}\int_{x-\xi}^{x+\xi}   (1+\gamma(y-x))(x-(y+\delta))\, dy   =\nonumber\\ 
    =&k^{\text{H}}\left( 2\delta\xi +  \frac{2}{3}\gamma \xi^3\right)
\end{align}

Similar calculations can be performed for $g_2(x)$. Starting with the interval $0\leq x \leq \xi$, we find

\begin{align}
    g_2(x)|_{x\in [0,\xi] }=&-k^{\text{H}}\int_{x-\xi}^{x+\xi}(1+\gamma(|y|-|x|))(x-(y+\sign(y)\delta))(y-x)\,dy=\nonumber\\
    =& -k^{\text{H}}\left( \int_{x-\xi}^{0}   (1+\gamma(-y-x))(x-(y-\delta))(y-x)\, dy   + \int_{0}^{\xi}   (1+\gamma(y-x))(x-(y+\delta))(y-x)\, dy \right)=\nonumber\\ 
    =&k^{\text{H}} \left(\delta(1-\gamma x)(\xi^2-x^2)+\frac{\delta\gamma +1}{3}2\xi^3 -\frac{2\gamma x}{3}(\xi^3-x^3)+\frac{\gamma}{2}(\xi^4-x^4)\right)
\end{align}
whereas for $\xi< x < L/2-\xi$,
\begin{align}
g_2(x)|_{x\in (\xi,L/2-\xi] }=&-k^{\text{H}}\int_{x-\xi}^{x+\xi}(1+\gamma(|y|-|x|))(x-(y+\sign(y)\delta))(y-x)\,dy=\nonumber\\
    =& -k^{\text{H}} \int_{x-\xi}^{z+\xi}   (1+\gamma(y-x))(x-(y+\delta))(y-x)\, dy =\nonumber\\ 
    =&k^{\text{H}}\frac{2}{3}(\gamma\delta+1)\xi^3 
\end{align}
Using the same same steps for the other intervals we finally obtain
the expressions of $g_1(x)$ and $g_2(x)$ reported in Eqs.~(15-16) in the Methods. As explained there, the functions $g_1(x)$ and $g_2(x)$  coincide with the functions $v_1(x)$ and $v_2(x)$ in Eq.~(3-4) of the main text up to a constant factor derived from the adimensionalization of the equations; furthermore, to obtain $v_2(x)$ we have to subtract from $g_2(x)$ the (constant) contribution arising from the SR repulsion between herders and targets $\alpha^{\text{SR}}$. 
\subsubsection{Higher order terms in the gradient expansion}
In this section we briefly consider consequences
of including higher order terms in the gradient expansion given in 
Eq.~(\ref{eq:gradient}). The field equations describing the dynamics of $\rh(x,t)$ and of $\rt(x,t)$ will then have the form

\begin{align}
    \partial_t \rt=& \nabla\cdot \left[{D}^{\text{T}}(\rt)\nabla\rt+\tilde k^{\text{T}}\rt\nabla\rh +\rt\sum_{i=3}^{Z}w_{i}\nabla^{(i-1)}\rh\right]
    \label{eqn:T_field_high_order}\\
    \partial_t \rh=& \nabla\cdot\left[{D}^{\text{H}}(\rh)\nabla\rh- v_1(x)\rh\rt-v_2(x)\rh\nabla\rt +\rh\sum_{i=3}^{Z}v_{i}(x)\nabla^{(i-1)}\rt \right]
     \label{eqn:H_field_high_order}
\end{align}
where $Z$ is maximum order of the expansion we consider, and $w_i$ and $v_i(x)$ are the corresponding couplings.
%$v_i(x)$ and $w_i$ are the corresponding coupling functions. 
Notice that, as argued before, the coupling functions $w_i$ have to be constant since all kernels related to the interactions in the target's dynamics are translationally invariant. On the other hand, the decision-making interaction kernel in the herders' dynamics is not translationally invariant; this yields the $x$-dependence of $v_i(x)$.

As shown in the main text, for the continuum equation system up to linear order in the gradients, there is no homogeneous steady state.
Interestingly, this statement holds for \textit{any} order $Z$ of the gradient expansion. To see this, we evaluate 
Eqs.~(\ref{eqn:T_field_high_order}) and (\ref{eqn:T_field_high_order})
at $\rh=\rh_0$, $\rt=\rt_0$, yielding
%A key point worth to notice is that, as discussed in the main text, the homogeneous state is not a steady state of the above field equations, for \textit{any} order $N$ of the gradient expansion: in fact, when evaluated at the homogeneous state $\rh=\rh_0$, $\rt=\rt_0$, we have
%
\begin{align}
    &\rt_0\sum_{i=3}^{Z}w_{i}\nabla^{(i-1)}\rh_0=0\\
    &\rh_0\sum_{i=3}^{Z}v_{i}(x)\nabla^{(i-1)}\rt_0=0, \quad \forall Z\geq3
\end{align}
The above relations imply that Eq.~(6) of the main text holds true at any order $Z$ of the gradient expansion. We conclude that the "shepherding" inhomogeneities that we observe are {\em not} just a by-product of the order of truncation of the gradient expansion of 
Eq.~\ref{eq:gradient}; rather they result from the decision-making itself.

\subsection{Analysis in the absence of decision-making}
\label{sec:analysis_without_decision}
Without decision-making (i.e., $\gamma=0$, $\delta=0$), the interaction kernel
Eq.~(\ref{eq:kernel_DM}) becomes translationally invariant, that is,
\begin{equation}
    F^{\text{DM}}(x,y)|_{\gamma=0,\,\delta=0}=-k^{\text{H}}(x-y).
\end{equation}
As a result, the functions $g_1(x)$ and $g_2(x)$ appearing in Eq.~(\ref{eq:g1_g2}) reduce to constants, and 
we recover expressions similar to those found for the repulsive interaction kernel. Specifically,
\begin{align}
g_1(x)&=0\nonumber\\
g_2(x)&=\frac{2}{3}k^{\text{H}}\xi^3=   g_2^0
\end{align}
The resulting equations can be written as 
\begin{align}
\partial_t \rt=& \nabla\cdot \left[{D}^{\text{T}}(\rt)\nabla\rt+\tilde{k}^{\text{T}}\rt\nabla\rh \right]
    \label{eqn:T_field_1_0}
    \end{align}
    \begin{align}
    \partial_t \rh%|(\gamma=\delta=0)
    = &\nabla\cdot\left[D^{\text{H}}(\rh)\nabla\rh- v_2^0\rh\nabla\rt \right]
     \label{eqn:H_field_1_0}
\end{align}
where the density-dependent couplings $D^{\text{T}}(\rho^\text{T})$ (with $\text{A=H,T}$) and the constants $\tilde{k}^{\text{T}}$ and 
$v_2^0$ are defined in the Methods. 
The above equations involve nonreciprocal cross-diffusive couplings $\nabla\cdot[\rho^{\text{A}}\nabla\rho^{\text{B}}]$ akin of the nonreciprocal Cahn-Hillard model \cite{saha2020scalar,you2020nonreciprocity}. The new coupling $\nabla\cdot[v_1(x)\rt\rh]$ arising from herders' decision-making is absent, as expected at $\gamma=\delta=0$. 
%
%\al{for the definition of the above quantities  see Sec. REF of the Methods; an important point to notice is that all the couplings ($D^{\text{T}}$,$D^{\text{H}}$, $v_2^0$ and $\tilde{k}^{\text{T}}$) are strictly positive in our setup}. 
%The above equations involve nonreciprocal cross-diffusive couplings $\nabla\cdot[\rho^{\text{A}}\nabla\rho^{\text{B}}]$ that are common in scalar nonreciprocal field theories \cite{saha2020scalar,you2020nonreciprocity}. The new coupling $\nabla\cdot[v_1(x)\rt\rh]$ arising from herders' decision-making is absent, as expected at $\gamma=\delta=0$. 

We now perform a linear stability analysis (LSA) around the homogeneous, stationary state $\rh(x,t)=\rh_0$ and $\rt(x,t)=\rt_0$ which corresponds to the trivial solution of 
Eqs.~(\ref{eqn:T_field_1_0}) and (\ref{eqn:H_field_1_0}). To this end we introduce small fluctuations, i.e.,
\begin{align}
\rho^{\text{A}}(x,t)=\rho_0^{\text{A}}+\delta\rho^{\text{A}}(x,t)
%    &\rt=\rt_0 +\drt (x,t)\\
 %   &\rh=\rh_0 +\drh (x,t)
 \label{eq:fluc}
\end{align}
and linearize the field equations in these fluctuations, yielding
\begin{equation}
\partial_t \drt (x,t)=D^{\text{T}}(\rt_0)\nabla^2\drt (x,t)+ \rt_0\tilde{k}^{\text{T}}\nabla^2 \drh(x,t)
\label{eqn:H_linear_PDE}
\end{equation}
\begin{equation}
\partial_t \drh (x,t)=D^{\text{H}}(\rh_0)\nabla^2\drh (x,t) -v_2^0\rh_0 \nabla^2 \drt(x,t)
\label{eqn:T_linear_PDE}
\end{equation}

%We now perform  the stability analysis (LSA) around the homogeneous state $\rh(x,t)=\rh_0$ and $\rt(x,t)=\rt_0$ which is a stationary state solution of the field equations. To perform the LSA, we perturb the system around the homogeneous state
%\begin{align}
%   &\rt=\rt_0 +\drt (x,t)\\
%    &\rh=\rh_0 +\drh (x,t)
%\end{align}
%we can then assume that $\drt(x,t)$ and $\drh(x,t)$ are small perturbation and linearize  Eq. \al{refs}, obtaining

% where 
% \begin{align}
% &\tilde{D}^T=\rt_0\frac{k_{rep}}{3}\sigma^3 +D\\
% &\tilde{D}^H=\rh_0\frac{k_{rep}}{3}\sigma^3 +D\\
% &\tilde{K}^T=\rt_0(\frac{k_{rep}}{3}\sigma^3+\frac{k^T}{3}\lambda^3)\\
% &\tilde{K}^H=\rh_0(\frac{k_{rep}}{3}\sigma^3-\frac{2k^H}{3}\xi^3)
% \end{align}
We expand the fluctuations in a Fourier series, assuming that the growth rate $\Omega(q)$ is the same for both fields, that is
\begin{equation}
    \delta\rho^{\text{A}}(x,t) =\int dq\, e^{iqx+\Omega(q) t} \hat{\rho}^{\text{A}} (q)
    \label{eq:ansatz})
\end{equation}
Stability of the homogeneous, stationary state implies that the real part of the growth rate $\Omega(q)$,
$\text{Re}(\Omega)$, is negative for all nonzero wavenumbers $q$ (note that fluctuations at $q=0$ cannot occur anyway due to particle number conservation).
Inserting the ansatz (\ref{eq:ansatz}) into the linearized field equations \eqref{eqn:T_linear_PDE}, \eqref{eqn:H_linear_PDE} we obtain the eigenvalue problem 
%
%with ${\text{A}}=\{\text{T},\text{H}\}$; we also assume that the growth rate $\Omega(q)$ is the same for the two fields. We then plug the above relation in Eq. \eqref{eqn:T_linear_PDE},\eqref{eqn:H_linear_PDE}, and we can study the growth rate solving the following eigenvalues problem
%
\begin{equation}
    -q^2 \begin{bmatrix}
        {D}^{\text{T}}(\rt_0) & \tilde{k}^{\text{T}}\rt_0\\
        -v_2^0\rh_0 & {D}^{\text{H}}(\rh_0)
    \end{bmatrix}\textbf{v}(q) = \Omega(q) \textbf{v}(q)
    \label{eqn:eigenproblem}
\end{equation}
where the eigenvalues $\Omega(q)$ corresponding to the eigenvectors $\textbf{v}(q)$ are given by
\begin{equation}
 \Omega(q)=-q^2 \frac{1}{2}\left( \Omega_0 \pm \sqrt{\Delta}\right)
 \label{eqn:growth_rates}
 \end{equation}
 with 
 \begin{align}
 \Omega_0&={D}^{\text{T}}(\rt_0)+{D}^{\text{H}}(\rt_0) \nonumber\\
 \Delta&=\left(D^{\text{T}}(\rt_0)-D^{\text{H}}(\rt_0)\right)^2-4v_2^0\tilde{k}^{\text{T}}\rt_0\rh_0
 \label{eq:omega_0_delta}
 \end{align}

For parameter choices corresponding to our microscopic model, where all short-range interactions as well as the long-range herder-target interaction are repulsive, we have $D^{\text{A}}(\rho_0^{\text{A}})>0$ (for $\text{A}=\text{H,T}$) and 
$\tilde{k}^{\text{T}}>0$. Further, $v_2^0>0$ which, together with the negative sign of the corresponding term in Eq.~(\ref{eqn:H_field_1_0}), reflects the attraction of herders to the center-of-mass target. 

We first consider the case of equal number densities,
$\rho_0^{\text{H}}=\rho_0^{\text{T}}$, considered in the main text. In this case, we have $\Delta<0$ 
(for $v_2^0>0$, $\tilde{k}^{\text{T}}>0$)
such that the eigenvalues become imaginary. However, since $\Omega_0>0$, $\text{Re}(\Omega)$ is negative. Thus, the homogeneous steady state is stable.
This result of the LSA conforms with the outcome of our numerical simulations as shown in Fig.~3(c) of the main text.
Only by incorporating decision-making (that is, by introducing a coupling of the form 
$\nabla\cdot[v_1(x)\rt\rh]$ with $\gamma\neq 0$, $\delta\neq 0$) we achieve an inhomogeneous steady state characteristic of shepherding.

\begin{figure}
    \centering
    \begin{overpic}[width=.8\textwidth]{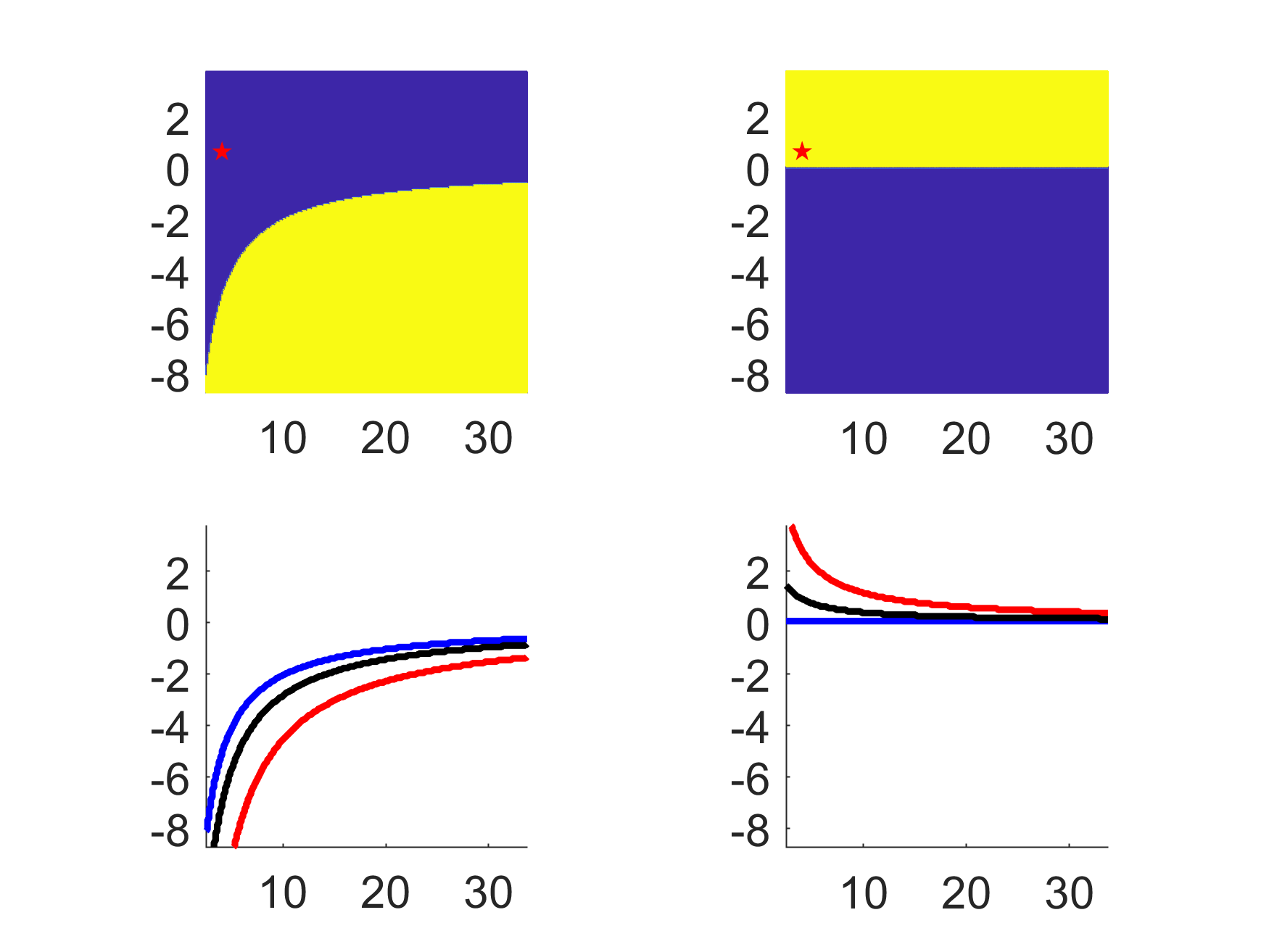}
    \put(10,72){\makebox(0,0){(a)}}
    \put(56,72){\makebox(0,0){(b)}} 
    \put(10,33){\makebox(0,0){(c)}}
    \put(56,33){\makebox(0,0){(d)}}
    
    \put(33,50){\makebox(0,0){$\text{Re}(\Omega)>0$}}
    \put(75,65){\makebox(0,0){$\text{Im}(\Omega)\neq0$}}
    
    \put(28,38){\makebox(0,0){$\tilde{k}^{\text{T}}$}}
    \put(75,38){\makebox(0,0){$\tilde{k}^{\text{T}}$}}
    \put(75,1){\makebox(0,0){$\tilde{k}^{\text{T}}$}}
    \put(28,1){\makebox(0,0){$\tilde{k}^{\text{T}}$}}

    \put(10,55){\makebox(0,0){\rotatebox{90}{$v_2^0$}}}
    \put(10,20){\makebox(0,0){\rotatebox{90}{$v_2^0$}}}
    \put(55,55){\makebox(0,0){\rotatebox{90}{$v_2^0$}}}
    \put(55,20){\makebox(0,0){\rotatebox{90}{$v_2^0$}}}

\end{overpic}
    \caption{Results of the LSA around the homogeneous, steady state ($\rt(x)=\rt_0$, $\rh(x)=\rh_0$) in the absence of decision-making (i.e., $\gamma=0$, $\delta=0$) in the parameter plane spanned by 
    $\tilde{k}^\text{T}$ and $v_2^0$. Parts (a) and (b) refer to the case $\rt_0=\rh_0=0.5\rho_0$ (with $\rho_0=0.5/\sigma$) considered in the main text, with the red star indicating the corresponding parameter combination $(\tilde{k}^\text{T}, v_2^0)$. Yellow regimes indicate where the real part of one of the growth rates is positive (a) or the imaginary part is nonzero (b). It is seen that the two regimes do not overlap. Parts (c) and (d) refer to the case of different densities (with the total density being the same as before). 
 Here we show the curves dividing regimes with 
 $\text{Re}(\Omega)>0$ from $\text{Re}(\Omega)\leq 0$ (c)
 and $\text{Im}(\Omega)\neq 0$ from $\text{Im}(\Omega)=0$)
(d) for three cases: $\rt_0=0.1\rho_0$ (red), $\rt_0=0.5\rho_0$ (blue), and $\rt_0=0.8\rho_0$ (black), with $\rh_0=\rho_0-\rt_0$.
In all cases, the real part is negative as long as $v_2^0>0$, which coincides with the condition for a nonzero imaginary part.}
 %      (a) and (b) respectively show in yellow the regions of the parameter space where one growth rate $\Omega$ has positive real part and non zero imaginary part  for the values of $\rt_0=\rh_0=0.5\rho_0$ used in the main text. The red star shows where are performed the simulations in the main text. 
% In  (c) (and (d)) we show, for values of $\rt_0=0.1\rho_0$ (red), $\rt_0=0.5\rho_0$ (blue), and $\rt_0=0.8\rho_0$ (black), with $\rh_0=\rho_0-\rt_0$, the curves dividing the regions with $Re(\Omega)>0$ from $Re(\Omega)\leq 0$, and $Im(\Omega)\neq 0$ from $Im(\Omega)=0$), respectively. In all the cases $\rho_0=0.5/\sigma$ as in the main text. We can see as long as $v_2^0>0$, $\Omega$ only has negative real part, whereas only for $v_2^0>0$ a non-zero imaginary part can be achieved.}
    \label{fig:LSA_g0d0}
\end{figure}

Similar considerations apply when we vary the densities
$\rt_0$, $\rh_0$ (keeping the total mass constant), or when we vary the constants $v_2^0$ and $\tilde{k}^{\text{T}}$
(keeping them positive). In all of these cases, we find 
$\text{Re}(\Omega)<0$. An overview of the results obtained by numerical solution of Eq.~(\ref{eqn:growth_rates}) is given in Fig. \ref{fig:LSA_g0d0}.

Still, it is interesting to investigate whether and at which coupling parameters the real part of the growth rate can become positive at all. Inspection of Eq.~(\ref{eqn:growth_rates}) shows that this can only occur
if $\Delta$ is positive and its square root exceeds the (positive) term $\Omega_0$. A necessary requirement is that either $\tilde{k}^{\text{T}}<0$ (i.e, targets are attracted by herders) or
$v_2^0<0$ (i.e, herders are repelled by targets), in contradiction to what we have assumed so far.
The corresponding parameter regions are indicated in Fig. \ref{fig:LSA_g0d0}.

Finally, regarding the imaginary part of the growth rate, $\text{Im}(\Omega)$, we see from Fig.~\ref{fig:LSA_g0d0} that there are no regions where both, $\text{Re}(\Omega)>0$ {\em and} $\text{Im}(\Omega)\neq 0$, since the respective conditions contradict each other. Therefore,
for the system without decision-making considered here,
we do not find traveling patterns that have been observed in nonreciprocal field theories of phase separating systems \cite{you2020nonreciprocity,saha2020scalar}.
A particularly prominent example featuring traveling patterns is the non-reciprocal Cahn-Hilliard (NRCH) model \cite{you2020nonreciprocity,saha2020scalar,brauns2024nonreciprocal}, and indeed, our present model (S21)-(S22) without decision making has important similarities with the NRCH. As in the NRCH, the present model is mass-conserving and the cross-couplings ($\tilde{k}^\text{T}$, $-v_2^0$) have different signs, yielding antagonistic (i.e., anti-reciprocal) couplings between different species.

However, there is a crucial difference with our model. In the NRCH, which is a generalized model of phase separation in binary mixtures, at least one of the intraspecies couplings is negative, reflecting attractive forces within agents of the same type. This leads to phase separation and the formation of a stable interface between the two phases (realized by a finite surface tension term in the NRCH) already for reciprocal cross-couplings. The underlying phase separation in the NRCH is a crucial prerequisite for the appearance of traveling patterns at high nonreciprocity \cite{brauns2024nonreciprocal}.

In contrast, in our model \eqref{eqn:T_field_1_0}-\eqref{eqn:H_field_1_0}, the prefactors related to intraspecies couplings are both positive due to the underlying repulsive (excluded-volume) interactions. This precludes the occurrence of traveling patterns (see Fig. 9 in \cite{brauns2024nonreciprocal}). We expect, however, that such patterns could occur when we add, e.g., adhesion between the targets.
We will later show (see Sec.~\ref{sec:pattern_design} of the SI) that traveling patterns can indeed occur for appropriate choices of the decision-making coupling term.

%\clearpage
%\subsection{Role of $v_1(x)$ and $v_2(x)$ in determining the sharpness of the herder-target interface}
%
\subsection{Sharpness of the herder-target interface}
\label{sec:sharpness}
As shown in Fig. 4(d) of the main text, 
our continuum theory in presence of decision-making predicts the maximum sharpness of the target-herder interface, $\sigma/\Delta$, at large values of $\gamma$ and small values of $\delta$. In contrast, at the agent-based level the maximum is found at large values of both, $\gamma$ and $\delta$. To explain this difference between the two levels of description we here 
present an argument based on the functions $v_1(x)$ and 
$v_2(x)$ entering the field equations Eqs. (3-4) of the main text in presence of decision-making. As we will see, these coupling functions have, to some extent, competing roles.
%
%explore discuss the role that the couplings $v_1(x)$ and $v_2(x)$ have at the continuum level, which, as we shall see, can be competing roles to some extent. 

To this end, we consider the following simplified scenario. 
We assume that the targets have already arranged in a confined situation, that is, $\rt(x)$ can be described by a hyperbolic-tangent profile, compatible with what is observed at long times in Fig. 2(d). Assuming this profile to be static
(i.e., $\partial_t\rt=0$) we can focus on the dynamics of herders alone (once this is understood, we can infer the consequences for the sharpness of the interface).

\begin{figure*}[!htb]
\vspace{1cm}

\begin{overpic}[width=.8\textwidth]{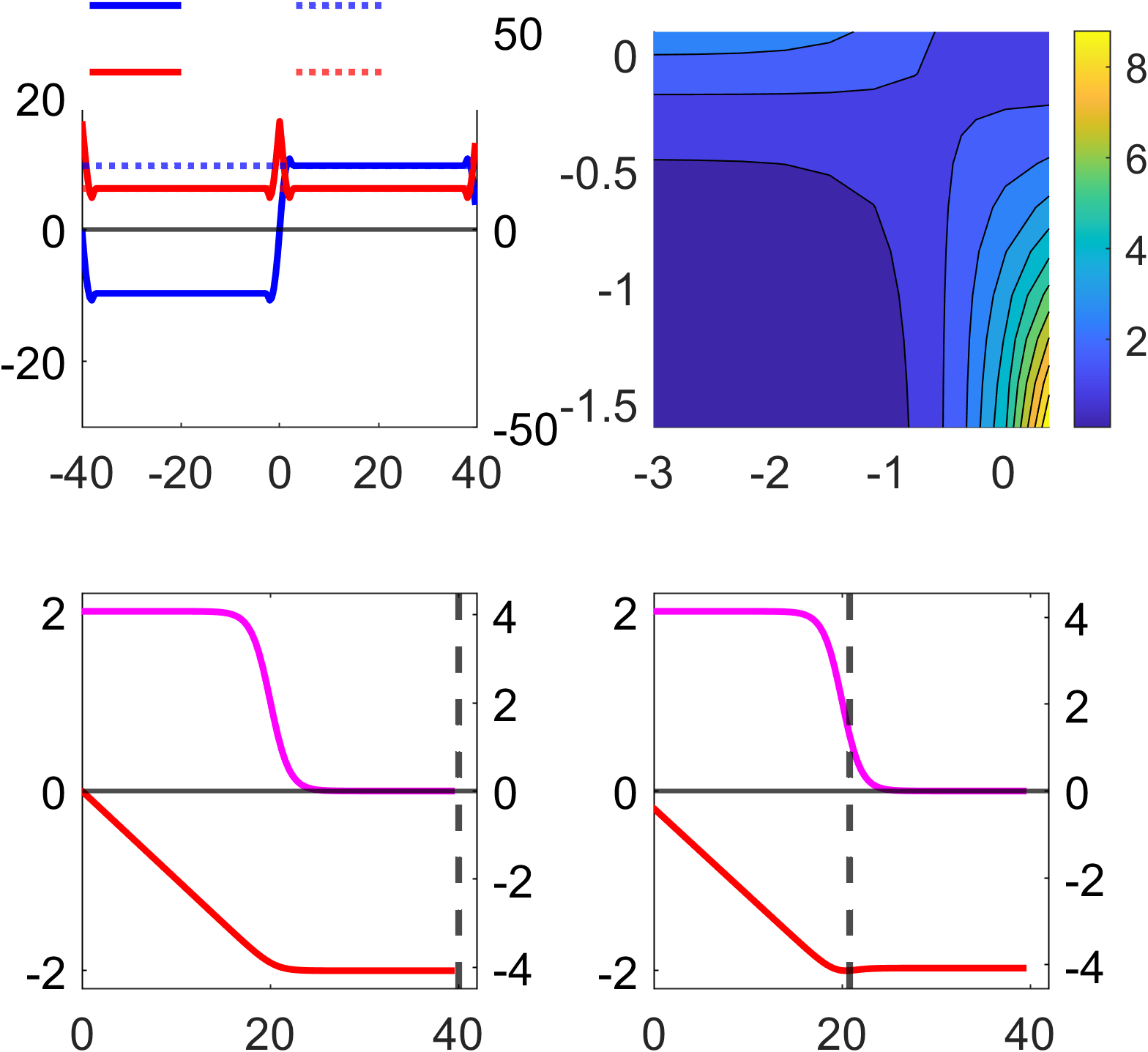}
    \put(24,45){\makebox(0,0){$x/\sigma$}}
    \put(24,-2){\makebox(0,0){$x/\sigma$}}
    \put(74,-2){\makebox(0,0){$x/\sigma$}}
    
    \put(3,92){\makebox(0,0){(a)}}
    \put(50,92){\makebox(0,0){(b)}}
    \put(3,42){\makebox(0,0){(c)}}
    \put(50,42){\makebox(0,0){(d)}}

    \put(19,92){\makebox(0,0){$v_1(x)$}}
    \put(19,86){\makebox(0,0){$v_2(x)$}}
    \put(35,86){\makebox(0,0){$\tilde{v}_2$}}
    \put(35,92){\makebox(0,0){$\tilde{v}_1$}}

    \put(74,92){\makebox(0,0){$\tilde{v}_1/\tilde{v}_2$}}
    \put(74,45){\makebox(0,0){$log_{10}(\gamma)$}}
    \put(49.5,67){\makebox(0,0){\rotatebox{90}{$log_{10}(\delta)$}}}
    \put(43,77.5){\makebox(0,0){\rotatebox{90}{$v_2,\,\tilde{v}_2$}}}
    \put(0,75){\makebox(0,0){\rotatebox{90}{$v_1,\,\tilde{v}_1$}}}

    \put(0,20){\makebox(0,0){\rotatebox{90}{$\rt(x)$}}}
    \put(46,20){\makebox(0,0){\rotatebox{-90}{$V(x)$}}}

    \put(51,20){\makebox(0,0){\rotatebox{90}{$\rt(x)$}}}
    \put(97,20){\makebox(0,0){\rotatebox{-90}{$V(x)$}}}
\end{overpic}
\vspace{1cm}
\caption{(a) The functions $v_1(x;\gamma,\delta)$
and $v_2(x;\gamma,\delta)$ defined
 in the Methods, and their approximated amplitudes $\tilde{v}_1(\gamma,\delta)$ and $\tilde{v}_2(\gamma,\delta)$; the dependence on $\gamma$ and $\delta$ was omitted in the legend for the sake of visibility.  (b) The ratio of the amplitudes of $v_1$ and $v_2$, $\tilde{v}_1/\tilde{v}_2$ as a function of the control parameters $\gamma$ and $\delta$. As explained in Sec.~\ref{sec:sharpness}, $\tilde{v}_1/\tilde{v}_2$ determines the steepness $\sigma/\Delta$ [Fig. 4(d) of the main text]. 
 (c-d) Representative hyperbolic-tangent profile of $\rt(x)$ (magenta line) and respective effective potential $V(x)$ (red line) as defined in Eq.~(\ref{eqn:potential}) for (c) $\tilde{v}_1=0.1$ and $\tilde{v}_2=0$, and (d) $\tilde{v}_1=0.1$, $\tilde{v}_2=0.2$. Dashed black lines indicate the position of the minimum of $V(x)$. Increasing $\tilde{v}_2$, the position of the minimum of $V(x)$ moves towards regions where $\rt$ assumes a larger value. }
\label{fig:v1v2_competing}
\end{figure*}

As a further simplification of the herder dynamics, we approximate the function $v_1(x)$ by $\tilde{v}_1\sign(x)$ and 
$v_2(x)=\tilde{v}_2$, as motivated by the actual shape of these functions shown in Fig.~\ref{fig:v1v2_competing}(a). The definition of the (positive) amplitudes $\tilde{v}_1$ and $\tilde{v}_2$ is also indicated in 
Fig.~\ref{fig:v1v2_competing}(a). Neglecting short-ranged contributions contained in $D^{\text{H}}(\rh)$, the herder profile then evolves according to 
    \begin{equation}
    \partial_t \rh= \nabla\cdot[D'\nabla\rh-\tilde{v}_1\sign (x)\rh\rt- \tilde{v}_2\rh\nabla\rt ]
     \label{eqn:H_field_v1v2conflicting}
\end{equation}
%
%which is Eq \al{ref} of the Main Manuscript where we neglected the $x$ dependence of the function $v_2(x)=\tilde{v}_2$, simplified the function $v_1(x)$ to be a $\tilde{v}_1\sign(x)$ function, and neglected the dependence on $\rh$ of the diffusion coefficient $D^{\text{H}}(\rh)$; the graphical definition of the amplitudes $\tilde{v}_1$ and $\tilde{v}_2$, with respect to the functions $v_1(x)$ and $v_2(x)$ can be seen in Fig. \ref{fig:v1v2_competing}(a).
%Being interested in the sharpness at the steady state, we assume  $\rt(x)$ to be a hyperbolic-tangent profile, compatibly with what observed for long times in the Main \al{ref}, and we neglect the time evolution of $\rt$, i.e. $\partial_t\rt=0$.
where $D'=D\frac{\tau}{\sigma^2}$ is the adimensionalised diffusion coefficient.
Focusing on the region $x\geq 0$, 
we find from Eq.~(\ref{eqn:H_field_v1v2conflicting})
that the term involving $\tilde{v}_1$ drives the herders "to the right" (i.e., larger values of $x$), whereas the term involving $\tilde{v}_2$ drives the herders towards higher values of $\rt(x)$, which means to move to the "left". In this sense, the two terms have conflicting roles.

If the term $\propto \tilde{v}_1$ dominates (i.e., $\tilde{v}_2=0$, or $\tilde{v}_1\gg\tilde{v}_2$), the herders will all move to the right-most region and push the targets to the left. This will support the system in developing a sharp 
herder-target interface. However, if we increase $\tilde{v}_2$ from small values, the herders have to find a balance between  the need to move right according to $\tilde{v}_1$, and to move left according to $\tilde{v}_2$. In this case, the herders will distribute themselves around the interface. Therefore, and since the herders exert a long-range repulsion on the targets, the interface will spread out, causing a decrease of the sharpness. 

The preferred localization of herders depending on $\tilde{v}_1$
and $\tilde{v}_2$ can be seen more explicitly when we solve the one-dimensional Fokker-Planck
Eq.~(\ref{eqn:H_field_v1v2conflicting}), yielding the steady-state solution [\cite{zwanzig2001nonequilibrium}]
%
%the right of the targets, pushing them to the left, hence creating a sharp herder-target interface.
%However, if we increase $\tilde{v}_2$, the herders will move finding a balance between  the need to move right according to $\tilde{v}_1$, and to move left according to $\tilde{v}_2$. In this case, the herders will distribute themselves around the target inteface, and since the herders will exert repulsion on the targets, the interface will spread.
%This is well captured also if we look at the shape of the steady state of Eq. \ref{eqn:H_field_v1v2conflicting}: it is known that such a Fokker-Planck equation has a steady state of the form \al{ref}

% \begin{equation}
%     \rt=\rho_0(x)=\left\{\begin{aligned}
%        & c&\quad |x|\leq x_1\\
%        & c - c\frac{x-x_1}{x_2-x_1}&\quad x_1\leq|x|< x_2\\
%        & 0 &\quad |x|>x_2
%     \end{aligned}\right.
%     \label{eqn:rt_piecewise}
% \end{equation}

\begin{align}
    &\rh(x)\sim \exp\left[-V(x)/D'\right]\nonumber\\
    &\equiv \exp\left[\left(\tilde{v}_2\rt(x) +\tilde{v}_1\int_0^x\rt(x')\,dx'\right)/D'\right]
    \label{eqn:potential}
\end{align}
where $V(x)$ may be interpreted as an effective potential. The minima of $V(x)$ indicate where the herders tend to accumulate. 
In Fig.~\ref{fig:v1v2_competing}(c-d) we show that, by increasing $\tilde{v}_2$ relative to $\tilde{v}_1$, the minimum of $V(x)$ moves towards positions with higher values of $\rt(x)$. Specifically, the minimum then occurs at the reflection point of the target profile, i.e., the center of the interface. Herders accumulating around this minimum disturb the distribution of targets (through herder-target repulsion), which effectively leads to a ``spreading'' of the target-herder interface. We conclude that the ratio $\tilde{v}_1/\tilde{v}_2$ 
should be a good indicator of the sharpness, $\sigma/\Delta$.

To close the argument we need to investigate 
how $\tilde{v}_1/\tilde{v}_2$ depends on $\gamma$ and $\delta$. The results are shown in 
Fig.~\ref{fig:v1v2_competing}(b). It is seen that the largest values of $\tilde{v}_1/\tilde{v}_2$ indeed are reached for large $\gamma$ and small $\delta$ values. This is completely consistent with the corresponding dependencies 
of $\sigma/\Delta$ shown in Fig. 4(d) of the  main text.

% \begin{figure}[!htb]
% \begin{overpic}[width=.5\textwidth]{Appendix/v1v2.png}
%                     \put(40,51){\makebox(0,0){(a)}}
%                     \put(95,51){\makebox(0,0){(b)}}
%                     \end{overpic}
%     \caption{ \al{check notazione, forse rimetti pannello su sharpness} (a) $v_1(\gamma,\delta,x)$ and $v_2(\gamma,\delta,x)$, and their approximated amplitudes $\tilde{v}_1(\gamma,\delta)$ and $\tilde{v}_2(\gamma,\delta)$; the dependence on $\gamma$ and $\delta$ was omitted in the legend for the sake of visibility.  (b) The ratio of the amplitudes of $v_1$ and $v_2$, $\tilde{v}_1/\tilde{v}_2$; this quantity drives the steepness $\sigma/\Delta$ [Fig. \ref{fig:PDE_colormap}(f)] \al{Aggiungi snapshot}.}
%     \label{fig:v1v2}
% \end{figure}

%\clearpage

%\section{Pattern design beyond \al{a simple shepherding task}}
\section{Pattern design beyond simple containment}
\label{sec:pattern_design}
In the main text we discussed how our continuum framework, beyond describing the specific shepherding task illustrated in Fig.~1 of the main text
could also be used to model pattern formation in more general decision-making systems (see Fig. 5 of the main text). To this end we propose to vary the spatial profile
of the function $v_1(x)$ that multiplies the product of $\rt$ and $\rh$ in Eq. (3) of the main text. In other words, we consider this function as an {\em adjustable} control input whose shape can be varied to generate a desired pattern. Below we first discuss the explicit choices for $v_1(x)$ underlying the continuum results in Fig.~5 of the  main text. Second, we describe strategies to translate these continuum rules to the agent-based level of description, yielding the microscopic results in Fig.~5 of the main text.
\subsection{Continuum: choices for $v_1(x)$}
\label{sec:general_v1}
%\section{Choices of the $v_1(x)$ function to reproduce Fig. 5}
%
In all of the four cases considered in Fig.~5 of the main text, the function $v_1(x)$ is a square wave with different periods or phase shifts, while 
$v_2(x)$ is set to a positive constant, i.e., $v_2(x)=\tilde v_2>0$.
With these choices, the herder dynamics is given by
%
%As shown in Fig CITA of the Main Manuscript, by understanding the role and the meaning of $v_1(x)$, we can design different $v_1(x)$ functions to reproduce a variety of behaviours. We consider for simplicity field equations analogous to Eq. \al{ref} of the Main Manuscript, where we approximated $v_2(x)=\tilde v_2>0$. Namely, for the herders
%
    \begin{equation}
    \partial_t \rh= \nabla\cdot[D^{\text{H}}(\rh)\nabla\rh-v_1(x)\rh\rt- \tilde{v}_2\rh\nabla\rt ]
     \label{eqn:H_field_v1_choice}
\end{equation}
The target dynamics remains unchanged, its evolution equation is given by (see Eq. (4) in the main text)
\begin{align}
    \partial_t \rt=& \nabla\cdot \left[{D}^T(\rt)\nabla\rt+\tilde{k}^T\rt\nabla\rh \right].
    \label{eqn:T_field}
\end{align}

The concrete choices for $v_1(x)$ in Eq.~(\ref{eqn:H_field_v1_choice}) corresponding to the four cases in Fig.~5 (main text) are as follows:
\begin{itemize}
    \item \textbf{Containment}\\
This case corresponds to the shepherding task discussed in the main part: herders need to collect the targets in a goal region around $x=0$. In the present study we have {\em derived} $v_1(x)$ 
from our agent-based model, yielding the expressions given in Eq. (15) of the Methods. As a simplified ansatz, we can use 
% \textcolor{red}{(Why do we need the sin-function in the argument
% of the sign-function?)}
%
%   This is the same task analyzed in the Main Manuscript, where we derive the function $v_1(x)$ as of Eq. \al{ref} from the agent-based equations \al{ref} where the herders need to collect the targets in a goal region around $x=0$. $v_1$ can be approximated as a square wave, that we right in the following way
%
    \begin{equation}
        v_1(x)=\tilde{v}_1 \sign(x)=\tilde{v}_1 \sign\left(\sin \left(\frac{2\pi x}{L}\right)\right)
        \label{eq:v1_containment}
    \end{equation}
Here, the appearance of the sin-function in the argument could be replaced by a simpler function that is uneven in $x$; e.g., just by $x$ itself. The ansatz (\ref{eq:v1_containment}) is particularly suitable for comparison with the case of static patterns discussed below.
%\al{the presence of the $\sin$ function  is not necessary in this specific case, but it helps with the comparison with respect to the $v_1(x)$ function needed to generate static patterns presented below}.  
Irrespective of this detail, Eq.~(\ref{eq:v1_containment}) generates currents of $\rh$ to the "right" (to the "left") if a non-zero $\rt(x)$ is observed for $x> 0$ ($x<0$), eventually generating an accumulation of the herders around $x=\pm L/2$. The herder-target repulsion then pushes the targets toward the goal region around $x=0$.

\item \textbf{Expulsion}\\
In another application, herders may want to push targets {\em away} from a predefined (potentially dangerous) region. This can be realized by reversing the rule given in Eq.~(\ref{eq:v1_containment}), such that 
%
%As we can see from Fig \al{ref}, we can recreate the case were the herders have to expel the targets from a \textit{safe} region in $x=0$. We can design $v_1(x)$ as
%
    \begin{equation}
        v_1(x)=-\tilde{v}_1\sign\left( \sin\left(\frac{2\pi x}{L}\right)\right)
        \label{eq:v1_expulsion}
    \end{equation}
This choice of $v_1(x)$ generates currents of $\rh$ to the "right" (to the "left") if a non-zero $\rt(x)$ is observed for $x< 0$ ($x>0$), eventually leading to the accumulation of the herders around $x=0$. Targets are then expelled towards the boundary.

Notice that, given the periodicity of the domain, containment around $x^*$ is equivalent to the expulsion from $x^*\pm L/2$: indeed the $v_1$-functions for containment [Eq.~(\ref{eq:v1_containment})] and expulsion [Eq.~(\ref{eq:v1_expulsion})] are the same but translated by $L/2$. This means that the two tasks are mathematically equivalent on this level of description.
\end{itemize}
The two cases discussed so far show that a change of sign of the function $v_1(x)$ around the position $x^*$ produces containment of targets around $x^*$ if $(x-x^*)v_1(x)\geq 0$ 
(for all $x\neq x^{*}$), or expulsion from the region around $x^*$ if $(x-x^*)v_1(x)\leq 0$ (for all $x\neq x^{*}$).
This suggests that more sophisticated patterns can be generated by including more zeros into $v_1(x)$.
An example is given in the third case considered in Fig.~5 (main text):
%
%What we understand from the two above cases is that, for every zero of the function $v_1$, say $x^*$, we can have either containment around $x^*$ if $(x-x^*)v_1(x)\geq 0$, or expulsion from $x^*$ if if $(x-x^*)v_1(x)\leq 0$. We can then achieve more sophisticated patterns including more zeros in the $v_1(x)$ function

\begin{itemize}
\item \textbf{Static patterns}\\
We consider a system where herders want to collect targets at several regions centered at positions $x_j^*$ in the domain, where $j$ is a positive integer going from $1$ to the number of regions where to collect the targets. As a consequence, the targets will be expelled by the other regions of the domain.
%while these targets are expelled from the remaining regions.
In a one-dimensional setup, the desired resulting state would be characterized by a stripe pattern.
As an example, we choose $x^*_{1,2}=\pm L/3$ and $x^{*}_3=0$ as the positions around which the targets shall be accumulated.
This can be realized with
%    We can recreate the case where the herders collect the targets in different goal regions centered around some points $x^*$ spread across the domain (or, equivalently, to expel them from a group of safe regions). For example, to create the case where the herders need to collect the targets around the positions in $x^*=\pm L/3$ and in $x^*=0$, we can use the following $v_1(x)$ function
    \begin{equation}
        v_1(x)=\tilde{v}_1\sign\left( \sin\left(\frac{6\pi x}{L}\right)\right)
    \end{equation}
    whose zeros $x^{*}_{1,2,3}$ satisfy the condition $(x-x^*_j)v_1(x)\geq 0$ for $x$ in a neighborhood of $x^*_j$ ($j=1,2,3)$.
    %\textcolor{red}{(for all $x\neq x^{*})$}. 
\end{itemize}
%
%
%The final case is considered since it is the simplest possible choice for the coupling function $v_1(x)$:
\begin{itemize}
\item \textbf{Traveling patterns}\\
Finally, it is interesting to consider the simplest choice 
for the function $v_1(x)$, namely
    \begin{equation}
        v_1(x)=\tilde{v}_1=const
    \end{equation}
 Indeed, as shown in Fig.~5 in the main text, a suitable choice for the constant $\tilde{v}_1$ produces a traveling pattern, where the herders persist to push the targets across the domain. A corresponding linear stability analysis predicting the occurrence of time-dependent motion is given in Sec.~\ref{sec:LSA_v1v2}. 
 %\textcolor{red}{It turns out that both constants have to be positive (CHECK!!! Do they just have to have the same sign??)}
 
 Intuitively, we can understand the travelling patterns as follows: If $\tilde{v}_1>0$ (and $v_2^0>0$), the herders will move towards the "right" as long as they observe a non-zero $\rt(x)$. This will eventually lead to the accumulation of the herders at the "right" of the targets. However, this cannot be a steady state since the coupling $-\tilde{v}_2\rh\nabla\rt$ pushes the herders towards the "left" where the targets are concentrated. This, in turn, will push the  target even more towards the "left" 
 as they try to escape the herders 
 [due to the coupling $\tilde{k}^{\text{T}} \rt\nabla\rh$ in the target dynamics, Eq.~(\ref{eqn:T_field})]. As a result, we observe traveling waves to the "left".
 
 % puts the herders back towards the target region (i.e., to the "left"). Targets then try to escape the herders 
 % (due to the coupling $\tilde{k}^{\text{T}} \rt\nabla\rh$ in the target dynamics), yielding a non-zero target density for all $x$. 
 % Then the chase-escape cycle starts again, yielding traveling waves to the "left".
 
% generating a chase-escape motion to the "left" of herders chasing targets (due to the coupling $\tilde{v}_2\rh\nabla\rt$) and the targets escaping by the herders (due to the coupling $\tilde{k}^{\text{T}} \rt\nabla\rh$).  
 %   It is shown in Fig. \al{ref} that this choice, in a suitable range of parameters, produces travelling patterns, where the herders push the targets across the domain creating travelling patterns. Below we report the linear stability analysis whose results are compatible with the emergence of such patterns. 
 %   If $\tilde{v}_1>0$, in all the circumstances, the herders will move "right" as long as they observe a non-zero $\rt(x)$. This will eventually lead to the accumulation of the herders at the "right" of the targets, generating a chase-escape motion to the "left" of herders chasing targets (due to the coupling $\tilde{v}_2\rh\nabla\rt$) and the targets escaping by the herders (due to the coupling $\tilde{k}^{\text{T}} \rt\nabla\rh$).  
\end{itemize}

\subsection{Agent dynamics: decison-making rules}
\label{sec:ab_pattern_design}
%\subsection{From $v_1(x)$ to the distrbuted control rules for the herders}
We now show how to "translate" the design rules represented by the different functions $v_1(x)$ discussed in the preceding section towards the microscopic level of description. To this end we properly design
the feedback control input $\textbf{u}_i$ and combine these expressions with the agent-based Langevin equations given in Eqs.~(7-8)  of the Methods.
To design the control input, we use the same general strategy as in the shepherding case considered in the main text. In particular, the herder's decision-making involves two key elements, namely (i) a selection rule for targets (whose sensitivity is tuned by $\gamma$)
yielding $\textbf{T}_i^{*}$, and (ii) a planning of the trajectory (controlled by $\delta$). The recipes described below are the basis of the agent-based numerical results shown in Fig.~5 of the main text.
All other parameters of the simulations are the same as those used for the simulations used to produce in Fig.~3 of the main text.
%%design distributed microscopic control input for the herders to reproduce the long time  behaviours in \al{ref} starting from the results obtained for different shaphes of the function $v_1(x)$ at the continuum level. We show that such behaviours  can be obtained designing the distributed control law $u_i$ for herder $i$ using the same structure used in the Main, where  herders' strategy to consist of  two key elements: the selection rule of the target to influence ($\gamma$) and the trajectory planning on how to influence the target ($\delta$).

\begin{itemize}
    \item \textbf{Containment}\\
    This is the same task analyzed in the main text (see Fig.~3) with the difference that we here employ a rectangular geometry for the goal region, different from the geometry used in the main text (see Sec. \ref{sec:rec_geom} for the discussion on this point). Specifically, we use the following control input for the herders 
\begin{equation}
    \textbf{u}_i=-\left( \textbf{H}_i-\left(\textbf{T}_i^{*,\text{cont}}+\delta \,\sign(\textbf{T}_{i,x} ^{*,\text{cont}})\right) \right)
    \label{eq:ui_containment}
\end{equation}
where
\begin{equation}
\textbf{T}_i^{*,\text{cont}} = \frac{\sum_{a\in N_{i,\xi}}e^{\gamma (|\textbf{T}_{a,x}|-|\textbf{H}_{i,x}|)} \textbf{T}_{a}}{\sum_{a\in N_{i,\xi}}e^{\gamma (|\textbf{T}_{a,x}|-|\textbf{H}_{i,x}|)}},
\label{eq:selection_containment}
\end{equation}
These rules eventually generate an accumulation of the herders around $x=\pm L/2$, pushing the targets into the region around $x=0$.
\item \textbf{Expulsion}\\
As discussed above at the continuum level (see Sec. \ref{sec:general_v1}), containment around a given position $x^*$ is equivalent to the expulsion from $x^*\pm L/2$ (which are the points with largest distance from $x^*$) due to the periodicity of the domain. As a consequence, to generate expulsion from $x=0$ we design the selection rule so as to select the target with the largest distance from $x^*=L/2$ (which coincides with $x^*=-L/2$). This yields
\begin{equation}
\textbf{T}_{i}^{*,\text{exp}} (\gamma)= \frac{\sum_{a\in N_{i,\xi}}e^{\gamma \left( (L/2-|\textbf{T}_{a,x}|)-(L/2-|\textbf{H}_{i,x}|)\right)} \textbf{T}_{a}}{\sum_{a\in N_{i,\xi}}e^{ \gamma \left( (L/2-|\textbf{T}_{a,x}|)-(L/2-|\textbf{H}_{i,x}|)\right)}} = \frac{\sum_{a\in N_{i,\xi}}e^{-\gamma (|\textbf{T}_{a,x}|-|\textbf{H}_{i,x}|)} \textbf{T}_{a}}{\sum_{a\in N_{i,\xi}}e^{-\gamma (|\textbf{T}_{a,x}|-|\textbf{H}_{i,x}|)}}=\textbf{T}_i^{*,\text{cont}}(-\gamma)
\end{equation}
The third and fourth member of these equations show that selection of target the largest distance from $x^*=L/2$ is equivalent to the target selection rule Eq.~(\ref{eq:ui_containment}) we used for containment, yet with negative $\gamma$.
Indeed, the choice $\gamma<0$ implies selection of the target with the smallest (rather than the largest) distance from the origin. Regarding the trajectory planning, following the same reasoning, we want the herders to place themselves {\em between} the selected target and the origin. 
%since we want to push targets away from $x=0$: 
Therefore, we can again use the same formulation as in the containment case [Eq.~(\ref{eq:ui_containment})], but now the herders place themselves with a shift $-\delta \,\sign(\textbf{T}_{i,x}^*)$. The control input then becomes
\begin{equation}
    \textbf{u}_i=-\left( \textbf{H}_i-(\textbf{T}_i^*-\delta \,\sign\textbf{T}_{i,x}^{*,\text{cont}}(-\gamma) \right)
\end{equation}
with $0<\delta<\lambda$, and $1/\xi\ll\gamma$.

\item \textbf{Static patterns}\\
To create a static stripe-like pattern, we follow the strategy described on the continuum level where $v_1(x)$ was chosen as a square wave whose reflection points correspond to the centers of containment regions. In other words, the task of creating a stripe patterns is divided into several containment tasks, each involving one ``accumulation point'' $x^*$. 
This motivates to formulate the selection rule as follows: If one herder, say herder $i$, observes a target $a$ in its sensing region, it first identifies the accumulation points $x^*$ to which this herder and the target are the closest (among the set of accumulation points). We refer to these two special accumulation points as $x^*_i$ and $x^*_a$, respectively. The selection rule can then be written as
\begin{equation}
\textbf{T}_i^{*} = \frac{\sum_{a\in N_{i,\xi}}e^{\gamma (|\textbf{T}_{a,x}-x_a^*|-|\textbf{H}_{i,x}-x_i^*|)} \textbf{T}_{a}}{\sum_{a\in N_{i,\xi}}e^{\gamma (|\textbf{T}_{a,x}-x_a^*|-|\textbf{H}_{i,x}-x_i^*|)}}
\end{equation}
For the trajectory planning, we accordingly require the herder to place itself at the back of the target with respect to $x_a^*$. We note that $x^*_a$ can vary during the motion of all agents (contrary to the fixed goal regions considered in the other cases). Therefore, we write the control input as
\begin{equation}
    \textbf{u}_i=-\left( \textbf{H}_i-\textbf{H}_i^*(\gamma,\delta) \right)
\end{equation}

where $\textbf{H}_i^*$ is the desired position for herder $i$ defined as
\begin{equation}
    \textbf{H}_i^*(\gamma,\delta) = \frac{\sum_{a\in N_{i,\xi}}e^{\gamma (|\textbf{T}_{a,x}-x_a^*|-|\textbf{H}_{i,x}-x_i^*|)} (\textbf{T}_{a} +\delta\sign(\textbf{T}_{a,x}-x_a)) } {\sum_{a\in N_{i,\xi}}e^{\gamma (|\textbf{T}_{a,x}-x_a^*|-|\textbf{H}_{i,x}-x_i^*|)}}
\end{equation}
Notice that this strategy requires the herders to know the positions of the multiple accumulation points $x^*$.
\item \textbf{Travelling patterns}\\
As discussed in the continuum part, numerical simulations of the field equations with $v_1(x)=\tilde{v}_1$ (as well as the corresponding linear stability analysis) reveal for suitable choices of the constants $\tilde{v}_1$ and $\tilde{v}_2$ a travelling pattern. Specifically, for $\tilde{v}_1>0$, the coupling $\tilde{v}_1\rt\rh$ generates a persistent travelling wave moving to the "left". 
On the agent-based level, we can create such a situation
with a selection rule that favors the target with the largest value of the $x$ coordinate in the sensing region of herder $i$  
\begin{equation}
\textbf{T}_i^{*,\text{trav patt}} = \frac{\sum_{a\in N_{i,\xi}}e^{\gamma (\textbf{T}_{a,x}-\textbf{H}_{i,x})} \textbf{T}_{a}}{\sum_{a\in N_{i,\xi}}e^{\gamma (\textbf{T}_{a,x}-\textbf{H}_{i,x})}},
\end{equation}
We combine this rule with the requirement that a herder always places itself to the ``right'' of the target, in the $x>0$ direction. This is achieved by writing the  distributed control input for herder $i$ as 
\begin{equation}
    \textbf{u}_i=-\left( \textbf{H}_i-(\textbf{T}_i^{*,\text{trav patt}}+\delta) \right)
\end{equation}
\end{itemize}

\subsection{Linear stability analysis for the case of constant coefficients $\tilde{v}_1$ and $\tilde{v}_2$}
\label{sec:LSA_v1v2}
Putting the functions $v_1(x)$ and $v_2(x)$ in the full herder dynamics to constants, we arrive at the coupled PDEs
\begin{align}
    \partial_t \rt=& \nabla\cdot \left[{D}^T(\rt)\nabla\rt+\tilde{k}^T\rt\nabla\rh \right]
    \label{eqn:T_field_v1v2Const}\\
    \partial_t \rh= &\nabla\cdot\left[{D}^H(\rh)\nabla\rh- \tilde{v}_1\rh\rt-\tilde{v}_2\rh\nabla\rt \right]
     \label{eqn:H_field_v1v2Const}
\end{align}
We first note that 
%these equations can be trivially fulfilled with a solution of the form 
a trivial steady state solution for the above equations is given by $\rt(x)=\rt_0$, $\rh(x)=\rh_0$, corresponding to a homogeneous steady state. This is in stark contrast to the case of a space-dependent function $v_1(x)$ that does not allow for such a steady state, as discussed in the main text (see Eq. (6)).

We now perform a linear stability analysis (LSA) around the homogeneous solution, following the procedure outlined in Sec.~\ref{sec:analysis_without_decision} (where we considered the system without decision-making).
%
%First, we notice that the homogeneous state $\rt(x)=\rt_0$; $\rh(x)=\rh_0$ is a steady state, compatibly with Eq. \al{ref} of the Main Manuscript since $v_1(x)=\tilde{v}_1$ is constant. We can then perform the stability analysis around the homogeneous state. Specifically, we start from the state 
%\begin{align}
%    &\rt=\rt_0 +\drt (x,t)\\
%    &\rh=\rh_0 +\drh (x,t)
%\end{align}
Introducing small fluctuations $\delta\rho^{\text{A}}(x,t)$ with
$\text{A}=\text{H,T}$ [see Eq.~(\ref{eq:fluc})] and linearizing Eqs.~(\ref{eqn:H_field_v1v2Const}) 
we obtain
\begin{equation}
\partial_t \drt (x,t)=D^{\text{T}}(\rt_0)\nabla^2\drt (x,t)+ \rt_0\tilde{k}^{\text{T}}\nabla^2 \drh(x,t)
\label{eqn:H_linear_PDE_v1v2}
\end{equation}
\begin{equation}
\partial_t \drh (x,t)=D^{\text{H}}(\rh_0)\nabla^2\drh (x,t) -\tilde{v}_2\rh_0 \nabla^2 \drt(x,t) -\tilde{v}_1\left(\rt_0\nabla\drh(x,t) +\rh_0 \nabla\drt(x,t)\right)
\label{eqn:T_linear_PDE_v1v2}
\end{equation}
Next, we expand the fluctuations in a Fourier series 
[see Eq.~(\ref{eq:ansatz})]. Insertion into
Eqs.~(\ref{eqn:T_linear_PDE_v1v2}) and
(\ref{eqn:H_linear_PDE_v1v2}) brings us to the eigenvalue problem
%
%We assume the perturbation can be written in the form
%\begin{equation}
%    \delta\rho^A(x,t) =\int dq\, e^{iqx+\Omega(q) t} \hat{\rho}^A(q)
%\end{equation}
%where $A=\{T,H\}$; we also assume that the growth rate $\Omega(q)$ is the same for the two fields. We then plug the above relation in Eq. \eqref{eqn:T_linear_PDE_v1v2},\eqref{eqn:H_linear_PDE_v1v2}, and we can study the growth rate solving the following eigenvalues problem
%
\begin{equation}
    -q \begin{bmatrix}
        q D^{\text{T}}(\rt_0) & q \tilde{k}^{\text{T}}\rt_0\\
        -q\tilde{v}_2\rh_0+i\tilde{v}_1\rh_0  &  qD^{\text{H}}(\rh_0)  +i\tilde{v}_1\rt_0
    \end{bmatrix}\textbf{v}(q) = \Omega(q) \textbf{v}(q)
    \label{eqn:eigenproblem_v1v2}
\end{equation}
where we note the different dependency on the wavenumber $q$ as compared to the uncontrolled case [see Eq.~(\ref{eqn:eigenproblem})].
A further difference is that the two matrix elements 
involving $\tilde{v}_1$ are complex (due to the product of densities appearing in the related current). As a consequence, if $\tilde{v}_1\neq 0$, the growth rate
$\Omega(q)$ has always an imaginary part. 

\begin{figure}[htb]
\vspace{1cm}

\begin{overpic}[width=.888\textwidth]{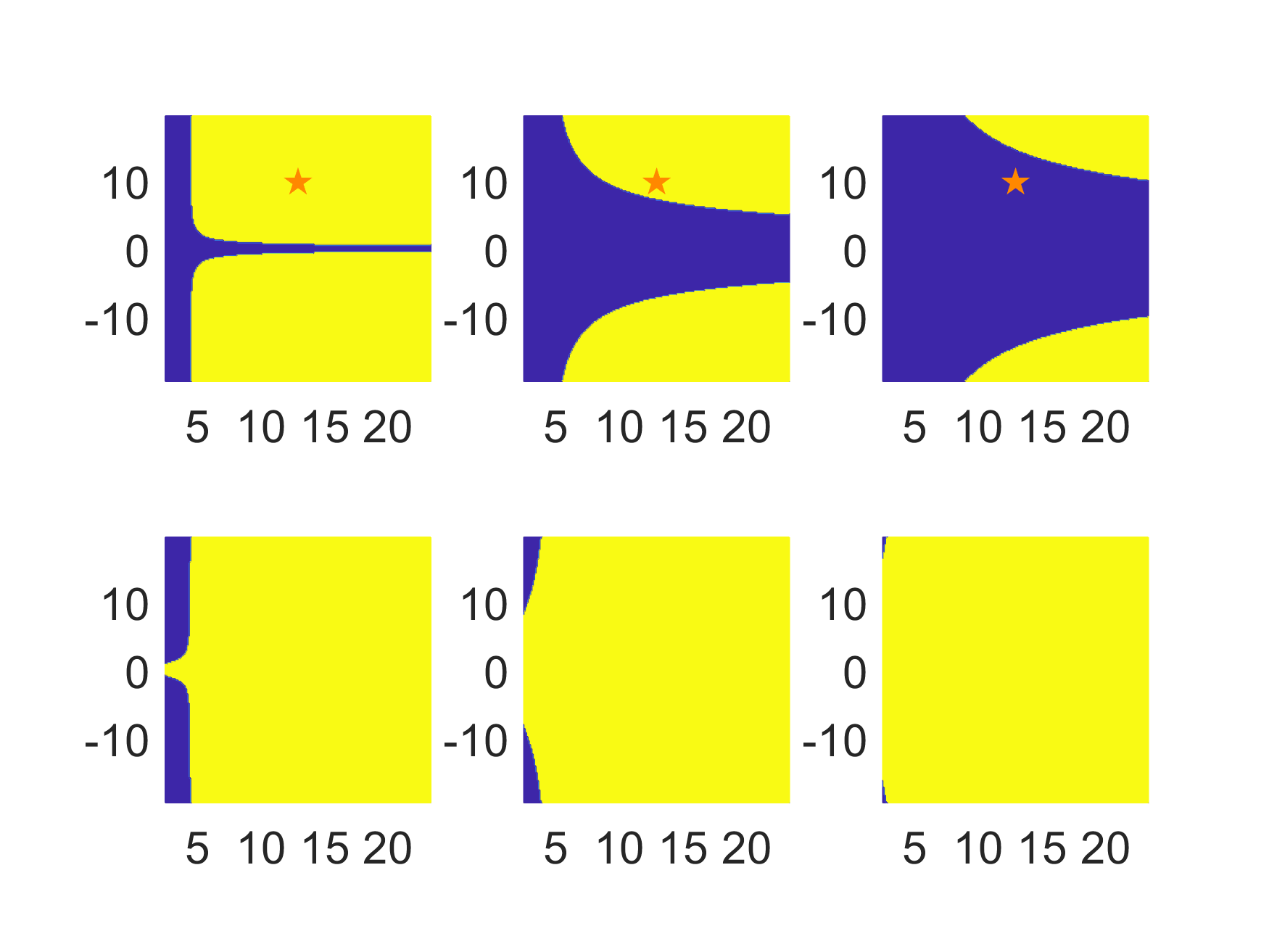}

    \put(10,67){\makebox(0,0){(a)}}
    \put(39,67){\makebox(0,0){(b)}}
    \put(68,67){\makebox(0,0){(c)}}

    \put(10,34){\makebox(0,0){(d)}}
    \put(39,34){\makebox(0,0){(e)}}
    \put(68,34){\makebox(0,0){(f)}}

    \put(24,38){\makebox(0,0){$\tilde k^{\text{T}}$}}
    \put(52,38){\makebox(0,0){$\tilde k^{\text{T}}$}}
    \put(81,38){\makebox(0,0){$\tilde k^{\text{T}}$}}
    \put(24,5){\makebox(0,0){$\tilde k^{\text{T}}$}}
    \put(52,5){\makebox(0,0){$\tilde k^{\text{T}}$}}
    \put(81,5){\makebox(0,0){$\tilde k^{\text{T}}$}}
    
    \put(7,55){\makebox(0,0){\rotatebox{90}{$\tilde v_1$}}} 
    \put(7,22){\makebox(0,0){\rotatebox{90}{$\tilde v_1$}}} 
   
    \put(1,55){\makebox(0,0){\rotatebox{0}{$\Huge \tilde v_2>0$}}} 
    \put(1,22){\makebox(0,0){\rotatebox{0}{$\Huge \tilde v_2<0$}}} 

    \begin{tikzpicture}[overlay]
        \draw[->,ultra thick] (3.5,0.3) -- (13.,0.3); %rtl
    \end{tikzpicture}
    
    \put(51,0){\makebox(0,0){$\Huge q$}}

\end{overpic}

\caption{LSA of the field equations~\ref{eqn:H_field_v1v2Const} and \ref{eqn:T_field_v1v2Const} with respect to a homogeneous steady state. Yellow regions indicate
where one of growth rates obtained from Eq.~\ref{eqn:eigenproblem_v1v2} has a positive real part. We present results for both $\tilde v_2>0$ (a-c) and $\tilde v_2<0$ (d-f),  and for different values of the wavenumber $q$, increased from left to right with values $q=\frac{2\sigma\pi}{L}[1,10,20]$. In all cases, the corresponding imaginary part is non-zero.
(a-c) $\tilde v_2>0$: this is the case discussed in the main text, where herders are attracted to the targets on the field level. The results show that we cannot achieve a positive real part of the growth rates for $\tilde v_1=0$, consistently  with Fig. \ref{fig:LSA_g0d0}.
(d-f) $\tilde v_2<0$: in this case herders and targets both repel each other. The regions of instability increase with $q$, differently from the case $\tilde v_2>0$.
The parameter values used in (a-c) are the same as those used in
Fig.~5(l) of the main text (see Methods), except for the two varied parameters $\tilde k^{\text{T}}$ and $\tilde v_1$. The precise values related to Fig.~5(l) of the main text correspond to the the orange star. To reproduce panels (d-f), the same values are used, except for $k^{\text{H}}$ whose sign needs to be inverted. %\al{COMMENT: in the case $\tilde v_2<0$, the instability is mainly driven by the repulsion between herders and targets; indeed, the larger $q$ the more values of the two parameters here varied provide an unstable homogeneous state; in the case $\tilde v_2<0$, the instability is more likely to be driven by $v1$, as it dies for larger $q$ values (the repulsion provides a $q^2$ term in the Fourier space, whereas the $v1$ term a $q$ dependence).}}
}
\label{fig:LSA_v1v2}
\end{figure}

To decide about the (linear) stability of the homogeneous steady state, we have to consider the real part of the growth rate, 
$\text{Re}(\Omega)(q)$. We have performed extensive numerical investigations for a range of the parameters $\tilde{v}_1$
and the strength of herder-target repulsion, $\tilde{k}^\text{T}$.
Results for three wavenumbers $q$ and two values of $\tilde{v}_2$ (one positive, one negative) are presented in
Fig.~\ref{fig:LSA_v1v2}. In the yellow regions,
$\text{Re}(\Omega)(q)>0$ which, together with the nonzero imaginary parts, signals a temporal (oscillatory) instability. 

We first consider the case $\tilde{v}_2>0$
(top row of Fig.~\ref{fig:LSA_v1v2}), where herders are attracted to the targets (as assumed throughout the rest of this paper). 
The results show that, for sufficiently strong herder-target repulsion $\tilde{k}^{\text{T}}$ (such as the value considered in the main text, indicated by an orange star), an increase of $|\tilde{v}_1|$ from zero can produce a temporal instability. 
An exception is the special case $\tilde{v}_1=0$ where, consistent with our results in Fig.~\ref{fig:LSA_g0d0}, no temporal instability occurs.
We further observe from Fig.~\ref{fig:LSA_v1v2}(a-c) that
the (yellow) regimes indicating positive growth rates shrink with increasing wavenumber (and completely vanish for large enough $q$). This can be understood from the fact that the instability is driven by $\tilde{v}_1$ (providing  a term linear in $q$ in Eq.~(\ref{eqn:eigenproblem_v1v2}) which competes with
all of the other terms that provide a $q^2$-contribution in Fourier space (which, as discussed in Sec.~\ref{sec:analysis_without_decision}, would otherwise make the homogeneous state stable).

Turning to the case $\tilde{v}_2<0$ (where herders are, on average, repelled by targets), the situation is different, as revealed 
by the diagrams in the bottom row of Fig.~\ref{fig:LSA_v1v2}).
Here, it is seen that the system is temporarily unstable even for
$\tilde{v}_1=0$. This already indicates that for $\tilde{v}_2<0$,
the instability is driven by the repulsion between targets and herders (modulated by $\tilde{k}^{\text{T}}$), represented by a term
proportional to $q^2$ in Eq.~(\ref{eqn:eigenproblem_v1v2}).
Accordingly, the larger $q$, the larger the region of the parameters where the homogeneous steady state is unstable.

%We can see in Fig. \ref{fig:LSA_v1v2}  that, in this case, it is possible to have \textit{oscillatory} instabilities, namely growth rates $\Omega(q)$ with both positive real part and non-zero imaginary part; this result is consistent with the observation of travelling patterns in the simulations proposed in Fig. \al{ref} of the Main Manuscript. 

Finally, we recall that travelling patterns are one of the key effects observed in nonreciprocal systems, e.g., in the Cahn-Hillard equations for phase-separating systems if the nonreciprocity in the couplings between the fields is sufficiently pronounced \cite{you2020nonreciprocity,saha2020scalar,fruchart2021non}. 
% However, as we discussed previously, if we derive the mean field equations from a microscopic model whose agents interact through  linear forces, we cannot achieve such an oscillatory instability , unless we assume that one of the two species is \textit{cohesive} (or, in the Cahn-Hilliard model case, that at least one specie is phase-separating or supercritical); note that this is a requirement for the \textit{reciprocal} interactions within the system, that has nothing to do with nonreciprocity.
Here we obtain such an oscillatory instability even the case that there is no source of phase separation (such as attraction between particles of one species).

%the single populations are non-phase separating  (e.g. not relying on some reciprocal intra-specie interaction, such as attraction, as discussed above in Section \al{ref}).

% \begin{figure}[htb]
% \vspace{1cm}

% \begin{overpic}[width=.45\textwidth]{Appendix/RE_v1v2_Const.png}
% \end{overpic}
% \begin{overpic}[width=.45\textwidth]{Appendix/IM_v1v2_Const.png}
% \end{overpic}

% \caption{\al{Update figure.} Growth rates for Eq. \ref{eqn:T_field_v1v2const} and \ref{eqn:H_field_v1v2const}. In (a) and (b) we respectively plot the real and imaginary part of one of the growth rates for different values of the $v_1^0$ coupling. We can see that, according to the LSA, we can have an oscillatory instability.}
% \label{fig:LSA_v1v2}
% \end{figure}

\clearpage

\section{List of Supplementary videos}

\begin{enumerate}
    \item Supplementary Information Video 1: numerical simulations of the agent-based equations (7-8) of the main text, showing the emergence of containment starting from a homogeneous configuration. The parameters $\gamma$ and $\delta$ are set to the largest considered values, namely $\gamma=10/\sigma$, $\delta=\lambda/2$. Blue diamonds represent the herders, magenta dots represent the targets.
    \item Supplementary Information Video 2: Numerical simulations of the field equations (3-4) of the main text showing the emergence of containment at the continuum level starting from a homogeneous state. The parameters $\gamma$ and $\delta$ are set to the largest considered values, namely $\gamma=2.5/\sigma$, $\delta=\lambda/2$. The blue and magenta lines respectively represent $\rh$ and $\rt$.
    \item Supplementary Information Video 3: numerical simulations of the field equations Eq. \eqref{eqn:T_field_v1v2Const} and Eq. \eqref{eqn:H_field_v1v2Const} showing the emergence of traveling patterns from a perturbed homogeneous state. The numerical values of the parameters are presented in the Methods. The blue and magenta lines represent $\rh$ and $\rt$, respectively.
    \item Supplementary Information Video 4: agent-based numerical simulations showing the emergence of traveling patterns (see Section \ref{sec:ab_pattern_design}). Blue diamonds represent the herders, magenta dots represent the targets.
\end{enumerate}

\bibliography{biblio}